\documentclass[twoside,twocolumn,9pt]{article}
\usepackage{extsizes}
\usepackage[super,sort&compress,comma]{natbib} 
\usepackage[version=3]{mhchem}
\usepackage[left=1.5cm, right=1.5cm, top=1.785cm, bottom=2.0cm]{geometry}
\usepackage{balance}
\usepackage{mathptmx}
\usepackage{sectsty}
\usepackage{graphicx} 
\usepackage{bm}
\usepackage{subcaption}
\usepackage{lastpage}
\usepackage[format=plain,justification=justified,singlelinecheck=false,font={stretch=1.125,small,sf},labelfont=bf,labelsep=space]{caption}
\usepackage{float}
\usepackage[normalem]{ulem}
\usepackage{fancyhdr}
\usepackage{fnpos}
\usepackage[english]{babel}
\usepackage{array}
\usepackage{droidsans}
\usepackage{charter}
\usepackage[T1]{fontenc}
\usepackage[usenames,dvipsnames]{xcolor}
\usepackage{setspace}
\usepackage{comment}
\usepackage[compact]{titlesec}
\usepackage{hyperref}
\usepackage{amsmath}
\usepackage{amsfonts}
\usepackage{ulem}
\usepackage{CJKutf8}
\usepackage{braket}
\usepackage{cuted}

\usepackage[percent]{overpic}

\usepackage{epstopdf}

\definecolor{cream}{RGB}{222,217,201}

\begin{document}
\begin{CJK}{UTF8}{min} 
\pagestyle{fancy}
\thispagestyle{plain}
\fancypagestyle{plain}{
\renewcommand{\headrulewidth}{0pt}
}


\makeFNbottom
\makeatletter
\renewcommand\LARGE{\@setfontsize\LARGE{15pt}{17}}
\renewcommand\Large{\@setfontsize\Large{12pt}{14}}
\renewcommand\large{\@setfontsize\large{10pt}{12}}
\renewcommand\footnotesize{\@setfontsize\footnotesize{7pt}{10}}
\makeatother

\renewcommand{\thefootnote}{\fnsymbol{footnote}}
\renewcommand\footnoterule{\vspace*{1pt}%
\color{cream}\hrule width 3.5in height 0.4pt \color{black}\vspace*{5pt}} 
\setcounter{secnumdepth}{5}

\makeatletter 
\renewcommand\@biblabel[1]{#1}            
\renewcommand\@makefntext[1]%
{\noindent\makebox[0pt][r]{\@thefnmark\,}#1}
\makeatother 
\renewcommand{\figurename}{\small{Fig.}~}
\sectionfont{\sffamily\Large}
\subsectionfont{\normalsize}
\subsubsectionfont{\bf}
\setstretch{1.125} 
\setlength{\skip\footins}{0.8cm}
\setlength{\footnotesep}{0.25cm}
\setlength{\jot}{10pt}
\titlespacing*{\section}{0pt}{4pt}{4pt}
\titlespacing*{\subsection}{0pt}{15pt}{1pt}

\fancyfoot{}
\fancyfoot[LO,RE]{\vspace{-7.1pt}\includegraphics[height=9pt]{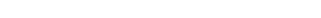}}
\fancyfoot[CO]{\vspace{-7.1pt}\hspace{13.2cm}\includegraphics{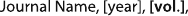}}
\fancyfoot[CE]{\vspace{-7.2pt}\hspace{-14.2cm}\includegraphics{head_foot/RF}}
\fancyfoot[RO]{\footnotesize{\sffamily{1--\pageref{LastPage} ~\textbar  \hspace{2pt}\thepage}}}
\fancyfoot[LE]{\footnotesize{\sffamily{\thepage~\textbar\hspace{3.45cm} 1--\pageref{LastPage}}}}
\fancyhead{}
\renewcommand{\headrulewidth}{0pt} 
\renewcommand{\footrulewidth}{0pt}
\setlength{\arrayrulewidth}{1pt}
\setlength{\columnsep}{6.5mm}
\setlength\bibsep{1pt}

\makeatletter 
\newlength{\figrulesep} 
\setlength{\figrulesep}{0.5\textfloatsep} 

\newcommand{\topfigrule}{\vspace*{-1pt}%
\noindent{\color{cream}\rule[-\figrulesep]{\columnwidth}{1.5pt}} }

\newcommand{\botfigrule}{\vspace*{-2pt}%
\noindent{\color{cream}\rule[\figrulesep]{\columnwidth}{1.5pt}} }

\newcommand{\dblfigrule}{\vspace*{-1pt}%
\noindent{\color{cream}\rule[-\figrulesep]{\textwidth}{1.5pt}} }

\makeatother

\twocolumn[
  \begin{@twocolumnfalse}
{\includegraphics[height=30pt]{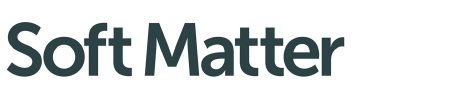}\hfill\raisebox{0pt}[0pt][0pt]{\includegraphics[height=55pt]{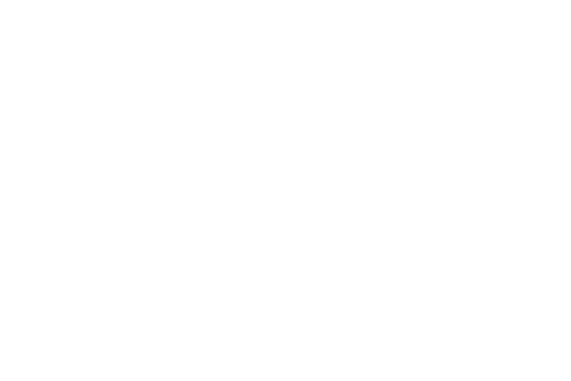}}\\[1ex]
\includegraphics[width=18.5cm]{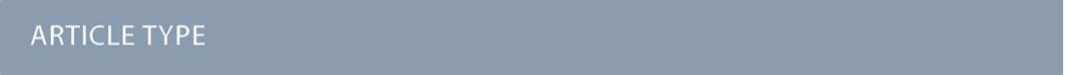}}\par
\vspace{1em}
\sffamily
\begin{tabular}{m{4.5cm} p{13.5cm} }

\includegraphics{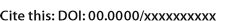} & \noindent\LARGE{\textbf{
Weakly Nonlinear Analysis of Jammed Solids}} \\
\vspace{0.3cm} & \vspace{0.3cm} \\

 & \noindent\large{Kota Noto\textit{$^{a}$}, Harukuni Ikeda\textit{$^{a}$}, Kuniyasu Saitoh\textit{$^{b}$}, and Hisao Hayakawa\textit{$^{a}$}} \\

\includegraphics{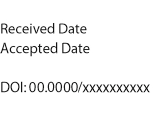} & \noindent\normalsize{
We investigate the structural evolution in jammed solids by analyzing the eigenmodes of a generalized Hessian matrix that incorporates spatial modulation via wave numbers. 
Unlike the conventional Hessian, this generalized formulation captures linearly unstable modes through a Fourier-based extension of the Hessian matrix, enabling us to study responses beyond the mechanically stable regime. 
We demonstrate that the excitation of unstable eigenmodes leads to structural rearrangements independent of the initial perturbation by the simulation. 
Furthermore, we derive a weakly nonlinear amplitude equation to describe the growth and saturation of these unstable modes, analogous to the Landau equation. 
Our framework provides a pathway to understand instability-driven configuration changes in disordered solids.

}

\end{tabular}

 \end{@twocolumnfalse} \vspace{0.6cm}

  ]

\renewcommand*\rmdefault{bch}\normalfont\upshape
\rmfamily
\section*{}
\vspace{-1cm}

\footnotetext{\textit{$^{a}$~Center for Gravitational Physics and Quantum Information,
Yukawa Institute for Theoretical Physics, Kyoto University, Kitashirakawa-Oiwake cho, Sakyo-ku, Kyoto 606-8502, Japan}}
\footnotetext{\textit{$^{b}$~Department of Physics, Faculty of Science, Kyoto Sangyo University, Motoyama, Kamigamo, Kita-ku, Kyoto 603-8555, Japan }}


\section{Introduction}
Jammed solids composed of repulsive particles, including powders, colloids, and emulsions, are rigid but mechanically fragile near the jamming transition~\cite{Jaeger96,Durian95,Wyart05,Baule18,Liu98,OHern2003JammingZeroTmp,Hecke2010,Liu2010,Zaccone2023,Hayakawa2025}. Due to fragility, their internal structures are easily altered by external forces, even when such forces are relatively weak. Understanding how the structure of these disordered materials evolves under applied forces is therefore a central issue in the physics of jammed solids.

A confined jammed solid in a box remains mechanically stable due to its rigidity. A key tool for analyzing such stability is the Hessian matrix, defined as the second derivative of the potential energy with respect to particle displacements. This matrix characterizes the local curvature of the energy landscape. For a mechanically stable configuration, its eigenvalues are non-negative, and the corresponding eigenmodes are commonly used to describe the mechanical response of disordered solids to perturbations.
In mechanically stable regimes, the response to external deformation, known as the non-affine deformation, can be expressed as a superposition of the eigenmodes of the Hessian matrix~\cite{C.Maloney2006AQS,C.Maloney2004Universal,Lemaitre2006SumRules,Zaccone2023,Hayakawa2025}. Such eigenmode analysis enables quantitative predictions of elastic moduli~\cite{C.Maloney2006AQS,C.Maloney2004Universal,Lemaitre2006SumRules,Zaccone2023,Ishima2023a,D.Ishima2023Eigenvlue,Hayakawa2025,Hara25,Mizuno2016}.
We note that when interparticle friction is included, the Hessian matrix is replaced by the corresponding Jacobian matrix~\cite{Ishima2023a,D.Ishima2023Eigenvlue}.

From a different angle, it is known that the smallest eigenvalue of the Hessian matrix approaches zero just before a plastic event during shear deformation~\cite{C.Maloney2004Universal,L.Gartner2016NonlinerPlasticMode,S.Bonfanti2019ElementaryPlasticEvents}. However, such precursors cannot be observed in harmonic systems~\cite{D.Ishima2023Eigenvlue,D.Xu2023DiscontinuousInstability}. 
This behavior has inspired the identification of {\it soft spots} based on quasi-localized low-frequency vibrational modes~\cite{L.Manning2011SoftSpots,E.Lerner2016Micromechanics,G.Kapteijns2020NQE,H.Mizuno2017ContinuumLimit}, which have been shown to correlate with impending plastic rearrangements.

In quasi-static simulations, such as athermal quasi-static (AQS) protocols~\cite{Malandro1998,C.Maloney2006AQS}, the system remains mechanically stable during Hessian analysis, provided the simulation box itself is not deformed.
However, when the simulation box undergoes deformation (e.g., under a nonzero shear strain), discontinuous changes of stress and energy can be observed. 
The analysis based on the Hessian matrix is useful to describe highly deformed systems, except for plastic events   ~\cite{C.Maloney2006AQS, Lemaitre2006SumRules, D.Ishima2023Eigenvlue, S.Bohy2012SoftSpher, C.Goodrich2014JammingInFiniteSytems, D.Xu2024Lowfrequency, S.Chakraborty2025Instabilities, S.Torquato2001Multiplicity}.
Recent studies have further shown that topological defects appearing in Hessian eigenmodes of the undeformed configuration are closely related to subsequent plastic rearrangements in amorphous solids~\cite{Wu2023}.
This raises a natural question: Can the Hessian matrix analysis be extended to regimes where the system becomes unstable?

Remarkably, some previous studies have introduced a generalized Hessian matrix by incorporating modulation via wave numbers, enabling the detection of unstable modes~\cite{S.Schoenholz2013Stability,R.Dennis2022Emergence}.

This analysis introduces a Bloch-like wave representation, inspired by crystalline solids, to characterize spatial modulations of disordered particle configurations. In this framework, structural disorder is contained within a unit cell, whereas modulations on length scales larger than the cell size are represented by a wave number.
See also trials to characterize amorphous solids using crystals~\cite{Otsuki23,Pashine23,J.Zhang2023Tessellated,Suzuki25}, and crystals including topological defects~\cite{Huang2025,Baggioli2022,P.Desmarchelier2024TopologicalCharacterization,Salman25,Nakai25}.
This approach reveals that the generalized Hessian, incorporating spatial modulation, is effective in describing non-uniform particle arrangements without requiring explicit consideration of box deformations.
However, a direct reconstruction of unstable structures from eigenmodes in such regimes has not yet been systematically explored.
In this sense, a systematic understanding of how unstable eigenmodes reconstruct the evolving structure is still lacking.
In the present framework, the Bloch-like representation is made well defined by periodically replicating a unit cell containing a disordered jammed configuration. Consequently, the resulting system possesses periodicity on scales larger than the unit cell, while retaining structural disorder within the unit cell.

Weakly nonlinear analysis is a powerful tool for pattern formation~\cite {Cross}, dissipative structures and weak turbulence~\cite{Manneville}, chemical oscillations~\cite{Kuramoto}, relatively dilute granular flows~\cite{Shakla09,Saitoh11}, and pattern dynamics of granular media~\cite{Aranson2006}.
Nevertheless, weakly nonlinear analysis for densely packed jammed systems was only applied to stable cases, such as Refs.~\cite{L.Gartner2016NonlinerPlasticMode,G.Kapteijns2020NQE,E.Lerner2016MicromechanicsOfNonliner,D.Richard2024ConnectingMicroAndMeso}.
This may be because the conventional Hessian analysis is mainly formulated
around mechanically stable configurations.
Here, we instead focus on configuration changes in jammed solids induced
by the growth of unstable modes after a small perturbation.

Here, we combine the wave-number-dependent generalized Hessian of Refs.~~\cite{S.Schoenholz2013Stability,R.Dennis2022Emergence} with a weakly nonlinear analysis to characterize the growth and saturation of finite-wave-number unstable modes~\cite{Cross,Manneville,Kuramoto,Shakla09,Saitoh11,Aranson2006}.
Thus, this study aims to investigate structural changes associated with eigenmodes of the generalized Hessian that have negative eigenvalues,
focusing on the time evolution of these unstable modes.
We consider a periodic structure constructed by replicating a unit cell with an internally disordered particle configuration, as in Ref.~\cite{S.Schoenholz2013Stability}.
While the overall system obeys periodic boundary conditions, the particles within each cell are allowed to deform differently from those in neighboring cells. 
This setup allows us to introduce well-defined wave numbers and examine spatially modulated instabilities. However, these wave numbers are defined based on a periodic replication of cells, and thus do not correspond, in a strict sense, to wave numbers in a fully disordered amorphous solid.
The advantage of our approach is that we need only one initial configuration without deformations, which is unlike a standard extension of the Hessian matrix~\cite{C.Maloney2006AQS,Lemaitre2006SumRules,D.Ishima2023Eigenvlue,S.Bohy2012SoftSpher,C.Goodrich2014JammingInFiniteSytems,S.Torquato2001Multiplicity} that requires updated configurations of particles under external deformations.

The remainder of this paper is organized as follows. Section~\ref{sec:ModelMethod} introduces the model, notation, and linear theory based on the generalized Hessian matrix. Section~\ref{sec:Nonlinear} develops the weakly nonlinear theory and derives the corresponding amplitude equation. 
Section~\ref{sec:Num} presents numerical validations and compares the theoretical predictions with simulation results. 
Finally, Section~\ref{sec:Conclusion} summarizes the main findings and discusses possible directions for future work.
The theoretical details are explained in the Appendices.

\section{Model and linear theory}\label{sec:ModelMethod}
In this section, we first introduce the model and notation used throughout the paper.
We then formulate the linear stability analysis based on a Fourier representation over cell positions and the generalized Hessian matrix~\cite{S.Schoenholz2013Stability}.
This framework allows us to characterize unstable modes through the dispersion relation and to derive the corresponding linear response.

\begin{figure}[h]
    \centering
    \includegraphics[trim=80 50 70 50, clip,width=0.8\linewidth]{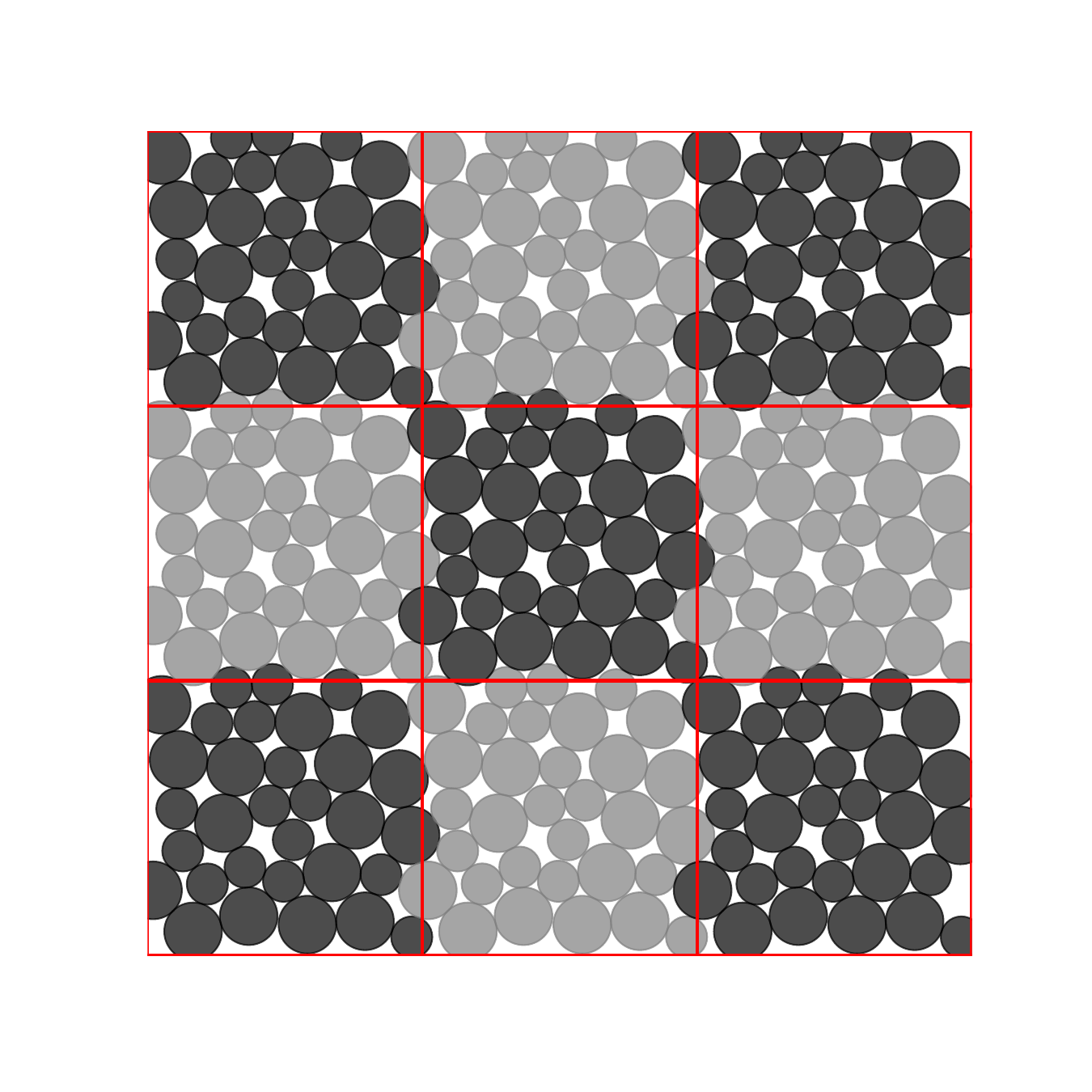}
    \caption{An example of a tile structure consisting of a 3×3 array of cells.  
Each cell consists of a packing of 32 particles.
}
    \label{fig:tile}
\end{figure}

\subsection{Outline of the model}

We consider two-dimensional jammed solids composed of an equal number of bidisperse, equal-mass circles with a diameter ratio of $d_S : d_L = 1 : 1.4$, chosen to prevent crystallization~\cite{R.Speedy1999GlassTransition}.
We assume that all grains are frictionless and thermal fluctuations are negligible.
We first prepare a disordered jammed configuration containing $N$
particles in a unit cell and periodically replicate this configuration
(see Fig.~\ref{fig:tile}).
Thanks to this mesoscopic periodic-cell construction, we introduce a Fourier representation over cell positions to describe long-wavelength modulations in a Bloch-like manner, while retaining the disordered structure within each cell.
We analyze the stability of this cell structure and identify unstable modes.
We denote the position of particle $i \in 1, 2, \dots, N$ in the cell $\bm{\mu}=(\mu_x,\mu_y) \in \mathbb{Z}^2 $as $\bm{r}_i^{\bm{\mu}}$.
The corresponding particle
\(\bm r_i^{\bm\nu}\), with \(\bm\nu\neq\bm\mu\), is regarded as a mirror
particle in another cell.
The cell indices $\bm{\mu}$ form a square lattice.
Each cell contains $N$ particles and its linear size is $L$.
We assume that large-scale modulations can be described by trigonometric functions satisfying periodic boundary conditions.

Let us consider the overdamped dynamics given by
\begin{equation}\label{overdamped_dyn}
   \eta \dot{\bm{r}}_i^{\bm{\mu}}  = -  \frac{\delta V}{\delta \bm{r}_i^{\bm{\mu}}}  ,
\end{equation}
where $\eta$ is the viscosity of the background fluid controlling the time scale of the overdamped dynamics, and $V$ is the total potential energy. 
We have used $\delta V/\delta \bm{r}_i^{\bm{\mu}}$ to stress the differentiation with respect to $\bm{r}_i$ in a specific $\bm{\mu}$.
The potential energy is written as 
\begin{equation}\label{potential}
V:
=\sum_{\langle\!\langle\bm{\zeta}\bm{\xi}\rangle\!\rangle}\sum_{\langle k,\ell \rangle}
U(r_{k \ell}^{\bm{\zeta}\bm{\xi}})
\end{equation}
with the pairwise potential $U(r_{k\ell}^{\bm{\zeta}\bm{\xi}})$.
The double bracket
\(\langle\!\langle\bm\zeta\bm\xi\rangle\!\rangle\)
denotes a cell pair
\(\{\bm\zeta,\bm\xi\}\in\mathcal N_{\rm c}\),
where \(\mathcal N_{\rm c}\) is the set of cell pairs such that either
\(\bm\zeta=\bm\xi\) or the distinct cells
\(\bm\zeta\) and \(\bm\xi\) are in contact.
For each such cell pair,
$\langle k,\ell\rangle$ denotes the corresponding contacting particle pairs,
with self-pairs excluded when $\bm{\zeta}=\bm{\xi}$.
The relative distance is defined as
$
r_{k\ell}^{ \bm{\zeta}\bm{\xi}}
:=\left|\bm r_k^{\bm\zeta}-\bm r_\ell^{\bm\xi}\right|$.
Since we consider particles with short-range repulsive interactions characterized by the contacts, the contributions from different cells, i.e., $\bm{\zeta}\ne \bm{\xi}$, are only from particles located on the boundaries of the cells $\bm{\zeta}$ and $\bm{\xi}$.

In this study, we assume that particles at positions $\bm{r}_i^{\bm{\mu}}$ and $\bm{r}_j^{\bm{\nu}}$ interact via a harmonic potential:
\begin{equation}\label{harmonic_pot}
U(r_{ij}^{\bm\mu\bm\nu})
:=
\begin{cases}
\displaystyle
\frac{\kappa}{2}
\left(
d_{ij}-r_{ij}^{\bm\mu\bm\nu}
\right)^2
\Theta\!\left(
d_{ij}-r_{ij}^{\bm\mu\bm\nu}
\right),
&
\{\bm\mu,\bm\nu\}\in\mathcal N_{\rm c},
\\[6pt]
0,
&
\{\bm\mu,\bm\nu\}\notin\mathcal N_{\rm c},
\end{cases}
\end{equation}
where \(d_{ij}:=(d_i+d_j)/2\) is the average diameter of particles \(i\) and \(j\), \(\kappa\) is the stiffness constant, and \(\Theta(x)\) is the Heaviside step function, defined as \(\Theta(x)=1\) for \(x\ge0\) and \(\Theta(x)=0\)
otherwise.
Note that most of the interacting pairs belong to the same cell, and only particles located on the boundaries of the cell interact with particles in different cells. 

In this study, we perform pressure-controlled simulations.
See Appendix~\ref{app:preparation} for details of the protocol used to prepare the initial configurations.
Throughout this paper, Greek indices $\bm{\mu,\nu,\zeta,\xi}$ denote cell labels, Latin indices $i,j,k,\ell$ denote particle labels within a cell, and Latin indices $a,b$ denote Cartesian components. 
Boldface symbols denote vectors in real and reciprocal space.

\subsection{Generalized Hessian and Fourier Representation}

Let displacement from the force-balanced configuration be $\delta \bm{r}_i^{\bm{\mu}}(t)$.
The linearized equation of eqn.~\eqref{overdamped_dyn} around an equilibrium configuration is given by
\begin{equation}\label{eq:motion_eq_v2}
\eta \frac{d}{dt} \delta \bm{r}_i^{\bm{\mu}}(t) = -\sum_{\bm{\nu},j} \mathsf{H}_{ij}^{\bm{\mu}\bm{\nu}} \delta \bm{r}_j^{\bm{\nu}}(t),
\end{equation}
where the summation over \(\bm\nu\) is restricted to cell pairs
\(\{\bm\mu,\bm\nu\}\in\mathcal N_{\rm c}\),
and 
$\mathsf{H}_{ij}^{\bm{\mu}\bm{\nu}}$ is defined as
\begin{equation}
\mathsf{H}_{ij}^{\bm{\mu}\bm{\nu}}:=\left(
\frac{\delta^2 V}{\delta \bm{r}_i^{\bm{\mu}} \delta \bm{r}_j^{\bm{\nu}}}
\right)_0 .
\end{equation}
The explicit form of the Hessian matrix is provided in Appendix~\ref{Ap:Hij}.
Let us introduce the Fourier transform over cell positions as
\begin{equation}\label{eq:Fourier_transf}
\hat{\bm{r}}_i(\bm{k},t) := \sum_{\bm{\mu}} e^{i \bm{k} \cdot \bm{R}^{\bm{\mu}}} \delta \bm{r}_i^{\bm{\mu}}(t)
\end{equation}
with inverse transform
\begin{equation}\label{eq:Fourier_Inv_transf}
\delta \bm{r}_i^{\bm{\mu}}(t) := \frac{1}{\Omega_{\mathrm{BZ}}}\int_{BZ}   d^2\bm{k} e^{-i \bm{k} \cdot \bm{R}^{\bm{\mu}}} \hat{\bm{r}}_i(\bm{k},t),
\end{equation}
where $\bm{R}^{\bm{\mu}}$ is the center position of the cell $\bm{\mu}$, chosen as $\bm R^{\bm\mu}=L(\mu_x\bm e_x+\mu_y\bm e_y)$ and $\bm e_a$ denotes the unit vector in the $a$-direction.
Here, we have introduced the wave vector $\bm{k}=k_x\bm{e}_x+k_y\bm{e}_y$. 
The integral $\int_{BZ}$ indicates integration over the first Brillouin zone (a single primitive cell in $\bm{k}$-space), and $\Omega_{\mathrm{BZ}}:= (2\pi)^2/L^2$ is the Brillouin zone area.

Substituting eqn.~\eqref{eq:Fourier_transf} into eqn.~\eqref{eq:motion_eq_v2}, with the aid of the decoupling approximation, we obtain an approximate equation (see Appendix~\ref{app:Fourier_4} for the derivation):
\begin{align}\label{eq_of_Fourier2}
\eta \frac{d}{dt} \hat{\bm{r}}_i(\bm{k},t) \approx  -\sum_j \mathsf{D}_{ij}(\bm{k}) \hat{\bm{r}}_j(\bm{k},t),
\end{align}
where $\sf{D}_{ij}(\bm{k})$ is the generalized Hessian matrix:~\cite{S.Schoenholz2013Stability}
\begin{align}\label{D_{ij}}
\mathsf D_{ij}(\bm k)
&:=
\sum_{\bm\Delta_{\bm{\mu\nu}}\in\mathcal N}
\mathsf H_{ij}^{ \bm 0,-\bm\Delta_{\bm{\mu\nu}}}\,e^{i\bm k\cdot \bm L_{\bm\Delta_{\bm{\mu\nu}}}}
\notag\\
&=
\sum_{\bm\Delta_{\bm{\mu\nu}}\in\mathcal N}
\mathsf H_{ij}^{ \bm 0,-\bm\Delta_{\bm{\mu\nu}}}
+
\sum_{\bm\Delta_{\bm{\mu\nu}}\in\mathcal N}
\mathsf H_{ij}^{ \bm 0,-\bm\Delta_{\bm{\mu\nu}}}\Bigl(e^{i\bm k\cdot \bm L_{\bm\Delta_{\bm{\mu\nu}}}}-1\Bigr) .
\end{align}
Here, \(\bm\Delta_{\bm{\mu\nu}}:=\bm\mu-\bm\nu\) denotes the relative cell-offset
vector used in the Fourier representation. After shifting the reference cell
\(\bm\mu\) to \(\bm0\), the interacting cell \(\bm\nu\) is located at
\(\bm\nu-\bm\mu=-\bm\Delta_{\bm{\mu\nu}}\). Therefore, using the assumed
translational invariance of the Hessian blocks,
\(\mathsf H_{ij}^{\bm\mu\bm\nu}\) is represented as
\(\mathsf H_{ij}^{\bm0,-\bm\Delta_{\bm{\mu\nu}}}\), which couples particle \(i\)
in the reference cell \(\bm0\) to particle \(j\) in the cell
\(-\bm\Delta_{\bm{\mu\nu}}\). The summation is performed over the relative offsets
\(\bm\Delta_{\bm{\mu\nu}}\in\mathcal N\), not over the absolute cell labels.
The set \(\mathcal N\) is defined as
\begin{align}
    \mathcal N &:= \{\bm 0\}\cup\mathcal N_1 \cup\mathcal N_2,
\end{align}
where we have introduced
\begin{align}\label{def:N_1}
\mathcal N_1 &:= \{\pm \bm e_x,\ \pm \bm e_y\}, \\
\mathcal N_2 &:= \{\bm e_x+\bm e_y,\ \bm e_x-\bm e_y,\ -\bm e_x+\bm e_y,\ -\bm e_x-\bm e_y \}.
\end{align}
For each \(\bm\Delta_{\bm{\mu\nu}}\), the corresponding
cell-translation vector is defined as
\(\bm L_{\bm\Delta_{\bm{\mu\nu}}}:=L\bm\Delta_{\bm{\mu\nu}}\).
With this notation, the term with \(\bm\Delta_{\bm{\mu\nu}}=\bm0\) represents the
in-cell contribution, where both particles belong to the reference cell. The
terms with \(\bm\Delta_{\bm{\mu\nu}}\neq\bm0\) represent inter-cell contributions arising
from contacts across the cell boundaries, which correspond to the usual
periodic-boundary contributions at \(\bm k=\bm0\). For finite \(\bm k\),
these inter-cell contributions acquire the phase factor
\(e^{i\bm k\cdot\bm L_{\bm\Delta_{\bm{\mu\nu}}}}\), which encodes the relative phase
between neighboring cells.
Although eqns.~\eqref{eq_of_Fourier2} and \eqref{D_{ij}} are based on the decoupling approximation, this treatment does not introduce significant errors, as shown in Appendix~\ref{app:Fourier_4}.

This formulation is inspired by Bloch's theorem, which is strictly valid for
crystalline systems.
In the present case, its use is only justified approximately under the following assumptions:
(i) The disordered medium can be partitioned into quasi-periodic cells.
(ii) The particle configuration in a cell can be replicated in the other cells.
(iii) Inter-cell
inhomogeneity
is sufficiently weak so that off-diagonal wave-vector interactions can be neglected.
(iv) The linear response is dominated by a single $\bm{k}$ mode. 
Assumptions (i) and (ii) are idealizations and are not expected to hold
exactly in realistic amorphous solids.

While the generalized Hessian $\sf{D}_{ij}(\bm{k})$ is not derived from first principles for disordered systems, we find numerically that it successfully captures instability and mode growth for perfectly replicated systems. 
Thus, we regard it as an effective, empirical tool for analyzing emergent collective behavior.

Using the generalized Hessian matrix, the corresponding eigenvalue equation becomes:~\cite{S.Schoenholz2013Stability}
\begin{equation}
\mathsf{D}(\bm{k})\hat{\bm{\varepsilon}}_n(\bm{k})= \lambda_n(\bm{k}) \hat{\bm{\varepsilon}}_n(\bm{k}) ,
\label{eq:eigenEq}
\end{equation}
where $\lambda_n(\bm{k})$ and $\bm{\varepsilon}_n(\bm{k})$ are the $n$-th eigenvalue and eigenvector of $\mathsf{D}(\bm{k}):=[\mathsf{D}_{ij}]^{i,j=1,\cdots,N}$. 
For central-force interactions, $\mathsf D(\bm k)$ is Hermitian, $\mathsf D(\bm k)=\mathsf D^\dagger(\bm k)$, 
and therefore its eigenvectors $\hat{\bm\varepsilon}_n(\bm k)$ form an orthonormal basis satisfying $\hat{\bm\varepsilon}_n^\dagger(\bm k)\hat{\bm\varepsilon}_m(\bm k)=\delta_{nm}$ with Kronecker's delta, where we have used the standard Hermitian inner product $ \bm u^\dagger \bm v$ with the conjugate transpose $(\cdot)^\dagger$.
For each wave vector $\bm{k}$, the generalized Hessian matrix $\mathsf{D}(\bm{k})$ is a finite-dimensional Hermitian matrix of size $2N\times2N$. 
Its eigenvalue problem therefore yields a discrete set of \(2N\) eigenvalues. In the presence of floating particles, some of these eigenvalues correspond to zero modes associated with particles having no contacts. Throughout this study, we exclude such floating-particle zero modes from the spectrum of interest. The remaining eigenvalues are ordered in ascending order and are denoted again by \(\{\lambda_n(\bm{k})\}\). Accordingly, \(\lambda_1(\bm{k})\) denotes the lowest eigenvalue excluding the zero modes associated with floating particles.
Apart from symmetry-enforced degeneracies, such as the translational zero
modes at \(\bm{k}=\bm{0}\), eigenvalue degeneracies are not generically
expected at a fixed wave vector in the present two-dimensional disordered
system. The disordered reference structure does not possess crystalline
point-group symmetries that would enforce such degeneracies
\cite{Dresselhaus2008}.
A brief explanation is given in Appendix~\ref{app:symmetry}.
In the numerical examples analyzed in this study,
the lowest eigenvalue at each selected wave vector is numerically found to
be nondegenerate.

\subsection{Dispersion relation and unstable modes}\label{sec:dis}


\begin{figure}[thbp]
   \centering
    \includegraphics[width=0.49\textwidth]{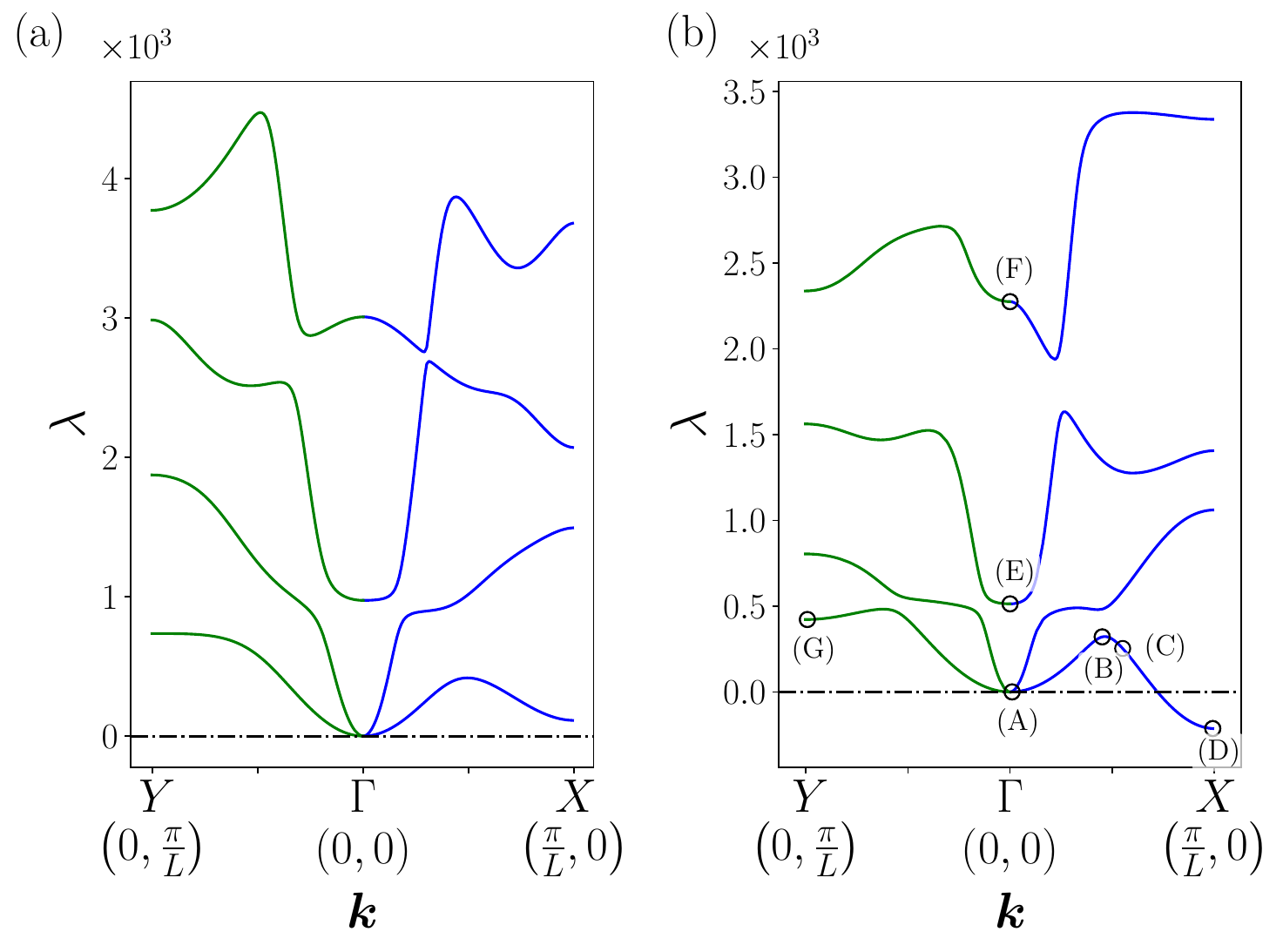}
    \caption{
The four lowest eigenvalue branches along the $k_x\bm{e}_x$-direction (blue) and the $k_y\bm{e}_y$-direction (green) for $P/\kappa=5\times10^{-4}$ and $N=128$. All branches are non-negative in (a), whereas the lowest branch becomes negative near the $X$ point in (b).
    The labeled points (A)--(G) in panel (b) indicate the wave
numbers at which the corresponding eigenvectors are shown in
Fig.~\ref{fig:eigenvec} as panels (a)--(g), respectively.
        }
    \label{fig:dispersion}
\end{figure}

By considering a perfectly replicated set of an infinite number of cells (see Fig.~\ref{fig:tile}), we can evaluate the dispersion relation—that is, the relationship between the eigenvalue $\lambda_n(\bm{k})$ and the wave vector $\bm{k}$.
Although the eigenvalues of the standard Hessian matrix cannot be negative at a mechanically stable configuration,
negative eigenvalues $\lambda_n(\bm{k})$ at $\bm{k}\ne \bm 0$ are allowed in eqn.~\eqref{eq:eigenEq}, provided that $\lim_{|\bm{k}|\to 0}\lambda_n(|\bm{k}|)\ge 0$. 
In this case, the eigenvalue equation describes 
unstable modes with finite spatial modulation, which may be useful for understanding structural changes in disordered solids under perturbations.\footnote{
The instability of the shortest scale may come from a mismatch of particle configuration between the nearest-neighboring cells, since we use a perfectly replicated configuration.
}

\begin{figure}[thbp]
   \centering
    \includegraphics[width=0.49\textwidth]{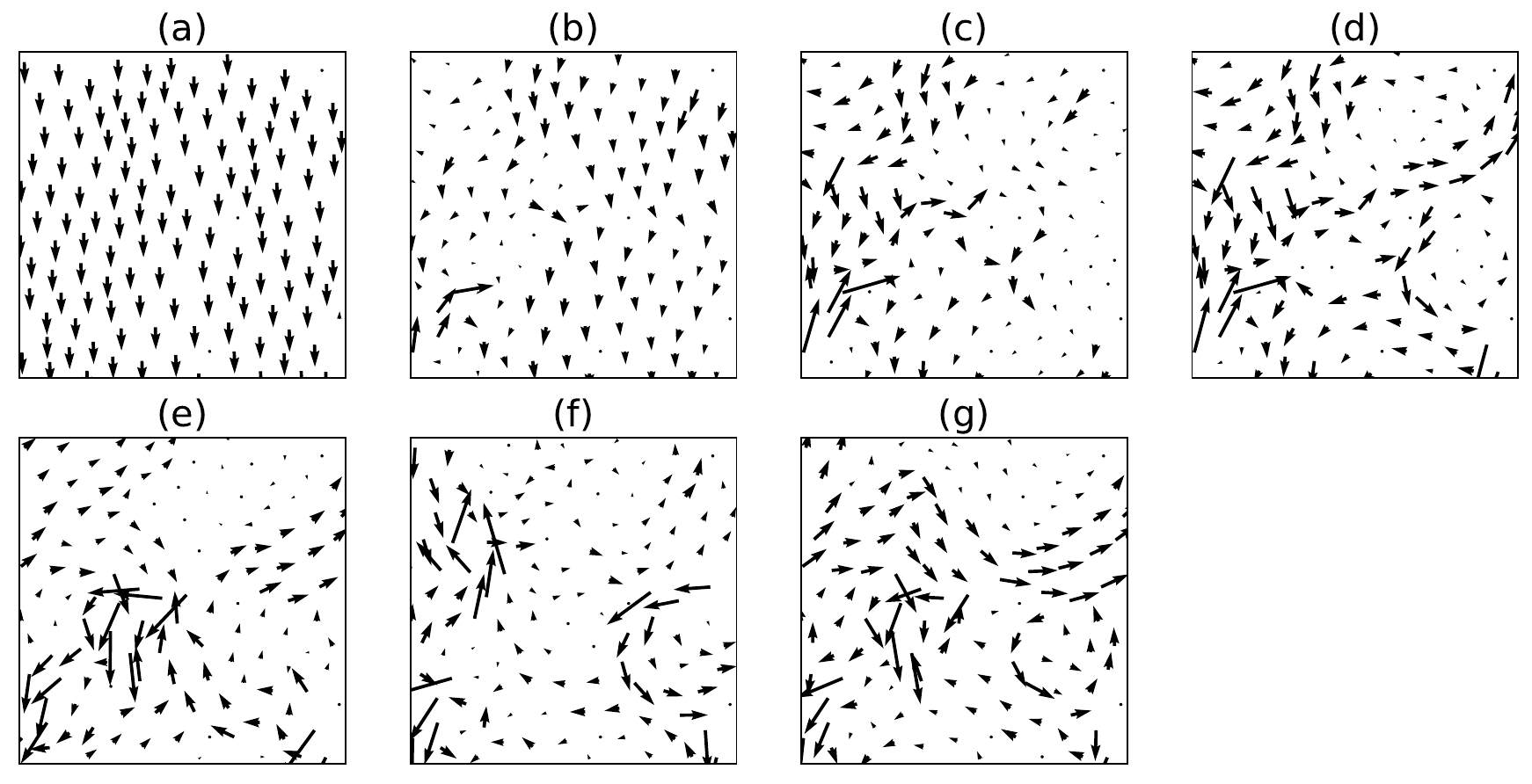}
    \caption{
    Eigenvectors corresponding to the labeled points (A)--(G) in Fig.~\ref{fig:dispersion}(b), shown here as panels (a)--(g), respectively. Panels (a)–(d) show the eigenvectors of the lowest branch in the $k_x\bm{e}_x$ direction at wave numbers (a) $k_x = 0$, (b) $k_x = 9\pi/20L$, (c) $k_x = 11\pi/20L$, and (d) $k_x = \pi/L$, respectively.
Panels (e) and (f) depict the eigenmodes corresponding to the third-lowest and fourth-lowest branches at $\bm{k} = 0$, i.e., the lowest nonzero eigenmode and the second-lowest nonzero eigenmode.
Panel (g) shows the eigenvector of the lowest branch in the $k_y\bm{e}_y$ direction at $k_y = \pi/L$.
        }
    \label{fig:eigenvec}
\end{figure}

Since we focus on the response to perturbations in a periodic structure, we draw the dispersion relation only within the first Brillouin zone. 
The dispersion relation generally depends on the direction of the wave vector.
Figure~\ref{fig:dispersion} shows the four lowest dispersion branches in the $k_x \bm{e}_x$-direction (\(\Gamma\)-\(X\) line, blue)  and in the $k_y\bm{e}_y$-direction (\(\Gamma\)-\(Y\) line, green) for a given packing at the pressure \( P = 5\times 10^{-4}\kappa  \) with $N=128$ particles in each cell. 
Figure~\ref{fig:dispersion}(a) presents the dispersion relation for a system in which all eigenvalues are non-negative. 
On the other hand, Fig.~\ref{fig:dispersion} (b) depicts the dispersion in which eigenvalues can be negative at some $\bm k$.
Both examples in Figs.~\ref{fig:dispersion}(a) and ~\ref{fig:dispersion}(b) exhibit positive curvature of the
lowest branch near \(\bm k=\bm0\). Other samples, however, may exhibit
negative curvature near \(\bm k=\bm0\). We also find cases in which an
eigenvalue becomes negative only over an intermediate wave-number range
and returns to positive values near the Brillouin-zone boundary.
Note that the modes originating from floating particles, defined here as particles with no contacts, remain as zero modes even at finite wave numbers.

Near $\bm k=\bm 0$, both the lowest and the second-lowest branches are phonon-like modes of translational origin. As the wave number varies, the eigenvectors change continuously, as seen in Figs.~\ref{fig:eigenvec} (a-d). 
In particular, crossing \(\lambda_n=0\) does not by itself produce an abrupt change in the eigenvector. More pronounced changes occur near avoided crossings with neighboring branches, where the branches approach one another closely and their eigenvectors hybridize.
In these regions, the branch exhibits avoided crossing behavior with
neighboring branches, accompanied by eigenvector hybridization and an
exchange of mode character.

Figure \ref{fig:eigenvec}(b) lies near an avoided crossing on the lowest branch, where the eigenvector exhibits hybridization between translational and anomalous mode character.

Through this hybridization, anomalous-mode character present at
\(\bm k=\bm0\) can be transferred continuously to the lowest branch at
finite \(\bm k\), which may become unstable at larger wave vectors.
Figures \ref{fig:eigenvec} (e) and \ref{fig:eigenvec}(f) show the lowest and second-lowest nonzero eigenmodes at $\bm k=\bm 0$, respectively. 
The unstable mode at the $X$ point ($\bm{k} = (\pi/L,0)$) shown in Fig.~\ref{fig:eigenvec}(d) strongly resembles the mode in Fig.~\ref{fig:eigenvec}(f). This resemblance reflects two key effects: mode hybridization and the spatial reshaping of the eigenvector field. 
This reshaping is necessary to satisfy the Bloch boundary condition at a finite $\bm{k}$, which introduces an inter-cell phase modulation of $e^{i\bm{k}\cdot \bm{R}^{\bm{\mu}}}$.
By contrast, the lowest eigenmode at the \(Y\) point
\((\bm{k}=(0,\pi/L))\) shown in Fig.~\ref{fig:eigenvec}(g)
is similar to the mode shown in Fig.~\ref{fig:eigenvec}(e).

Schoenholz \textit{et al.}\cite{S.Schoenholz2013Stability} demonstrated that instability is most likely observed when $PL^4/ (\kappa d_s^4) \sim O(1)$, while its probability decreases as $PL^4/(\kappa d_s^4)$ increases.
Therefore, to observe instability, it is necessary to choose an appropriate combination of \(L\) and pressure, although the theory assumes $L/d_s\gg 1$.
In principle, increasing \(L/d_s\) while keeping
\(PL^4/(\kappa d_s^4)\) of order unity requires a corresponding reduction
of \(P\). Since simulations at sufficiently small pressures are limited by numerical resolution, the accessible cell size is limited in the present numerical study.
The quantity $PL^4/(\kappa d_s^4)$ is related to the transverse length scale \cite{L.Silbert2005LengthScale}, and this condition implies that stability results from a competition between transverse waves and anomalous modes (non-phononic, non-Debye extended modes)~\cite{OHern2003JammingZeroTmp,Wyart05,L.Silbert2009NormalMode}, as well as finite-size effects~\cite{C.Goodrich2012FiniteSizeScaling,C.Goodrich2014JammingInFiniteSytems}.
This observation reflects sample-specific anisotropy: for a fixed packing, the dispersion curves differ between $\bm{k}\parallel \bm e_x$ and $\bm{k}\parallel \bm e _ y$. While such anisotropy is expected in small disordered systems~\cite{M.Tsamados2009LocalElasticity}, we confirmed that it can persist in individual realizations even at larger $N$, especially in higher branches. We stress that this does not necessarily imply a nonzero anisotropy in the ensemble-averaged response. Rather, dispersion relations are useful because they expose direction-dependent features that may be obscured in more aggregated measures.

These negative eigenvalues imply that perturbations along the corresponding eigenmodes grow exponentially within the linearized dynamics.
This motivates the following analysis of the linear response.

\subsection{Linear response to unstable modes}\label{subsec:response}
We now consider the time evolution of a small perturbation within the linearized dynamics.
Let us consider \(\hat{\bm{r}}(\bm{k},t):=(\hat{\bm{r}}_1(\bm{k},t),\hat{\bm{r}}_2(\bm{k},t),\cdots, \hat{\bm{r}}_N(\bm{k},t))^\mathsf{T}\), where the superscript $\mathsf{T}$ denotes the transpose.
Using the spectral decomposition of $\hat{\bm{r}}(\bm{k},t)$ with the assumption of nondegenerate eigenmodes, we may write
\begin{equation}
 \hat{\bm{r}}(\bm{k},t) = \sum_n a_n(\bm k)\hat{\bm{\varepsilon}}_{n}(\bm{k})e^{-\lambda_{n}(\bm{k})t/\eta} ,
\label{eq:u_eigen}
\end{equation}
where the expansion coefficients are
\begin{equation}
a_n(\bm k):=\hat{\bm\varepsilon}_n^\dagger(\bm k)\,\hat{\bm r}(\bm k,0)
 .
\end{equation}
Here and in the following, the summation excludes the zero modes associated with floating particles.
The expression eqn.~\eqref{eq:u_eigen} describes the linear response as long as all eigenvalues are non-negative, $\lambda_n(\bm k)\ge 0$. In this case, each mode either decays or remains neutral in time. 
In contrast, 
if the lowest eigenvalue excluding the floating-particle zero modes satisfies \(\lambda_1(\bm k)<0\) at some wave vector \(\bm k\), the corresponding mode grows exponentially as \(e^{-\lambda_1(\bm k)t/\eta}\).
In such a case, the linearized equation \eqref{eq:motion_eq_v2} is no longer sufficient at later times, because the exponentially growing mode eventually invalidates the small-perturbation assumption.

\section{Weakly nonlinear theory}
\label{sec:Nonlinear}

\subsection{General framework}

As shown in Sec.~\ref{subsec:response}, unstable modes identified through the generalized Hessian matrix exhibit exponential growth in the linear regime. 
However, this growth is expected to be suppressed due to nonlinear effects. 
To capture this transition, we derive a weakly nonlinear amplitude equation.

Before proceeding with the detailed discussion, 
we note that the present weakly nonlinear analysis is qualitatively different from the standard weakly nonlinear analysis
~\cite{Cross,Manneville,Kuramoto,Shakla09,Saitoh11,Aranson2006}. 
In standard weakly nonlinear analysis, large-scale modes with small $k$ can be unstable, whereas small-scale modes with large $k$ are usually stable. In contrast, our system exhibits the opposite behavior, which is stable at large scales and unstable at small scales.
This is possible if the system has a large wavenumber (Ultraviolet) cutoff at small scales, with $\pi/L$ playing a role in the UV cutoff.

We define $\bm{k}_c$ as the wave vector where the lowest eigenvalue $\lambda_1(\bm{k})$ reaches its minimum. 
Equivalently, this is the point where the linear growth rate, $-\lambda_1(\bm{k})/\eta$, reaches its maximum. 
The associated eigenvector \(\hat{\bm\varepsilon}_1(\bm k_c)\) is referred to as the critical mode.
In weakly nonlinear analysis, we assume that the growth rate $-\lambda_1(\bm{k})/\eta$ remains small, and thus, we can extract the slow dynamics associated with the unstable mode.
 We first approximate the displacement field
\(\delta \bm r^{\bm\mu}:=\{\delta \bm r_i^{\bm\mu}\}_{i=1}^N\)
by retaining only the critical Fourier component and projecting it onto the
critical mode. In the continuum
wave-vector representation, this single-mode approximation is written as
\begin{align}
    \hat{\bm r}(\bm k,t)
    &\approx
    \Omega_{\mathrm{BZ}}\big(
    A_c(t)
    \hat{\bm\varepsilon}_{1}(\bm k_c)
    \delta(\bm k-\bm k_c)
    +
    A_c^*(t)
    \hat{\bm\varepsilon}_{1}^{*}(\bm k_c)
    \delta(\bm k+\bm k_c)\big),
    \label{eq:critical_mode_ansatz_k}
\end{align}
where \(A_c(t)\) is the amplitude of the critical mode in real space.
Substituting eqn.~\eqref{eq:critical_mode_ansatz_k} into eqn.~\eqref{eq:Fourier_Inv_transf}, we obtain
\begin{equation}
   \delta \bm r^{\bm\mu}(t)
   \approx
   A_c(t)
   \hat{\bm\varepsilon}_{1}(\bm k_c)
   e^{-i\bm k_c\cdot\bm R^{\bm\mu}}
   + c.c. ,
   \label{eq:intK}
\end{equation}
where \(c.c.\) denotes the complex conjugate term.

Expanding the overdamped dynamics, eqn.~\eqref{eq:motion_eq_v2}, up to cubic order in the displacement field $\{\delta \bm{r}_i^{\bm\mu}\}_{i=1}^N$ around the force-balanced configuration, we obtain
\begin{align}\label{non_linear_eq}
\eta \frac{d}{dt} \delta \bm{r}_i^{\bm{\mu}} &\approx -\sum_{\bm{\nu}}\sum_j \mathsf{H}_{ij}^{\bm{\mu\nu}} \delta \bm{r}_j^{\bm{\nu}} -
\frac{1}{2} \sum_{\bm{\nu},\bm{\zeta}}\sum_{j,k} \mathsf{T}_{ijk}^{\bm{\mu\nu\zeta}} \delta \bm{r}_j^{\bm{\nu}} \delta \bm{r}_k^{\bm{\zeta}} \notag\\
&\quad 
- \frac{1}{6} \sum_{\bm{\nu},\bm{\zeta},\bm{\xi}} \sum_{j,k,l} \mathsf{M}_{ijk\ell}^{\bm{\mu\nu\zeta \xi}} \delta \bm{r}_j^{\bm{\nu}} \delta \bm{r}_k^{\bm{\zeta}}  \delta \bm{r}_\ell^{\bm{\xi}} .
\end{align}
Here, each sum over a cell index is restricted to cells that form
a pair in \(\mathcal N_{\rm c}\) with the reference cell
\(\bm{\mu}\). Specifically, the summations are taken over cell
indices satisfying
\(\{\bm{\mu},\bm{\nu}\}\in\mathcal N_{\rm c}\),
\(\{\bm{\mu},\bm{\zeta}\}\in\mathcal N_{\rm c}\), and
\(\{\bm{\mu},\bm{\xi}\}\in\mathcal N_{\rm c}\),
whenever the corresponding indices appear.
The tensors
$\mathsf{T}_{ijk}^{\bm{\mu\nu\zeta}}$ and $\mathsf{M}_{ijk\ell}^{\bm{\mu\nu\zeta \xi}}$ 
are the third- and fourth-order derivatives of the total potential
energy, respectively, evaluated at the force-balanced configuration, i.e.
\begin{align}
\mathsf{T}_{ijk}^{\bm{\mu\nu\zeta}}:&=\left(\frac{\delta^3 V}{\delta \bm{r}_i^{\bm{\mu}}\delta \bm{r}_j^{\bm{\nu}}\delta \bm{r}_k^{\bm{\zeta}}}\right)_0 , \\
\mathsf{M}_{ijk\ell}^{\bm{\mu\nu\zeta \xi}}:&=\left(\frac{\delta^4 V}{\delta \bm{r}_i^{\bm{\mu}}\delta \bm{r}_j^{\bm{\nu}}\delta \bm{r}_k^{\bm{\zeta}} \delta \bm{r}_\ell^{\bm{\xi}}}\right)_0 .
\end{align}

Using 
eqn.~\eqref{eq:critical_mode_ansatz_k}, we project eqn.~\eqref{non_linear_eq} onto the eigenvector
$\hat{\bm{\varepsilon}}_1(\bm{k}_c)$.
Imposing the solvability condition, we multiply both sides by
$e^{i\bm{k}_c\cdot\bm{R}^{\bm{\mu}}}$, sum over the cell index $\bm{\mu}$,
and contract with $\hat{\bm{\varepsilon}}_1^{\dagger}(\bm{k}_c)$
.
The details of this projection are presented in
Appendix~\ref{app:projection}.
Under the resonance condition at the critical wave vector $\bm{k}_c$, quadratic terms proportional to $A_c^2$ and $|A_c|^2$, as well as non-resonant cubic terms such as $A_c^3$, do not contribute
to the projection onto the critical mode. 
Thus, the
contributions involving \(\mathsf{T}_{ijk}\) vanish upon projection,
and the leading resonant nonlinear contribution is proportional to
\(|A_c|^2A_c\).
This yields the following projected equation for the amplitude:
\begin{equation}
\label{eq:Stuart_c}
\eta \dot{A}_c(t) \approx \lambda_c^\prime A_c(t) - \frac{1}{6 } \chi_c |A_c(t)|^2A_c(t) ,
\end{equation}
where $\dot{A}_c(t):=dA_c(t)/dt$, $\lambda^\prime_c:=-\lambda_c$.

We now decompose the linear coefficient \(\lambda_c\) and the nonlinear
coefficient \(\chi_c\). It is useful to separate their in-cell and inter-cell
contributions. The linear coefficient is written as
\begin{equation}\label{lambda_c}
\lambda_c
=
\lambda_c^{\rm in}
+
\lambda_c^{\rm inter},
\end{equation}
where
\begin{align}
\lambda_c^{\rm in}
&:=
\sum_{i,j}
\hat{\bm\varepsilon}_{1,i}^*(\bm k_c)
\mathsf H_{ij}^{\bm0,\bm0}
\hat{\bm\varepsilon}_{1,j}(\bm k_c),
\\
\lambda_c^{\rm inter}
&:=
\sum_{\bm\Delta_{\bm{\mu\nu}}\in\mathcal N\setminus\{\bm0\}}
\sum_{i,j}
\hat{\bm\varepsilon}_{1,i}^*(\bm k_c)
\mathsf H_{ij}^{\bm0,-\bm\Delta_{\bm{\mu\nu}}}
\hat{\bm\varepsilon}_{1,j}(\bm k_c)
e^{i\bm k_c\cdot\bm L_{\bm\Delta_{\bm{\mu\nu}}}} .
\end{align}
For the nonlinear coefficient, we similarly decompose
\begin{equation}
\chi_c
=
\chi_c^{\rm in}
+
\chi_c^{\rm inter}.
\end{equation}
To keep track of the cell indices in the cubic nonlinear term, we use the relative cell-offset vectors
$\bm\Delta_{\bm\mu\bm\zeta}:=\bm\mu-\bm\zeta,\bm\Delta_{\bm\mu\bm\xi}:=\bm\mu-\bm\xi$ . 
After shifting the reference cell \(\bm\mu\) to \(\bm0\), the cells \(\bm\nu\), \(\bm\zeta\), and \(\bm\xi\) are represented as \(-\bm\Delta_{\bm\mu\bm\nu}\), \(-\bm\Delta_{\bm\mu\bm\zeta}\), and \(-\bm\Delta_{\bm\mu\bm\xi}\), respectively.
To write these contributions compactly, we define
\begin{align}
\mathcal C_{jk\ell}
(\bm\Delta_{\bm\mu\bm\nu},\bm\Delta_{\bm\mu\bm\zeta},\bm\Delta_{\bm\mu\bm\xi})
&:=
\chi_{jk\ell}^{(1)}
+
\chi_{jk\ell}^{(2)}
+
\chi_{jk\ell}^{(3)},
\end{align}
where
\begin{align}
\chi_{jk\ell}^{(1)}
&:=
\hat{\bm\varepsilon}_{1,j}(\bm k_c)
\hat{\bm\varepsilon}_{1,k}(\bm k_c)
\hat{\bm\varepsilon}_{1,\ell}^*(\bm k_c)
e^{i\bm k_c\cdot\bm L_{\bm\Delta_{\bm\mu\bm\nu}}}
e^{i\bm k_c\cdot\bm L_{\bm\Delta_{\bm\mu\bm\zeta}}}
e^{-i\bm k_c\cdot\bm L_{\bm\Delta_{\bm\mu\bm\xi}}},
\\
\chi_{jk\ell}^{(2)}
&:=
\hat{\bm\varepsilon}_{1,j}(\bm k_c)
\hat{\bm\varepsilon}_{1,k}^*(\bm k_c)
\hat{\bm\varepsilon}_{1,\ell}(\bm k_c)
e^{i\bm k_c\cdot\bm L_{\bm\Delta_{\bm\mu\bm\nu}}}
e^{-i\bm k_c\cdot\bm L_{\bm\Delta_{\bm\mu\bm\zeta}}}
e^{i\bm k_c\cdot\bm L_{\bm\Delta_{\bm\mu\bm\xi}}},
\\
\chi_{jk\ell}^{(3)}
&:=
\hat{\bm\varepsilon}_{1,j}^*(\bm k_c)
\hat{\bm\varepsilon}_{1,k}(\bm k_c)
\hat{\bm\varepsilon}_{1,\ell}(\bm k_c)
e^{-i\bm k_c\cdot\bm L_{\bm\Delta_{\bm\mu\bm\nu}}}
e^{i\bm k_c\cdot\bm L_{\bm\Delta_{\bm\mu\bm\zeta}}}
e^{i\bm k_c\cdot\bm L_{\bm\Delta_{\bm\mu\bm\xi}}} .
\end{align}
It should be noted that $\chi_{jk\ell}^{(1)}$, $\chi_{jk\ell}^{(2)}$, and $\chi_{jk\ell}^{(3)}$ depend on $\bm{\Delta}_{\bm\mu\bm\nu}$, $\bm{\Delta}_{\bm\mu\bm\zeta}$, and $\bm{\Delta}_{\bm\mu\bm\xi}$, although we omit their dependence in the above equations.
The in-cell contribution is
\begin{align}
\chi_c^{\rm in}
&:=
\sum_{i,j,k,\ell}
\hat{\bm\varepsilon}_{1,i}^*(\bm k_c)
\mathsf M_{ijk\ell}^{\bm0,\bm0,\bm0,\bm0}
\mathcal C_{jk\ell}(\bm0,\bm0,\bm0),
\end{align}
and the inter-cell contribution is
\begin{align}
\chi_c^{\rm inter}
&:=
\sum_{\substack{
\bm\Delta_{\bm\mu\bm\nu},\bm\Delta_{\bm\mu\bm\zeta},\bm\Delta_{\bm\mu\bm\xi}\in\mathcal N\\
(\bm\Delta_{\bm\mu\bm\nu},\bm\Delta_{\bm\mu\bm\zeta},\bm\Delta_{\bm\mu\bm\xi})\neq(\bm0,\bm0,\bm0)
}}
\sum_{i,j,k,\ell}
\hat{\bm\varepsilon}_{1,i}^*(\bm k_c)
\mathsf M_{ijk\ell}^{\bm0,-\bm\Delta_{\bm\mu\bm\nu},-\bm\Delta_{\bm\mu\bm\zeta},-\bm\Delta_{\bm\mu\bm\xi}}\notag \\
& \quad \times  
\mathcal C_{jk\ell}(\bm\Delta_{\bm\mu\bm\nu},\bm\Delta_{\bm\mu\bm\zeta},\bm\Delta_{\bm\mu\bm\xi}).
\end{align}

Equation~\eqref{eq:Stuart_c} resembles the Stuart--Landau equation
, although its coefficients contain inter-cell contributions.
When $\lambda^\prime_c>0$ and $\chi_c>0$, the nonlinear term suppresses the linear instability, and the amplitude saturates at a finite value.
Equation~\eqref{eq:Stuart_c}, in such a case, can be solved as
\begin{align}\label{eq:A(t)}
&A_c(t) = \frac{A_{c,0}}{\sqrt{(1-\frac{\beta}{\alpha}|A_{c,0}|^2)e^{-2\alpha t}+\frac{\beta}{\alpha}|A_{c,0}|^2 }}, 
\end{align}
for the initial condition  $A_c(0) = A_{c,0}
$, where $\alpha: = \lambda_c^\prime/\eta$ and $\beta: = \chi_c/6\eta $.
Denoting the long-time saturated amplitude by
$A_{c,\infty}:=\lim_{t\to\infty}A_c(t),$ the corresponding nontrivial steady-state solution satisfies
\begin{equation}\label{A_infty}
|A_{c,\infty}|^2 = \frac{6\lambda^{\prime}_c }{\chi_c}.
\end{equation}
Thus, the weakly nonlinear theory predicts finite-amplitude saturation of the unstable mode.

\subsection{Two cell system}

The dispersion relation is defined for a periodically replicated system with
an infinite number of cells. However, to examine the real-space response of
individual particles to an unstable edge mode, we consider the minimal finite
realization of the corresponding modulation. Specifically, we prepare two
identical cells aligned along the modulation direction and impose periodic
boundary conditions at the outer edges. This two-cell geometry realizes the
edge-mode modulation while allowing the particles in the two cells to evolve
independently after perturbation.

Let \(\bm e_\parallel\) denote the unit vector along the modulation direction,
and let \(\mu_\parallel\) denote the corresponding cell index. Thus,
\((\bm e_\parallel,\mu_\parallel)=(\bm e_x,\mu_x)\) for an \(x\)-directed
modulation and \((\bm e_\parallel,\mu_\parallel)=(\bm e_y,\mu_y)\) for a
\(y\)-directed modulation.
The allowed wave vectors in
the first Brillouin zone are \(\bm k=\bm0\) and the edge wave vector
\begin{equation}
\bm k_E:=\frac{\pi}{L}\bm e_\parallel .
\end{equation}
Since \(\bm k_E\) is the only nonzero allowed wave vector, an unstable mode in the present two-cell geometry is associated
with \(\bm k_c=\bm k_E\).

At the Brillouin-zone edge, \(-\bm k_E\) is equivalent to \(\bm k_E\) up to a
reciprocal lattice vector. Therefore, the edge mode is self-conjugate, and
the complex conjugate contribution in eqn.~\eqref{eq:intK} is not an
additional independent mode. We therefore reparametrize the redundant pair
of complex-conjugate contributions by a single real amplitude \(A_E(t)\).
For \(\bm k_E=(\pi/L)\bm e_\parallel\), the phase factor is
\begin{align}
e^{-i\bm k_E\cdot\bm R^{\bm\mu}}
&=
e^{-i\pi\mu_\parallel}
=
(-1)^{\mu_\parallel}
\label{eq:mux}
\end{align}
The corresponding eigenvector can also be chosen to be real. Hence, the
displacement field associated with the edge mode becomes
\begin{equation}
\delta\bm r^{\bm\mu}(t)
=
A_E(t)\hat{\bm\varepsilon}_1(\bm k_E)(-1)^{\mu_\parallel}.
\label{eq:Fourier_Inv_transf_Lat}
\end{equation}
Thus, the edge mode represents an alternating displacement pattern of neighboring cells with wavelength \(2L\), and a negative eigenvalue at \(\bm k_E\) indicates a mechanical instability of the periodically replicated system against this pattern.

Substituting eqn.~\eqref{eq:Fourier_Inv_transf_Lat} into
eqn.~\eqref{non_linear_eq} and projecting onto

the edge mode, we obtain a real amplitude equation of the form
\begin{equation}
\eta \dot A_E(t)
\approx
\lambda_E^\prime A_E(t)
-
\frac{1}{6}\chi_E A_E^3(t),
\label{eq:Stuart_E}
\end{equation}
where \(\lambda_E^\prime:=-\lambda_E\).
The linear coefficient \(\lambda_E\) is obtained by     specializing
eqn.~\eqref{lambda_c} at \(\bm k_c=\bm k_E\).
In contrast, because the edge mode is self-conjugate, the nonlinear
coefficient \(\chi_E\) must be obtained by directly projecting
eqn.~\eqref{non_linear_eq} onto the real edge-mode representation.
\begin{align}
\chi_E
&=
\chi_E^{\mathrm{in}}
+
\chi_E^{\mathrm{inter}},
\label{eq:chiE_split}
\end{align}
where the in-cell contribution is
\begin{align}
\chi_E^{\mathrm{in}}
&:=
\sum_{i,j,k,\ell}
\hat{\bm\varepsilon}_{1,i}(\bm k_E)
\mathsf M_{ijk\ell}^{\bm0,\bm0,\bm0,\bm0}
\hat{\bm\varepsilon}_{1,j}(\bm k_E)
\hat{\bm\varepsilon}_{1,k}(\bm k_E)
\hat{\bm\varepsilon}_{1,\ell}(\bm k_E).
\label{eq:chiE_in}
\end{align}
The inter-cell contribution is given by
\begin{align}
\chi_E^{\mathrm{inter}}
&:=
\sum_{\substack{
\bm\Delta_{\bm\mu\bm\nu},\bm\Delta_{\bm\mu\bm\zeta},\bm\Delta_{\bm\mu\bm\xi}\in\mathcal N\\
(\bm\Delta_{\bm\mu\bm\nu},\bm\Delta_{\bm\mu\bm\zeta},\bm\Delta_{\bm\mu\bm\xi})
\neq
(\bm0,\bm0,\bm0)
}}
\sum_{i,j,k,\ell}
\hat{\bm\varepsilon}_{1,i}(\bm k_E)
\mathsf M_{ijk\ell}^{\bm0,-\bm\Delta_{\bm\mu\bm\nu},-\bm\Delta_{\bm\mu\bm\zeta},-\bm\Delta_{\bm\mu\bm\xi}}
\notag\\
&\quad\times
\hat{\bm\varepsilon}_{1,j}(\bm k_E)
\hat{\bm\varepsilon}_{1,k}(\bm k_E)
\hat{\bm\varepsilon}_{1,\ell}(\bm k_E)
(-1)^{\bm e_\parallel\cdot\bm\Delta_{\bm\mu\bm\nu}}
(-1)^{\bm e_\parallel\cdot\bm\Delta_{\bm\mu\bm\zeta}}
(-1)^{\bm e_\parallel\cdot\bm\Delta_{\bm\mu\bm\xi}}.
\label{eq:chiE_inter}
\end{align}
Here, \(\bm\Delta\) denotes any one of the relative cell-offset vectors \(\bm\Delta_{\bm\mu\bm\nu}\), \(\bm\Delta_{\bm\mu\bm\zeta}\), and \(\bm\Delta_{\bm\mu\bm\xi}\). The scalar product \(\bm e_\parallel\cdot\bm\Delta\) denotes the integer component of the
cell-offset vector \(\bm\Delta\) along the direction of the edge wave vector
\(\bm k_E=(\pi/L)\bm e_\parallel\). The factor
\((-1)^{\bm e_\parallel\cdot\bm\Delta}\) is obtained from
\(e^{i\bm k_E\cdot\bm L_{\bm\Delta}}\). For example, if
\(\bm e_\parallel=\bm e_x\), offsets \(\bm\Delta=\pm\bm e_x\) have phase
\(-1\), whereas offsets \(\bm\Delta=\pm\bm e_y\) have phase \(+1\).
Thus, all cell offsets in \(\mathcal N\) are included; the dot product only
determines the phase associated with each offset.

At \(\bm k_E\), both the eigenvector and the phase factors can be chosen real.
Consequently, \(\lambda_E\) and \(\chi_E\) are real.
Moreover, we have numerically confirmed that $\chi_E$ is always positive for all samples considered.
We note that the definitions of $\lambda_E$ and $\chi_E$ themselves are not restricted to the two-cell system. 
The explicit expressions of $\lambda_E$ and $\chi_E$ are presented in Appendix~\ref{Ap:L_C}.

The long-time limit of the amplitude, eqn.~\eqref{A_infty} is reduced to $A_{E,\infty} = \sqrt{6 \lambda_E^\prime / \chi_E}$. 
This reduced amplitude equation predicts the initial growth and subsequent
saturation of the unstable mode at a finite amplitude, consistent with the
numerical behavior observed in Secs.~\ref{subsec:Div} and~\ref{subsec:Init}.

Within the same weakly nonlinear approximation, the potential-energy change per unit cell is written as
\begin{equation}
\delta V(t) \approx
-\frac{1}{2}\lambda_E^\prime A_E^2(t)
+\frac{1}{24}\chi_E A_E^4(t),
\label{eq:DeltaU_t}
\end{equation}
which gives the theoretical time evolution of the potential energy after substituting eqn.~\eqref{eq:A(t)}.

In the long-time limit, eqn.~\eqref{eq:DeltaU_t} reduces to
\begin{align}
\delta V_\infty
&\approx
-\frac{1}{2}\lambda_E^\prime A_{E,\infty}^2
+\frac{1}{24}\chi_E A_{E,\infty}^4
\notag \\
&\approx
-\frac{3{\lambda_E^\prime}^2}{2\chi_E}.
\label{eq:dU_LC}
\end{align}

We emphasize that this derivation assumes a single dominant unstable mode and neglects coupling to other modes or wavevectors.

\section{Numerical validation}\label{sec:Num}
This section is organized as follows.
In Sec.~\ref{subsec:Div}, we present the numerical response of unstable modes.
In Sec.~\ref{subsec:Init}, we examine the dependence on the initial perturbation.
In Sec.~\ref{subsec:AmpComp}, we compare these numerical results with the weakly nonlinear amplitude equation.

\subsection{Finite-cell setup and response to unstable perturbations}\label{subsec:Div}
To obtain the dispersion relations shown in Fig.~\ref{fig:dispersion}, we used a perfectly replicated system consisting of an infinite number of cells. 
However, such a replicated system is not suitable for analyzing the linear response to a perturbation, since this analysis requires tracking the motion of individual grains.
To address this, we consider a system composed of a finite number of cells, each initialized with an identical grain configuration.
We then solve the equations of motion for all grains under the applied perturbation.
In this setup, once the perturbation is applied, the identical grains in different cells (i.e., particles at $\bm{r}_i^{\bm{\mu}}$ and $\bm{r}_i^{\bm{\nu}}$ with $\bm{\mu}\ne \bm{\nu}$ ) are allowed to respond differently, enabling a more accurate assessment of the linear response.
Furthermore, this section employs quasi-static simulations.

The present method introduces the wave number $\bm{k}$ at the outset and thus provides direct access to unstable modes from an undeformed reference state. In contrast to the AQS-based analyses in Refs.~\cite{C.Maloney2006AQS,Lemaitre2006SumRules,D.Ishima2023Eigenvlue,S.Bohy2012SoftSpher,C.Goodrich2014JammingInFiniteSytems,S.Torquato2001Multiplicity}, it does not require constructing a sequence of incrementally deformed and relaxed configurations. Consequently, the targeted unstable modes can be captured by applying a single perturbation to the initial configuration.

\begin{figure*}[h]
  \centering
 \begin{minipage}[t]{0.48\textwidth}
   \centering
    \includegraphics[width=\linewidth]{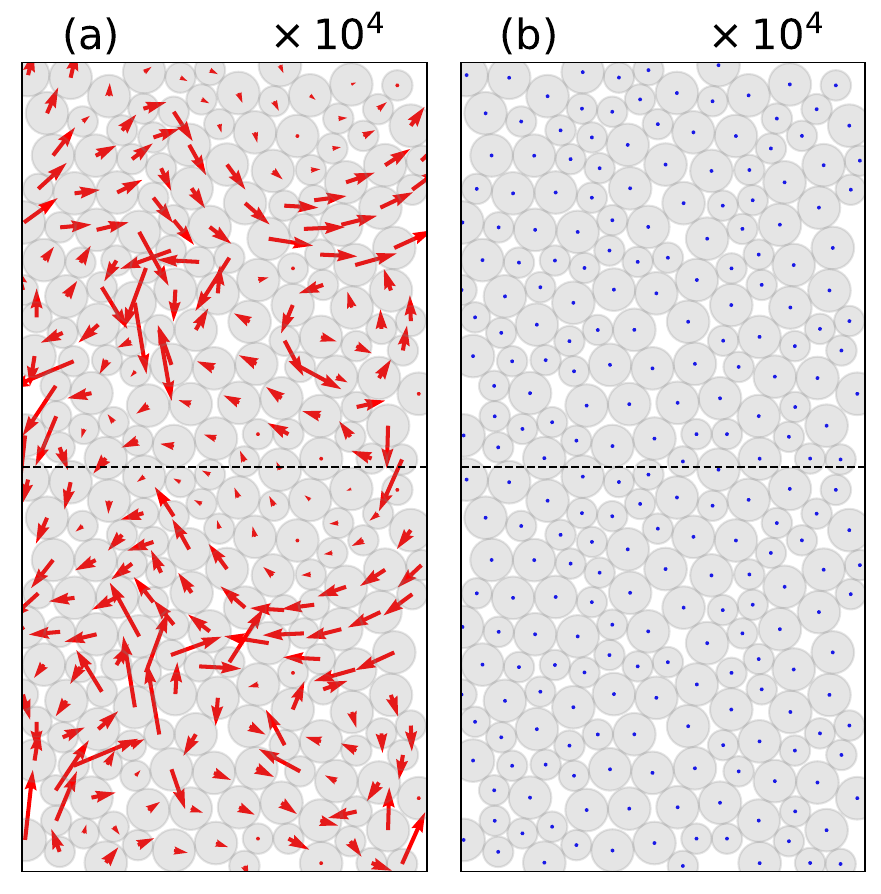}
    \caption{
    Displacements of grains after adding the perturbation along the direction of the eigenvector for the case with a positive eigenvalue, where red arrows in (a) indicate the initial perturbative displacements of grains and blue arrows in (b) exhibit the final displacements of grains after the relaxation.
    These figures demonstrate the stability of the initial configuration against the perturbation.
    }
    \label{fig:response_positive}
 \end{minipage}
\hfill
 \begin{minipage}[t]{0.48\textwidth}
    \centering
    \includegraphics[width=\linewidth]{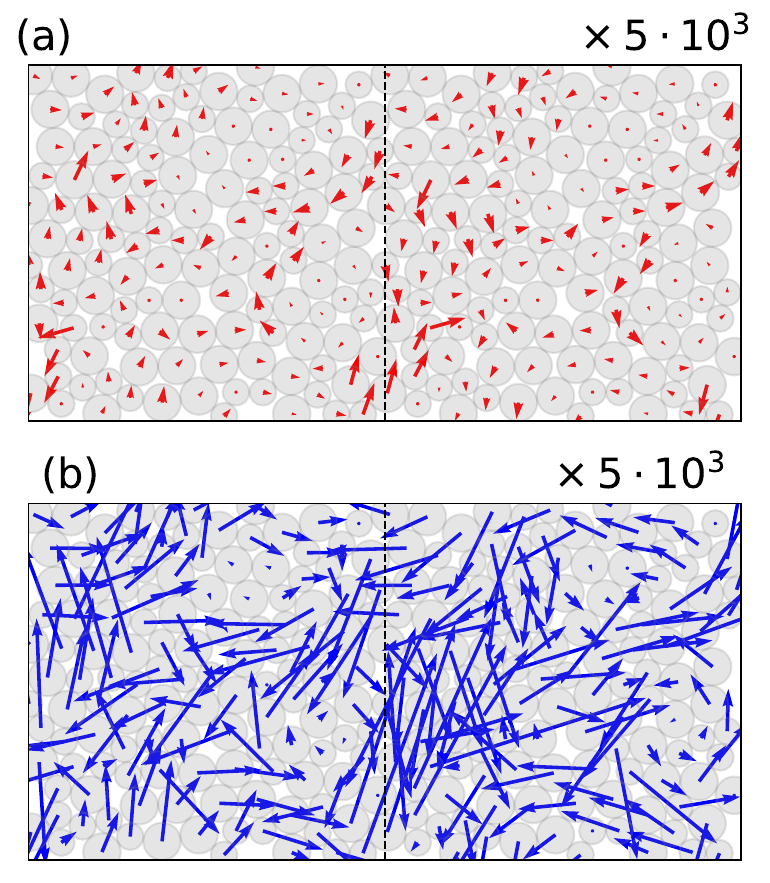}
    \caption{Displacements of grains after adding the perturbation along the direction of the eigenvector for the case with a negative eigenvalue, where red arrows in (a) indicate the initial perturbative displacements of grains and blue arrows in (b) exhibit the final displacements of grains after the relaxation.
    These figures demonstrate the instability of the initial configuration against the perturbation.}
    \label{fig:response_negative}
 \end{minipage}
\end{figure*}
Now, let us consider a case in which an unstable eigenmode is present.
For simplicity, we focus on the edge mode with
\(\bm k_E=(\pi/L)\bm e_\parallel\), where \(\bm e_\parallel\) denotes the
modulation direction. To realize this deformation in the particle-level
dynamics, we prepare two identical cells aligned along \(\bm e_\parallel\).
Thus, \(\bm k_E=(\pi/L)\bm e_x\) or \((\pi/L)\bm e_y\) for modulations along
the \(x\)- or \(y\)-direction, respectively.
To examine the response associated with the eigenvalues, we apply a deformation along the eigenvector corresponding to the lowest eigenvalue~\cite{C.Schreck2011RepulsiveContact,H.Mizuno2020AnharmonicProperties}.
Using the notation 
$a_1=B\sqrt{N}$ associated with the smallest eigenvalue $\lambda_1(\bm{k})$, we apply the perturbation
\begin{align}
     \Delta \bm{r}^{\bm{\mu}}=
     B
     \sqrt{N}\hat{\bm{\varepsilon}}_1(\bm{k}_E) e^{-i\bm{k}_E\cdot \bm R^{\bm{\mu}}}.  
     \label{eq:def_eigen_pre}
\end{align}
The factor $\sqrt{N}$ ensures that the displacement per particle remains consistent across systems of different sizes.\footnote{Specifically, since a system of  $N$  particles has  $2N$  components in  $|\bm{\varepsilon}_1 \rangle$, satisfying the orthonormal condition, the magnitude of each component of the eigenvector is approximately  $1/\sqrt{2N}$. The factor $\sqrt{N}$, thus, compensates for this, ensuring that the displacement per particle remains of the same order regardless of system size.}
We follow the dynamics after applying the perturbation eqn.~\eqref{eq:def_eigen_pre}.
In the main text, we restrict our analysis to systems with a single negative eigenvalue at a given wave number. However, there also exist systems with multiple negative eigenvalues. In such cases, the system's response should be described by a superposition of the unstable modes.
We present additional results for this scenario in Appendix~\ref{Ap:multi}, if multiple branches with negative eigenvalues appear at a certain wavenumber. 

Figures ~\ref{fig:response_positive} and \ref{fig:response_negative} visualize the response to stable and unstable perturbations applied along the selected eigenvector (eqn.~\eqref{eq:def_eigen_pre}).
In panel (a), the arrows show the imposed displacement field, i.e., the difference between the particle configuration before the perturbation and that after the applied perturbation.
In panel (b), the arrows show the relaxed (final) displacement field, i.e., the difference between the configuration before the perturbation and that after relaxation.
We can see that the perturbed displacement disappears in the stable case (see Fig.~\ref{fig:response_positive}), while the perturbation grows but relaxes to finite displacements in the unstable case, as in Fig.~\ref{fig:response_negative}.  
After the relaxation, the system settles into a different stable configuration, distinct from the initial one, following the transient response.
This behavior demonstrates that the negative eigenmode is unstable but non-divergent.
The observed saturation of this unstable mode is attributed to nonlinear effects~\cite{M.vanDeen2016ContactChange, M.Deen2014ContactChange,J.Boschan2019Irreversibility}.
These scenarios correspond to the dispersion relation in Fig.~\ref{fig:dispersion} (b).
The perturbation in Fig.~\ref{fig:response_positive} (a) uses the eigenvector of the lowest branch at the Y point, whereas that in Fig.~\ref{fig:response_negative} (a) uses the eigenvector of the lowest branch at the X point. The deformation amplitude is $B=10^{-4} d_S$.

\begin{figure*}[h]
 \begin{minipage}[t]{0.48\textwidth}
    \centering
    \includegraphics[width=\linewidth]{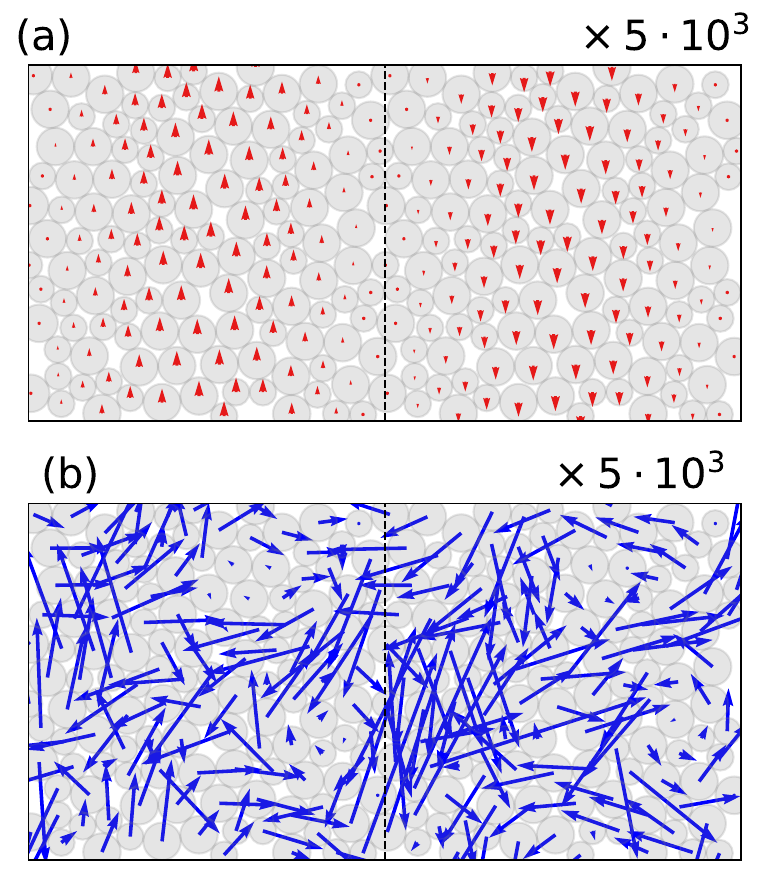}
    \caption{The response to a continuous sinusoidal transverse deformation in the case of negative eigenvalues. 
    The red arrows in (a) indicate the initial perturbative displacements of the particles, while the blue arrows in (b) represent the final displacements (perturbative displacements plus responses).
Figure (b) demonstrates that the initial configuration is unstable to the perturbation and that the displacements saturate over time.
Furthermore, it can be observed that, regardless of the initial perturbation, the same negative eigenmode as in Fig.~\ref{fig:response_negative}(b) becomes dominant.
 }
    \label{fig:wavechange_sin}
 \end{minipage}
\hfill
 \begin{minipage}[t]{0.48\textwidth}
    \centering
    \includegraphics[width=\linewidth]{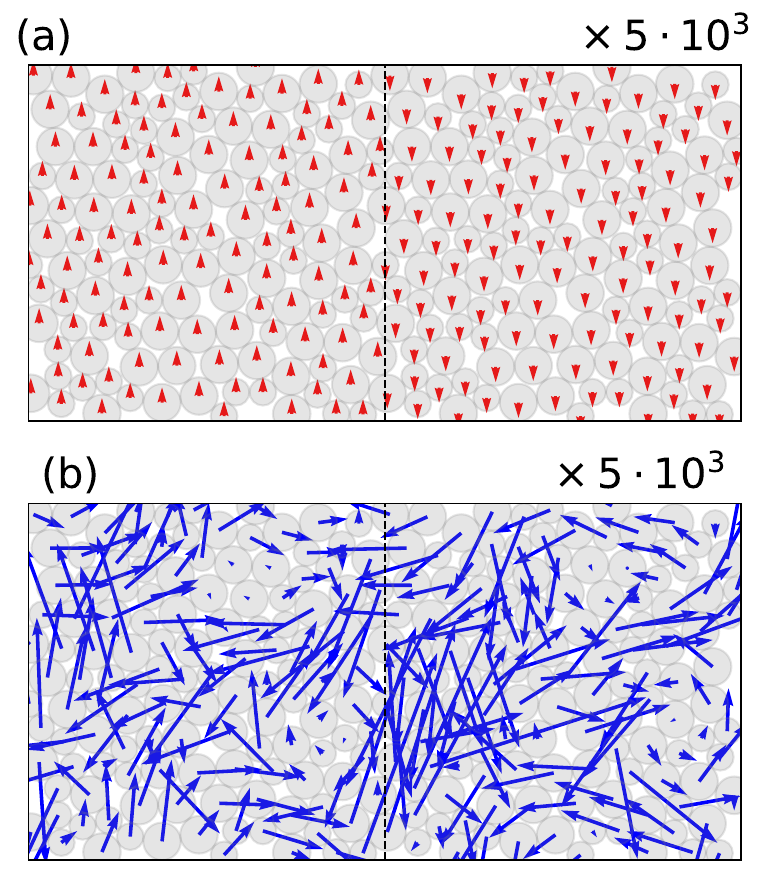}
    \caption{The response to a discontinuous rectangular-wave transverse deformation in the case of negative eigenvalues. The red arrows in (a) indicate the initial perturbative displacements of the particles, while the blue arrows in (b) represent the final displacements.
Figure (b) demonstrates that the initial configuration is unstable to the perturbation and that the displacements saturate over time.
Furthermore, it can be seen that, regardless of the initial perturbation, the same negative eigenmode as in Figs.~\ref{fig:response_negative}(b) and~\ref{fig:wavechange_sin}(b) becomes dominant.}
    \label{fig:wavechange_squ}
 \end{minipage}
\end{figure*}

\subsection{Initial condition dependence}\label{subsec:Init}

Although we have so far considered only initial deformations along the eigenvector corresponding to the lowest eigenvalue,
we expect that the system's response is essentially independent of the specific form of the initial perturbation, as suggested by eqn.~\eqref{eq:u_eigen}.
To clarify this point, we investigate how the growth of perturbative modulations in the particle configuration depends on the initial condition when the lowest eigenvalue is negative.

To this end, we apply two types of deformations along the x-direction (perturbations): a continuous transverse sinusoidal wave and a discontinuous transverse square wave, neither of which corresponds to an eigenmode.
The displacement of each particle is given by:
\begin{equation}
\Delta \bm{r}_{i,\rm sin}^{\bm{\mu}} = (0,B_{\text{sin}} \sqrt{N} \sin(k_x x^{\bm{\mu}}_i))^{\mathsf{T}},
\label{eq:sin_wave}
\end{equation}
\begin{equation}
\Delta \bm{r}_{i,\rm squ}^{\bm{\mu}} = (0,B_{\text{squ}} \sqrt{N} \text{sgn}(\sin(k_x x^{\bm{\mu}}_i)))^{\mathsf{T}} ,
\end{equation}
where the superscript $\mathsf{T}$ denotes the transpose.
These deformations are applied to a particle configuration in which only the lowest eigenvalue is negative.
Figures~\ref{fig:wavechange_sin} and~\ref{fig:wavechange_squ} show the resulting responses to the sinusoidal and square wave deformations for the same packing, respectively, as in Fig.~\ref{fig:response_negative}.
The simulation results confirm that the final displacement field is not
sensitive to the choice of initial modulation, consistent with our expectation.
Although a square wave may appear similar to a shear deformation at first glance, it is strictly a zigzag deformation due to the periodic boundary condition applied outside the two-cell system, and thus differs from a simple shear.
In the long-wavelength limit, however, where the number of cells per wavelength is increased, and the wave number tends to zero, such a deformation asymptotically approaches a shear deformation.

\subsection{Numerical validation of weakly nonlinear analysis}\label{subsec:AmpComp}

\begin{figure}[h]
    \centering
    \begin{subfigure}[b]{0.44\textwidth}
        \centering
        \begin{overpic}[width=\textwidth]
        {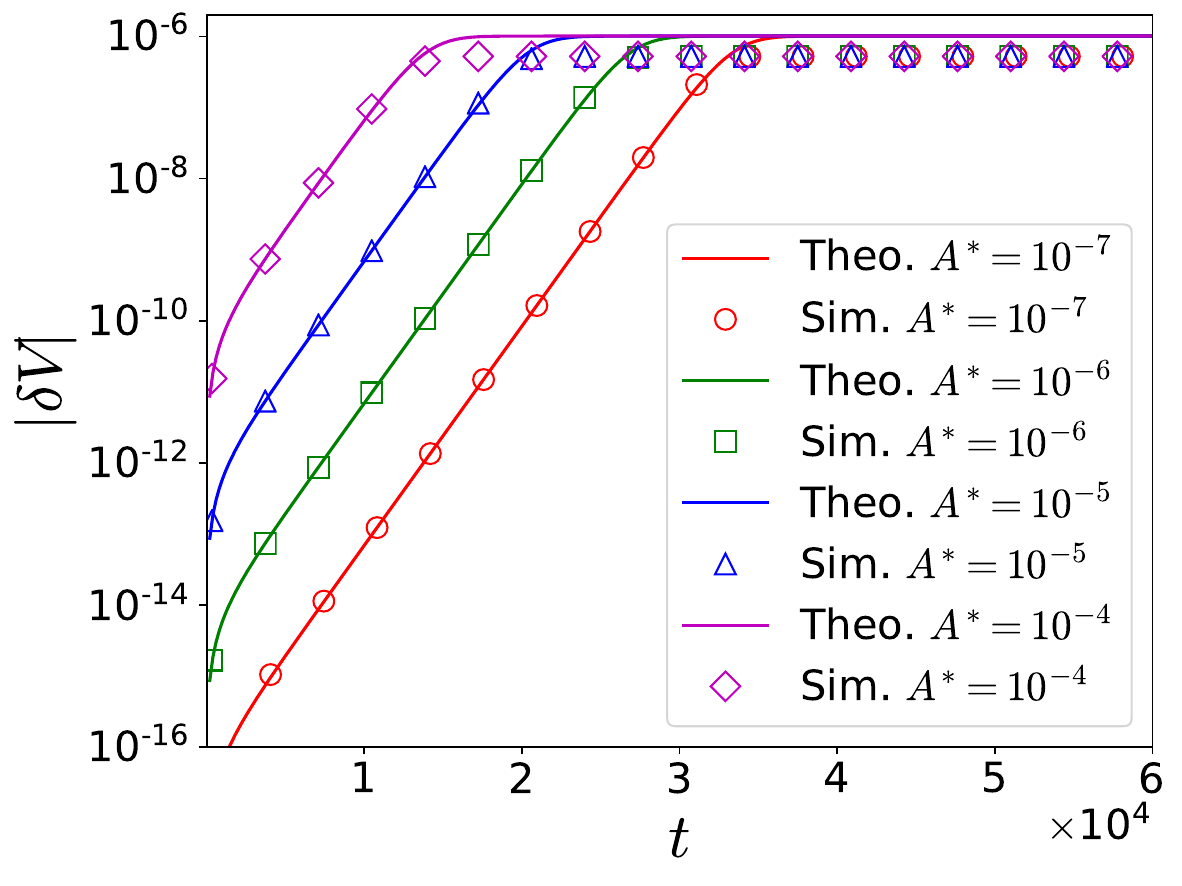}
        \put(5,77){\large\textbf{(a)}}
        \end{overpic} 
    \end{subfigure}
    \hfill
    \begin{subfigure}[b]{0.44\textwidth}
        \centering
        \begin{overpic}[width=\textwidth]
        {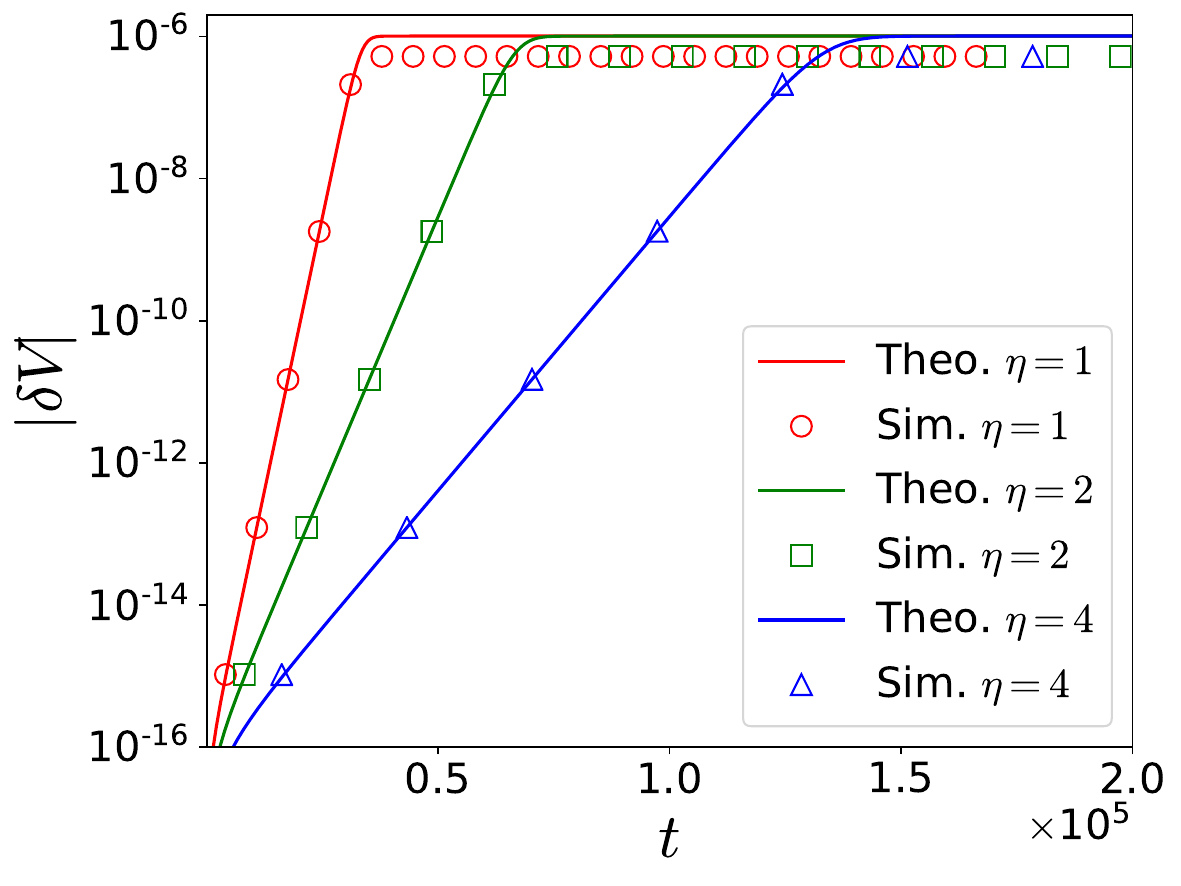}
        \put(5,77){\large\textbf{(b)}}
        \end{overpic} 
    \end{subfigure}
    \caption{
    Comparison between the weakly nonlinear theory and numerical simulations
    for a system with $N=32$ and $P=5\times10^{-4}\kappa$.
    The unstable mode lowers the potential energy, so $\delta V$ is negative.
    Therefore, the vertical axis shows the absolute value of the potential-energy change,
    $|\delta V|$, on a logarithmic scale.
    Simulation data below $|\delta V|=10^{-15}$ are omitted because this range
    is sensitive to the numerical output resolution.
    Solid lines represent the theoretical predictions based on the fourth-order
    expansion of the potential, while open symbols represent the simulation results. (a) Results for varying initial amplitude $A_{E,0}$ with fixed viscosity of background fluid corresponding to damping coefficient $\eta = 1$. 
    Note that $A^{*} = A_{E,0}/\sqrt{N}$.  (b) Results for varying $\eta$ with fixed $A_{E,0} = 10^{-7}\sqrt{N}$. 
    Although the relaxed potential-energy drop associated with the state transition
    is not reproduced exactly, the theory captures the characteristic growth
    behavior and the qualitative dependence on $A_{E,0}$ and $\eta$.
        }
    \label{fig:NL_Amp_etaChange}
\end{figure}

\begin{figure}[h]
    \centering
        \includegraphics[width=0.48\textwidth]{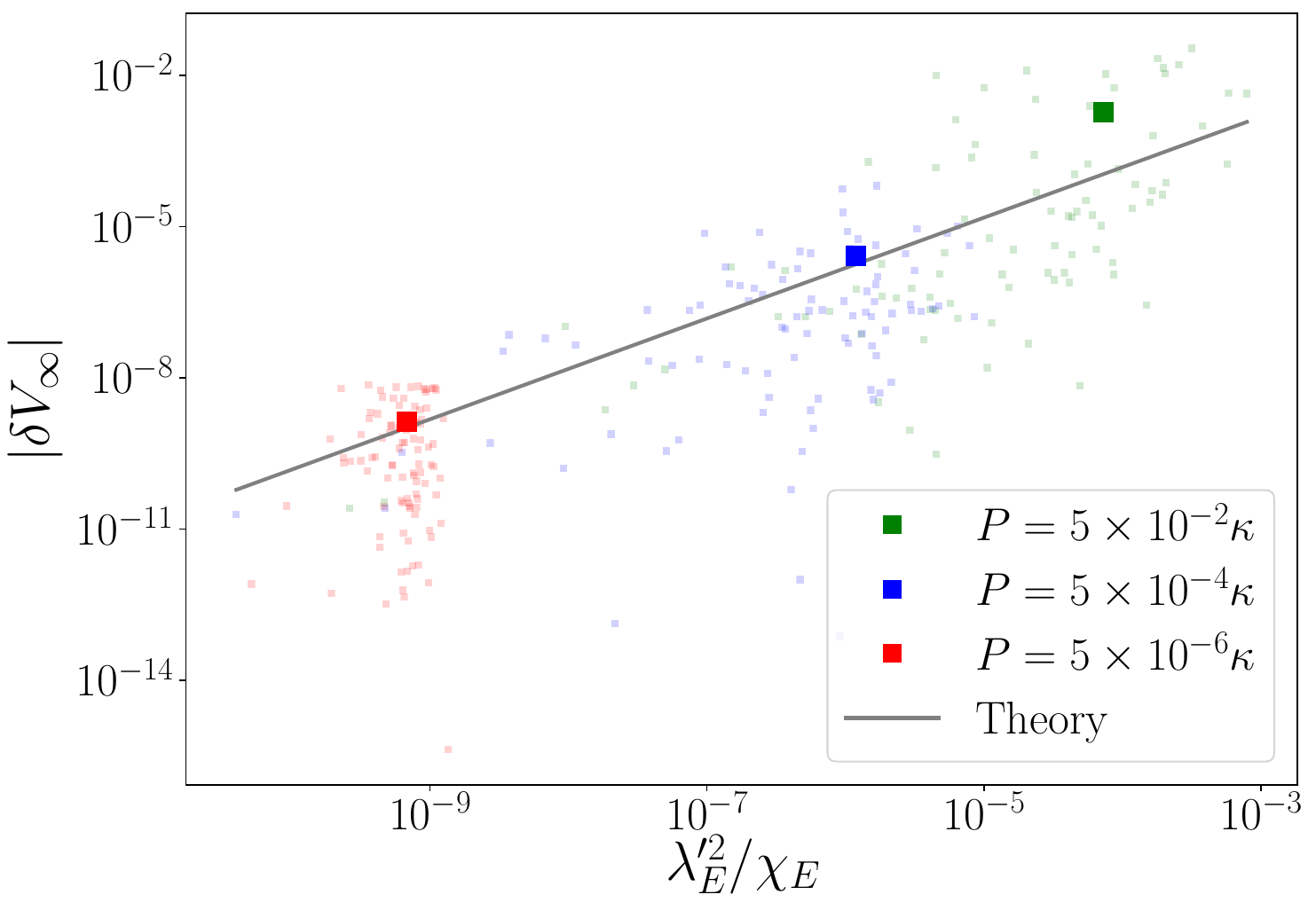}
    \caption{
A log--log plot of $\lambda_E^{\prime 2}/\chi_E$ versus the magnitude of the potential energy change $|\delta V_\infty|$ for systems with $N=128$ particles. 
The data are shown for pressures $P = 5\times 10^{-2}\kappa$ (green), $5\times 10^{-4}\kappa$ (blue), and $5\times 10^{-6}\kappa$ (red). 
For each pressure, we generated initial configurations until at least 50 samples with negative eigenvalues were obtained, and used only those samples
in the analysis.
The small transparent points represent individual samples, while the large filled square markers represent the arithmetic averages of both $\lambda_E^{\prime 2}/\chi_E$ and $|\delta V_\infty|$ over the retained samples. 
The solid line represents the theoretical prediction $|\delta V_\infty|=(3/2)\lambda_E^{\prime 2}/\chi_E$ given by eqn.~\eqref{eq:dU_LC}.
    }\label{fig:scatter_NL}
\end{figure}

Now, let us compare the results of our simulation with the weakly nonlinear analysis.
For simplicity, we restrict our interest to the two-cell system in this subsection.

Figures~\ref{fig:NL_Amp_etaChange}(a) and \ref{fig:NL_Amp_etaChange}(b) compare the theoretical time evolution of $|\delta V(t)|$ obtained from eqns.~\eqref{eq:A(t)} and \eqref{eq:DeltaU_t} with the simulation results. The vertical axis is shown on a logarithmic scale.
The solid lines represent the theoretical predictions, for which
$\lambda_E^\prime/\kappa=3.551629\times10^{-4}$ and
$\chi_E d_s^2/\kappa=7.839752\times10^{-1}$ are used, while the open symbols correspond to the simulation results.
In the simulations, we used a two-cell system and relaxed it using eqn.~\eqref{overdamped_dyn}. The theoretical curves describe a relaxation process governed by the amplitude equation.

Although the final potential values after relaxation, corresponding to transitions into distinct configurations, do not exactly match those predicted by theory, the weakly nonlinear theory well captures the characteristic growth behavior from the initial unstable state toward the final stable state without any fitting parameters.
The discrepancy in the long-time limit may be attributed to the fact that $\lambda_E^\prime$ and $\chi_E$ are evaluated at the saddle point, and thus cannot capture the full dynamics beyond the vicinity of that point.
Possible further instabilities or avalanche-like rearrangements during the relaxation process are also outside the scope of the present single-mode weakly nonlinear theory.

Note that the preceding analysis was conducted on a single sample. 
To assess the robustness of the results, we performed additional simulations for multiple samples, with the number of particles fixed and the pressure varied.
The resulting scatter plot is shown in Fig.~\ref{fig:scatter_NL},
while the filled squares represent the arithmetic means,
\begin{equation}
\overline{x} = \frac{1}{N_s} \sum_{s=1}^{N_s} x_s,
\qquad \overline{y} = \frac{1}{N_s}\sum_{s=1}^{N_s} y_s,
\end{equation}
where \(s\) is the index specifying the sample, and \(N_s\) is the number of samples with the introductions of
\(
x_s:=\lambda_{E,s}^{\prime 2}/{\chi_{E,s}}\) and
\(y_s:=|\delta V_{\infty,s}| .
\)
The data are distributed around the theoretical line, suggesting that the weakly nonlinear theory captures the characteristic scale of the potential-energy change statistically over many unstable configurations. 
The solid line, $y=3x/2$, represents the theoretical prediction given by eqn.~\eqref{eq:dU_LC}. 
Results for different system sizes are presented in Appendix~\ref{Ap:dV_Ndepend}.

Finally, we note that this analysis is based on weakly nonlinear overdamped dynamics and is therefore not expected to remain valid when $\alpha$ becomes too large. In particular, for sufficiently small $\eta$, inertial effects become non-negligible, and the dynamics can no longer be regarded as overdamped.

\section{Summary and discussion}\label{sec:Conclusion}

We have formulated a framework to analyze unstable dynamics in disordered solids by extending the Hessian analysis to include wave-number-dependent modulation. 
The introduction of a generalized Hessian matrix allows the identification of unstable eigenmodes that trigger structural rearrangements. 
Although the derivation of this generalized Hessian relies on uncontrolled approximations, our numerical results validate its usefulness in capturing instability and transient dynamics. 

By employing a weakly nonlinear analysis, we derived an amplitude equation
that describes the growth and saturation of unstable modes. The predicted
potential-energy drop and amplitude growth show good agreement with the
simulation results on average. Remaining discrepancies arise from
approximations inherent in the projection scheme and in the evaluation of
higher-order coefficients.

Our findings demonstrate a useful approach to analyzing instability-induced transitions in disordered systems, including the onset of plastic events and their possible connection to avalanche dynamics.

As a phenomenological approach to studying plastic deformation, one may consider the framework based on Eshelby’s inclusion theory~\cite{Eshelby1957,T.Mura1987Micromechanics,C.Weinberger2005LectureNote},
which models local rearrangements as elastic inclusions with quadrupole fields~\cite{C.Maloney2004Universal,C.Maloney2006AQS}.
Extensions of this elastic theory have shown that the emergence and alignment of Eshelby inclusions can manifest as a kind of discontinuous transition~\cite{Dasgupta2013,Lemaitre2021,Charan2023,Kumar2024,P.Desmarchelier2024TopologicalCharacterization}.
In contrast, our approach provides a theoretical prediction of the potential energy drop, which corresponds to the quantity determined by the eigenstrain in Eshelby’s theory, thereby offering a step toward quantitative predictions for Eshelby inclusions.
In future work, we aim to integrate our method with real-space theories such as Eshelby’s inclusion framework and extend it to driven systems under shear.

Another promising direction is the application of the present framework to tessellated granular metamaterials~\cite{Pashine23,J.Zhang2023Tessellated}. 
Recent experimental and numerical studies have shown that such materials can exhibit unconventional pressure-dependent elastic responses by controlling the structure of their constituent particle-filled cells.
In particular, large tessellated systems can be designed so that the shear modulus decreases with increasing pressure.  
The generalized Hessian introduced here provides a way to analyze spatially
modulated mechanical responses in periodically replicated disordered particle configurations.
It may therefore be useful for clarifying how cell-level structure and boundary
conditions determine the macroscopic response and stability of tessellated
granular metamaterials.

\section*{Author contributions}
N. Noto: Methodology, Software, Formal analysis, Investigation, Visualization, Writing – original draft, Writing – review $\&$ editing.
H. Ikeda: Writing – review $\&$ editing.
K. Saitoh: Writing – review $\&$ editing.
H. Hayakawa: Conceptualization, Supervision, Writing – original draft, Writing – review $\&$ editing.

\section*{Conflicts of interest}
There are no conflicts to declare.

\section*{Data availability}

The datasets supporting the findings of this study were generated by numerical simulations. The theoretical formulations and simulation parameters required to reproduce the results are provided in the manuscript. The numerical data underlying the figures are available from the corresponding author upon reasonable request.

\section*{Acknowledgements}
We thank Tan Van Vu for the fruitful discussions.
This work is partially supported by JSPS KAKENHI Grant Nos. JP26K06960, 22K03459, 23K13031, and 25H01401.
K.N. acknowledges Kyoto University for the support as a Research Assistant funded by the university's operating expenses.

\clearpage
\appendix
Let us explain the detailed description in the appendices as follows.
Appendix~\ref{app:preparation} explains the protocol to prepare an initial configuration of particles.
Appendix~\ref{Ap:Hij} provides an explicit expression for the Hessian matrix.
Appendix~\ref{app:Fourier_4} provides the derivations of the generalized Hessian matrix.
Appendix~\ref{app:symmetry} provides a brief explanation of symmetry-enforced degeneracy.
Appendix~\ref{app:projection} provides the details of the projection onto the critical mode.
Appendix~\ref{Ap:L_C} presents the explicit expressions for the coefficients in the nonlinear analysis with wavenumber dependence.
Appendix~\ref{Ap:multi} discusses cases in which multiple negative eigenvalues exist.
Appendix~\ref{Ap:dV_Ndepend} shows the system-size dependence of the relation between
$\lambda_E^{\prime 2}/\chi_E$ and $|\delta V_\infty|$.
Appendix~\ref{notations} presents a table for the notations used in this paper.

\section{Preparation of the initial configuration}\label{app:preparation}

We first prepare an initial configuration of grains by applying pressure control to a system with standard periodic boundary conditions. 
We replicate this cell and arrange the copies in a tiled structure, effectively creating an infinite number of them. 
In the numerical simulations, we mainly consider a finite two-cell system, where periodic boundary conditions are applied to the outer boundaries of the
composite system.

We examine cases where each cell contains $N = 8$ to $512$ particles, and the dimensionless pressure $P/\kappa$ is varied from $5 \times 10^{-2}$ to $5 \times 10^{-6}$.
Here, the contact pressure in the system is expressed as~\cite{M.Allen1987ComputerSimulation}:
\begin{equation}
P = - \frac{1}{2}L^{-2} \sum_{\langle i,j\rangle} \frac{r_{ij,a} r_{ij,a}}{r_{ij}} \frac{d V}{d r_{ij}} ,
\end{equation}
where $L$ is the linear size of the system, and $r_{ij,a}$ is the $a$-component of $\bm{r}_{ij}$ in the reference cell.
Here, the Einstein summation convention is assumed for repeated indices $a$, and $\langle i, j\rangle$ is the contacting pair of $i$ and $j$ particles.

In this study, particle configurations within each cell are generated using the following protocol:
Initially, particles with sufficiently small diameters are randomly placed.
Then, their diameters are gradually increased until the system reaches the target pressure.
After each diameter update, the particle configuration is energy-minimized
using the FIRE algorithm~\cite{E.Bitzek2006FIRE}.
Specifically, the diameter evolves via a common scaling factor $\mathcal{R}_s$ governed by the damped equation:
\begin{equation}\label{scale_R}
\frac{d^2}{dt^2}{\mathcal{R}}_s = -4\omega_0\frac{d}{dt}\mathcal{R}_s - \omega_0^2\left( \frac{P}{p} - 1 \right) \mathcal{R}_s ,
\end{equation}
where 
$\omega_0$ is set to $6 \times 10^{-2} \sqrt{\kappa/m}$,
and $p$ is the target pressure~\cite{G.Ellenbroek2009JammedFrictionlessDisks}.
The simulation proceeds until the maximum force on any particle is less than $10^{-9}\kappa d_S$ and the relative deviation of the pressure from the target value is less than $5 \times 10^{-5}$.
We adopt the same values for other FIRE parameters as in Ref.~\cite{E.Bitzek2006FIRE}.

\section{Explicit expression for \texorpdfstring{$\mathsf{H}_{ij}^{\bm{\mu}\bm{\nu}}$}{H_ij}}\label{Ap:Hij}

The first derivative of $V$ can be written as
\begin{align}
    \frac{\delta V}{\delta r_{i,a}^{\bm{\mu}}} &= \sum_{\langle\langle\bm{\zeta,\xi}\rangle\rangle}\sum_{\langle k,\ell\rangle} \frac{d U(r_{k\ell}^{\bm{\zeta}\bm{\xi}})}{d r_{k\ell}^{{\bm{\zeta}\bm{\xi}}}}\frac{\partial r_{k\ell}^{{\bm{\zeta}\bm{\xi}}}}{\partial r_{i,a}^{\bm{\mu}}}  .
\end{align}
Here, $r_{i,a}^{\bm\mu} $denotes the $a$-component of particle $i$ in cell \(\bm\mu\).
Accordingly, the Hessian matrix $H_{iajb}^{\bm{\mu}\bm{\nu}}$ is defined as
\begin{align}
    H_{iajb}^{\bm{\mu}\bm{\nu}} & := \frac{\delta^2 V}{\delta r_{i,a}^{\bm{\mu}} \delta r_{j,b}^{\bm{\nu}}}\notag \\&= \sum_{\langle\langle\bm{\zeta,\xi}\rangle\rangle}\sum_{\langle k,\ell\rangle} \frac{d^2 U(r_{k\ell}^{\bm{\zeta}\bm{\xi}})}{d r_{k\ell}^{\bm{\zeta}\bm{\xi}\,2}}\frac{\partial r_{k\ell}^{\bm{\zeta}\bm{\xi}}}{\partial r_{i,a}^{\bm{\mu}}}\frac{\partial r_{k\ell}^{\bm{\zeta}\bm{\xi}}}{\partial r_{j,b}^{\bm{\nu}}}\notag \\ & \quad +   \frac{d U(r_{k\ell}^{\bm{\zeta}\bm{\xi}})}{d r_{k\ell}^{\bm{\zeta}\bm{\xi}}}\frac{\partial^2 r_{k\ell}^{\bm{\zeta}\bm{\xi}}}{\partial r_{i,a}^{\bm{\mu}}\partial r_{j,b}^{\bm{\nu}}}.
\end{align}

Using the relation
\begin{equation}
    \frac{\partial r_{ij}^{\bm{\mu}\bm{\nu}}}{\partial r_{n,a}^{\bm{\zeta}}} = n_{ij,a}^{\bm{\mu}\bm{\nu}} ( \delta_{in}\delta_{\bm{\mu \zeta} }- \delta_{jn}\delta_{\bm{\nu \zeta} } ),
\end{equation}
we obtain
\begin{align}
    \frac{\delta^2 V}{\delta r_{i,a}^{\bm{\mu}} \delta r_{j,b}^{\bm{\nu}}} &= \sum_{\langle\langle\bm{\zeta,\xi}\rangle\rangle}\sum_{\langle k,\ell\rangle} \frac{d^2 U(r_{k\ell}^{\bm{\zeta}\bm{\xi}})}{d r_{k\ell}^{\bm{\zeta}\bm{\xi}\,2}}\delta_{k\ell,\bm{\zeta\xi}}^{i,\bm{\mu}} n_{k\ell,a}^{\bm{\zeta}\bm{\xi}}\delta_{k\ell,\bm{\zeta\xi}}^{j,\bm{\nu}} n_{k\ell,b}^{\bm{\zeta}\bm{\xi}} \notag \\ & \quad +   \frac{d U(r_{k\ell}^{\bm{\zeta}\bm{\xi}})}{d r_{k\ell}^{\bm{\zeta}\bm{\xi}}}\delta_{k\ell,\bm{\zeta\xi}}^{i,\bm{\mu}} \delta_{k\ell,\bm{\zeta\xi}}^{j,\bm{\nu}}  \frac{1}{r_{k\ell}^{\bm{\zeta}\bm{\xi}}}(\delta_{ab}-n_{k\ell,a}^{\bm{\zeta}\bm{\xi}}n_{k\ell,b}^{\bm{\zeta}\bm{\xi}}),
\end{align}
where 
$\delta_{ij,\bm{\mu\nu}}^{n,\bm{\zeta}}: = ( \delta_{in}\delta_{\bm{\mu \zeta} }- \delta_{jn}\delta_{\bm{\nu \zeta} } )$, and $n_{ij,a}^{\bm{\mu}\bm{\nu}}: = r_{ij,a}^{{\bm{\mu}\bm{\nu}}}/r_{ij}^{{\bm{\mu}\bm{\nu}}}$, with $\delta_{ij}$ denoting the Kronecker delta. 
We next consider the sum over contacting pairs $\langle k,\ell\rangle$.
For 
$\delta_{k\ell,\bm{\zeta\xi}}^{i,\bm{\mu}}
\delta_{k\ell,\bm{\zeta\xi}}^{j,\bm{\nu}}$ to be nonzero, the particle labels must satisfy
\begin{align}
     (i,j) \in \{(k,k),(k,\ell),(\ell,k),(\ell,\ell)\}.
\end{align}
Thus, for $j\neq i$, a nonzero contribution arises only from a contact
between particle $i$ in cell $\bm\mu$ and particle $j$ in cell $\bm\nu$.
Equivalently, we set $k=i$, $\ell=j$, $\bm\zeta=\bm\mu$, and
$\bm\xi=\bm\nu$.
This gives
\begin{align}
    \frac{\delta^2 V}{\delta r_{i,a}^{\bm{\mu}} \delta r_{j,b}^{\bm{\nu}}} 
    & = -\frac{d^2 U(r_{ij}^{\bm{\mu}\bm{\nu}})}{d r_{ij}^{\bm{\mu}\bm{\nu}\,2}} n_{ij,a}^{\bm{\mu}\bm{\nu}} n_{ij,b}^{\bm{\mu}\bm{\nu}} 
    \notag \\ & \quad -   \frac{d U(r_{ij}^{\bm{\mu}\bm{\nu}})}{d r_{ij}^{\bm{\mu}\bm{\nu}}} \frac{1}{r_{ij}^{\bm{\mu}\bm{\nu}}}(\delta_{ab}-n_{ij,a}^{\bm{\mu}\bm{\nu}}n_{ij,b}^{\bm{\mu}\bm{\nu}}).
\end{align}
On the other hand, when $j=i$, one may distinguish the cases
$\bm\mu=\bm\nu$ and $\bm\mu\neq\bm\nu$. Since the cell size is chosen to be
sufficiently large compared with the interaction range, a particle does not
interact with its own image in a neighboring cell. Therefore, the same-label
off-cell block $H_{ia,ib}^{\bm\mu\bm\nu}$ with $\bm\mu\neq\bm\nu$ is absent.
Thus, we only need to consider the self block with $\bm\mu=\bm\nu$.
This assumption does not exclude contacts between particle $i$ in the cell
$\bm\mu$ and other particles in neighboring cells. Such contacts contribute
to the diagonal self-block $H_{ia,ib}^{\bm\mu\bm\mu}$.
For the self block, a nonzero contribution arises from all contacts involving
particle $i$ in cell $\bm\mu$. Therefore, we set $k=i$ and $\bm\zeta=\bm\mu$,
while $\ell$ and $\bm\xi$ denote the particle and cell labels of the contacting
neighbor. This gives
\begin{align}
    \frac{\delta^2 V}{\delta r_{i,a}^{\bm{\mu}} \delta r_{i,b}^{\bm{\mu}}} 
     & = 
     \sum_{\bm{\xi}}\sum_{\ell \neq i} \frac{d^2 U(r_{i\ell}^{\bm{\mu}\bm{\xi}})}{d r_{i\ell}^{\bm{\mu}\bm{\xi}\,2}}n_{i\ell,a}^{\bm{\mu}\bm{\xi}} n_{i\ell,b}^{\bm{\mu}\bm{\xi}} \notag\\
     &\quad +   \frac{d U(r_{i\ell}^{\bm{\mu}\bm{\xi}})}{d r_{i\ell}^{\bm{\mu}\bm{\xi}}}\frac{1}{r_{i\ell}^{\bm{\mu}\bm{\xi}}}(\delta_{ab}-n_{i\ell,a}^{\bm{\mu}\bm{\xi}}n_{i\ell,b}^{\bm{\mu}\bm{\xi}}).
\end{align}

\onecolumn

\section{Derivation of eqn.~\eqref{eq_of_Fourier2}}\label{app:Fourier_4}

For simplicity, in the following theoretical derivation, we consider the limit \(L/d_s\gg1\) and take \(\mathcal N=\{\bm0\}\cup\mathcal N_1\), neglecting the second-nearest-neighbor contributions \(\mathcal N_2\). This simplification is used only in the theoretical derivation; the numerical calculations retain the contributions from \(\mathcal N_2\).
With the aid of eqn.~\eqref{eq:Fourier_transf}, eqn.~\eqref{eq:motion_eq_v2} can be rewritten as
 \begin{align}\label{eq7}
 \eta \frac{d}{dt}\hat{\bm{r}}_i(\bm{k},t) & =-\sum_{\bm{\mu}}\sum_{\bm{\nu}}\sum_j \mathsf{H}_{ij}^{\bm{\mu}\bm{\nu}}e^{i\bm{k}\cdot (\bm{R}^{\bm{\mu}}-\bm{R}^{\bm{\nu}})}\delta\bm{r}_j^{\bm{\nu}}(t)e^{i\bm{k}\cdot\bm{R}^{\bm{\nu}}} \notag\\
 & =
-\sum_{\bm\mu}
\sum_{\substack{
\bm\nu\\
\{\bm\mu,\bm\nu\}\in
\mathcal N_{\rm c}\setminus\{\{\bm\mu,\bm\mu\}\}
}}\sum_j \mathsf{H}_{ij}^{\bm{\mu}\bm{\nu}}e^{i\bm{k}\cdot (\bm{R}^{\bm{\mu}}-\bm{R}^{\bm{\nu}})}\delta\bm{r}_j^{\bm{\nu}}(t)e^{i\bm{k}\cdot\bm{R}^{\bm{\nu}}} - \sum_{\bm{\mu}}\sum_j \mathsf{H}_{ij}^{\bm{\mu}\bm{\mu}}\delta\bm{r}_j^{\bm{\nu}}(t)e^{i\bm{k}\cdot\bm{R}^{\bm{\mu}}}.
 \end{align}
Here, the first term excludes the same-cell pair
\(\{\bm\mu,\bm\mu\}\) from \(\mathcal N_{\rm c}\), while the
same-cell contribution is written separately in the second term.

To get eqn.~\eqref{eq_of_Fourier2}, we need to adopt the decoupling approximation, which can be justified if the Hessian matrix in each cell is almost independent of the location.
To demonstrate the validity of eqn.~\eqref{eq_of_Fourier2}, we trace the derivation of eqn.~\eqref{eq_of_Fourier2} step by step.

First, we rewrite the right-hand side of eqn.~\eqref{eq7} as
 \begin{align}\label{eq_Fourier}
 & \sum_{\bm\mu}
\sum_{\substack{
\bm\nu\\
\{\bm\mu,\bm\nu\}\in
\mathcal N_{\rm c}\setminus\{\{\bm\mu,\bm\mu\}\}
}}\sum_j \mathsf{H}_{ij}^{\bm{\mu}\bm{\nu}}e^{i\bm{k}\cdot (\bm{R}^{\bm{\mu}}-\bm{R}^{\bm{\nu}})}\delta\bm{r}_j^{\bm{\nu}}(t)e^{i\bm{k}\cdot\bm{R}^{\bm{\nu}}} \notag \\
 & =\sum_{\mu_x}\sum_{\mu_y}\sum_j 
 \left\{
 \mathsf{H}_{ij}^{(\mu_x,\mu_y)(\mu_x-1,\mu_y)}\delta \bm{r}_j^{(\mu_x-1,\mu_y)}(t)e^{i \bm{k}\cdot\bm{R}^{(\mu_x-1,\mu_y)}}e^{ik_xL}+\mathsf{H}_{ij}^{(\mu_x,\mu_y)(\mu_x+1,\mu_y)}\delta \bm{r}_j^{(\mu_x+1,\mu_y)}(t)e^{i \bm{k}\cdot\bm{R}^{(\mu_x-1,\mu_y)}}e^{-ik_xL}
\right. \notag\\
 &\left. \quad  + 
  \mathsf{H}_{ij}^{(\mu_x,\mu_y)(\mu_x,\mu_y-1)}\delta \bm{r}_j^{(\mu_x,\mu_y-1)}(t)e^{i \bm{k}\cdot\bm{R}^{(\mu_x,\mu_y-1)}}e^{ik_yL}+\mathsf{H}_{ij}^{(\mu_x,\mu_y)(\mu_x,\mu_y+1)}\delta \bm{r}_j^{(\mu_x,\mu_y+1)}(t)e^{i \bm{k}\cdot\bm{R}^{(\mu_x,\mu_y+1)}}e^{-ik_y L}
 \right\}
\end{align}
where we have used $|\bm{R}^\mu-\bm{R}^\nu|=L$ for the adjacent cells, with the notation $\bm{\mu}=(\mu_x,\mu_y)=\mu_x\bm{e}_x+\mu_y \bm{e}_y$ using the unit vector $\bm{e}_a$ in $a$- direction. 
Here, we take the limit $d_s/L \to 0$.
We have also introduced $k_x$ and $k_y$ as $x$-component and $y$-component of $\bm{k}$, respectively.
Noting that $e^{\pm i k_xL}$ and $e^{\pm ik_y L}$ are independent of the cell $\bm{\mu}=(\mu_x,\mu_y)$, we can rewrite eqn.~\eqref{eq_Fourier} as
\begin{align}\label{step2}
  & \sum_{\bm\mu}
\sum_{\substack{
\bm\nu\\
\{\bm\mu,\bm\nu\}\in
\mathcal N_{\rm c}\setminus\{\{\bm\mu,\bm\mu\}\}
}}\sum_j \mathsf{H}_{ij}^{\bm{\mu}\bm{\nu}}e^{i\bm{k}\cdot (\bm{R}^{\bm{\mu}}-\bm{R}^{\bm{\nu}})}\delta\bm{r}_j^{\bm{\nu}}(t)e^{i\bm{k}\cdot\bm{R}^{\bm{\nu}}} \notag \\ \quad 
 &=e^{ik_xL}\sum_{\bm{\mu},j}  \mathsf{H}_{ij}^{(\mu_x,\mu_y)(\mu_x-1,\mu_y)}\delta \bm{r}_j^{(\mu_x-1,\mu_y)}(t)e^{i \bm{k}\cdot\bm{R}^{(\mu_x-1,\mu_y)}}+
 e^{-k_xL}\sum_{\bm{\mu},j}\mathsf{H}_{ij}^{(\mu_x,\mu_y)(\mu_x+1,\mu_y)}\delta \bm{r}_j^{(\mu_x+1,\mu_y)}(t)e^{i \bm{k}\cdot\bm{R}^{(\mu_x+1,\mu_y)}}
 \notag\\
 &\quad +e^{ik_yL}\sum_{\bm{\mu},j}\mathsf{H}_{ij}^{(\mu_x,\mu_y)(\mu_x,\mu_y-1)}\delta \bm{r}_j^{(\mu_x,\mu_y-1)}(t)e^{i \bm{k}\cdot\bm{R}^{(\mu_x,\mu_y-1)}}
 +e^{-ik_yL}\sum_{\bm{\mu},j}\mathsf{H}_{ij}^{(\mu_x,\mu_y)(\mu_x,\mu_y+1)}\delta \bm{r}_j^{(\mu_x,\mu_y+1)}(t)e^{i \bm{k}\cdot\bm{R}^{(\mu_x,\mu_y+1)}} .
\end{align}
According to the periodic cell ansatz, we assume that $\mathsf{H}_{ij}^{\bm{\mu}\bm{\nu}}$ depends only weakly on $\bm{\mu}$ and $\bm{\nu}$.

Thus, we may rewrite
$\mathsf{H}_{ij}^{(\mu_x,\mu_y)}(\mp L,0):=\mathsf{H}_{ij}^{(\mu_x,\mu_y),(\mu_x\pm 1,\mu_y)}=\mathsf{H}_{ij}^{(\mu_x \mp 1,\mu_y)(\mu_x,\mu_y)} + \Delta H^{\mp}_x$ with 
$\Delta H^{\mp}_x := \mathsf{H}_{ij}^{(\mu_x,\mu_y)(\mu_x\mp1,\mu_y)} - \mathsf{H}_{ij}^{(\mu_x \pm 1,\mu_y)(\mu_x,\mu_y)}$.
Similary, $\mathsf{H}_{ij}^{(\mu_x,\mu_y)}(0,\mp L):=\mathsf{H}_{ij}^{(\mu_x,\mu_y),(\mu_x,\mu_y\pm 1)}=\mathsf{H}_{ij}^{(\mu_x,\mu_y \mp 1)(\mu_x,\mu_y)} + \Delta H^{\mp}_y$ with 
$\Delta H^{\mp}_y := \mathsf{H}_{ij}^{(\mu_x,\mu_y)(\mu_x,\mu_y\mp1)} - \mathsf{H}_{ij}^{(\mu_x ,\mu_y\pm 1)(\mu_x,\mu_y)}$.
Since the inhomogeneity is weak, we can ignore the terms $O(\Delta H^\mp_x)$, and $O(\Delta H^\mp_y)$. 
Thus, we may ignore $(\mu_x,\mu_y)$ dependence of $\mathsf{H}_{ij}^{(\mu_x,\mu_y)}(\mp L,0)$,and $\mathsf{H}_{ij}^{(\mu_x,\mu_y)}(0,\mp L)$.
Then we introdece $\mathsf{H}_{ij}(\mp L,0) := \mathsf{H}_{ij}^{(\mu_x,\mu_y)}(\mp L,0)$, and $\mathsf{H}_{ij}(0,\mp L) := \mathsf{H}_{ij}^{(\mu_x,\mu_y)}(0,\mp L)$.

When we adopt this approximation and eqn.~\eqref{eq:Fourier_transf}, we obtain
\begin{align}\label{1st_eq7}
     \sum_{\bm\mu}
\sum_{\substack{
\bm\nu\\
\{\bm\mu,\bm\nu\}\in
\mathcal N_{\rm c}\setminus\{\{\bm\mu,\bm\mu\}\}
}}\sum_j \mathsf{H}_{ij}^{\bm{\mu}\bm{\nu}}e^{i\bm{k}\cdot (\bm{R}^{\bm{\mu}}-\bm{R}^{\bm{\nu}})}\delta\bm{r}_j^{\bm{\nu}}(t)e^{i\bm{k}\cdot\bm{R}^{\bm{\nu}}}
 &\approx e^{ik_x L}\sum_j\mathsf{H}_{ij}(L,0)\sum_{\bm{\mu}} \delta \bm{r}_j^{(\mu_x-1,\mu_y)}(t)e^{i \bm{k}\cdot\bm{R}^{(\mu_x-1,\mu_y)}}+ e^{-ik_x L}\sum_j\mathsf{H}_{ij}(-L,0)\sum_{\bm{\mu}} \delta \bm{r}_j^{(\mu_x+1,\mu_y)}(t)e^{i \bm{k}\cdot\bm{R}^{(\mu_x+1,\mu_y)}}
\notag\\
&\quad +e^{ik_y L}\sum_{j}\mathsf{H}_{ij}(0,L)\sum_{\bm{\mu}}\delta \bm{r}_j^{(\mu_x,\mu_y-1)}(t)e^{i \bm{k}\cdot\bm{R}^{(\mu_x,\mu_y-1)}}
+e^{-ik_y L} \sum_j \mathsf{H}_{ij}(0,-L)\sum_{\bm{\mu}}\delta \bm{r}_j^{(\mu_x,\mu_y+1)}(t)e^{i \bm{k}\cdot\bm{R}^{(\mu_x,\mu_y+1)}}
\notag\\
&=e^{ik_x L}\sum_j \mathsf{H}_{ij}(L,0)\hat{\bm{r}}_j(\bm{k},t)+e^{-ik_xL}\sum_j\mathsf{H}_{ij}(-L,0)\hat{\bm{r}}_j(\bm{k},t) \notag\\
&\quad +e^{ik_y L}\sum_j \mathsf{H}_{ij}(0,L)\hat{\bm{r}}_j(\bm{k},t)+e^{-ik_yL}\sum_j \mathsf{H}_{ij}(0,-L)\hat{\bm{r}}_j(\bm{k},t)
.
\end{align}
Since $\mathsf{H}_{ij}^{\bm{\mu\mu}}$ depends on $\bm{\mu}$ weakly, we may use $\mathsf{H}_{ij}(0,0) : = \mathsf{H}_{ij}^{\bm{\mu\mu}}$. 
Then, the second term on the right-hand side of eqn.~\eqref{eq7} can be rewritten as
\begin{align}\label{2nd_eq7}
\sum_{\bm{\mu}}\sum_j\mathsf{H}_{ij}^{\bm{\mu\mu}}\delta\bm{r}_j(t)e^{i\bm{k}\cdot\bm{R^\mu}} & \approx \sum_j \mathsf{H}_{ij}(0,0) \sum_{\bm{\mu}}\delta\bm{r}_j(t)e^{i\bm{k}\cdot\bm{R^\mu}} \notag \\
& = \sum_j \mathsf{H}_{ij}(0,0) \hat{\bm{r}}_j(\bm{k},t).
\end{align}

Substituting Eqs.~\eqref{1st_eq7} and \eqref{2nd_eq7} into eqn.~\eqref{eq7}, we obtain
\begin{align}
&\sum_{\bm\mu}
\sum_{\substack{
\bm\nu\\
\{\bm\mu,\bm\nu\}\in
\mathcal N_{\rm c}\setminus\{\{\bm\mu,\bm\mu\}\}
}}\sum_j \mathsf{H}_{ij}^{\langle{\bm{\mu}\bm{\nu}}\rangle}e^{i\bm{k}\cdot (\bm{R}^{\bm{\mu}}-\bm{R}^{\bm{\nu}})}\delta\bm{r}_j(t)e^{i\bm{k}\cdot\bm{R}^{\bm{\nu}}} + \sum_{\bm{\mu}}\sum_j \mathsf{H}_{ij}^{\bm{\mu}\bm{\mu}}\delta\bm{r}_j(t)e^{i\bm{k}\cdot\bm{R}^{\bm{\mu}}}\notag \\
&\approx e^{ik_xL}\sum_j \mathsf{H}_{ij}(L,0)\hat{\bm{r}}_j(\bm{k},t)+e^{-ik_xL}\sum_j\mathsf{H}_{ij}(-L,0)\hat{\bm{r}}_j(\bm{k},t)
+e^{ik_y L}\sum_j \mathsf{H}_{ij}(0,L)\hat{\bm{r}}_j(\bm{k},t)+e^{-ik_yL}\sum_j \mathsf{H}_{ij}(0,-L)\hat{\bm{r}}_j(\bm{k},t) + \sum_j \mathsf{H}_{ij}(0,0) \hat{\bm{r}}_j(\bm{k},t) .
\end{align}

Using the cell-offset set $\mathcal N$ and the cell translation vector
$\bm L_{\bm\Delta}:=L\bm\Delta=L(\bm{\mu}-\bm{\nu})$ defined in the main text,
eqn.~\eqref{eq7} can be rewritten as
\begin{align}
\eta \frac{d}{dt}\hat{\bm r}_i(\bm k,t)
&\approx
-\sum_{\bm\Delta\in\mathcal N_1}
\sum_j
\mathsf H_{ij}(\bm L_{\bm\Delta})
e^{i\bm k\cdot \bm L_{\bm\Delta}}
\hat{\bm r}_j(\bm k,t)-\sum_j \mathsf H_{ij}(\bm0)\hat{\bm r}_j(\bm k,t) ,
\end{align}
where $\mathcal{N}_1$ is introduced in Eq.~\eqref{def:N_1}.
This can be rearranged as
\begin{align}
\eta \frac{d}{dt}\hat{\bm r}_i(\bm k,t)
&\approx
-\sum_j
\bigg[
\mathsf H_{ij}(\bm0)
+
\sum_{\bm\Delta\in\mathcal N_1}
\mathsf H_{ij}(\bm L_{\bm\Delta})
\bigg]
\hat{\bm r}_j(\bm k,t)
-\sum_j
\sum_{\bm\Delta\in\mathcal N_1}
\mathsf H_{ij}(\bm L_{\bm\Delta})
\Bigl(
e^{i\bm k\cdot \bm L_{\bm\Delta}}-1
\Bigr)
\hat{\bm r}_j(\bm k,t).
\end{align}

To connect this expression with the notation used in the main text, we identify
\begin{equation}
\mathsf H_{ij}(\bm0)
=
\mathsf H_{ij}^{\bm0,\bm0},
\qquad
\mathsf H_{ij}(\bm L_{\bm\Delta})
:=
\mathsf H_{ij}^{\bm0,-\bm\Delta},
\end{equation}
Here, \(\bm0\) denotes the reference cell, and
\(\bm L_{\bm\Delta}\) represents the displacement from the cell
\(-\bm\Delta\) to the reference cell \(\bm0\).
With this identification, the above expression is written as
\begin{align}
\eta \frac{d}{dt}\hat{\bm r}_i(\bm k,t)
&\approx
-\sum_j
\sum_{\bm\Delta\in\mathcal N}
\mathsf H_{ij}^{\bm0,-\bm\Delta}
e^{i\bm k\cdot\bm L_{\bm\Delta}}
\hat{\bm r}_j(\bm k,t)
\notag\\
&=
-\sum_j
\mathsf D_{ij}(\bm k)
\hat{\bm r}_j(\bm k,t),
\end{align}
where \(\mathcal N=\{\bm0\}\cup\mathcal N_1\). This is equivalent to
eqn.~\eqref{eq_of_Fourier2}.

\twocolumn

\section{Symmetry-enforced degeneracy}
\label{app:symmetry}

In this appendix, we briefly explain why spatial symmetry can enforce
eigenvalue degeneracies.

Suppose that the generalized Hessian $D(\bm{k})$ is invariant under a
symmetry operation $C$,
\begin{equation}
CD(\bm{k})C^{-1}=D(\bm{k}).
\end{equation}
If $\hat{\bm{\varepsilon}}$ is an eigenvector of $D(\bm{k})$ with
eigenvalue $\lambda$,
\begin{equation}
D(\bm{k})\hat{\bm{\varepsilon}}
=
\lambda
\hat{\bm{\varepsilon}},
\end{equation}
then applying the symmetry operation gives
\begin{align}
D(\bm{k})(C\hat{\bm{\varepsilon}})
&=
CD(\bm{k})\hat{\bm{\varepsilon}}
\\
&=
\lambda
(C\hat{\bm{\varepsilon}}).
\end{align}
Therefore, $C\hat{\bm{\varepsilon}}$ is also an eigenvector with the
same eigenvalue.

If $C\hat{\bm{\varepsilon}}$ is linearly independent of
$\hat{\bm{\varepsilon}}$, the two independent eigenvectors share the
same eigenvalue, and the eigenvalue is therefore degenerate.
This is the origin of symmetry-enforced degeneracies in crystalline
solids.

In crystals, point-group symmetries can generate such independent
symmetry-related eigenvectors, giving rise to degenerate eigenvalues
\cite{Dresselhaus2008}.
In contrast, the disordered configurations considered in the present
work possess no crystalline point-group symmetry.
Therefore, no symmetry operation exists that can enforce such
degeneracies.

\section{Projection onto the critical mode}
\label{app:projection}

In this appendix, we explicitly show how the projection onto the
critical mode removes the nonresonant nonlinear terms and leads to
Eq.~\eqref{eq:Stuart_c}.
Substituting Eq.~\eqref{eq:intK} into Eq.~\eqref{non_linear_eq},
we multiply both sides by
$e^{i\bm{k}_c\cdot\bm{R}^{\bm{\mu}}}$,
sum over the cell index $\bm{\mu}$,
and contract with
$\hat{\bm{\varepsilon}}_1^\dagger(\bm{k}_c)$,
as in the main text.

The key point is that this projection extracts only the Fourier
component with the critical wave vector.
Indeed, the cell summation satisfies
\begin{equation}
\sum_{\bm\mu}
e^{i\bm q\cdot\bm R^{\bm\mu}}
=
\begin{cases}
N_{\rm cell},
&
\bm q=\bm G,
\\
0,
&
\bm q\neq\bm G,
\end{cases}
\label{eq:wavevector_selection}
\end{equation}
where $N_{\rm cell}$ is the total number of cells,
$\bm G$ is a reciprocal lattice vector, and
$\bm q$ denotes the residual wave vector after multiplication by
$e^{i\bm k_c\cdot\bm R^{\bm\mu}}$.
Thus, only terms with no residual oscillating phase, modulo a reciprocal lattice vector, survive the summation over $\bm\mu$.

For simplicity, we retain only the phase factors carried by each contribution.
The quadratic term contains
\[
A_c^2e^{-2i\bm k_c\cdot\bm R^{\bm\mu}}
,\qquad
|A_c|^2
, \qquad 
A_c^{*2}e^{2i\bm k_c\cdot\bm R^{\bm\mu}}.
\]
After multiplication by
$e^{i\bm k_c\cdot\bm R}$,
these become
\[
A_c^2e^{-i\bm k_c\cdot\bm R^{\bm\mu}},
\qquad
|A_c|^2e^{i\bm k_c\cdot\bm R^{\bm\mu}},
\qquad
A_c^{*2}e^{3i\bm k_c\cdot\bm R^{\bm\mu}}.
\]
After multiplication by
$e^{i\bm k_c\cdot\bm R^{\bm\mu}}$,
they retain oscillating phase factors and therefore vanish upon the
summation over the cells.
Therefore, all quadratic contributions vanish after the projection.

Similarly, the cubic nonlinear term produces
\[
A_c^3e^{-3i\bm k_c\cdot\bm R^{\bm\mu}}
, \quad 
A_c^2A_c^*
e^{-i\bm k_c\cdot\bm R^{\bm\mu}}
, \quad 
A_cA_c^{*2}
e^{i\bm k_c\cdot\bm R^{\bm\mu}}
, \quad 
A_c^{*3}
e^{3i\bm k_c\cdot\bm R^{\bm\mu}}.
\]
After the projection, these become
\[
A_c^3e^{-2i\bm k_c\cdot\bm R^{\bm\mu}},
\quad
A_c^2A_c^*,
\quad
A_cA_c^{*2}
e^{2i\bm k_c\cdot\bm R^{\bm\mu}},
\quad
A_c^{*3}
e^{4i\bm k_c\cdot\bm R^{\bm\mu}}.
\]
The first, third, and fourth terms retain oscillating phase factors
after multiplication by
$e^{i\bm k_c\cdot\bm R^{\bm\mu}}$
and therefore vanish upon summation over the cells.
Only
\[
A_c^2A_c^*
=
|A_c|^2A_c
\]
contains no remaining phase factor after the projection.
Therefore, it survives the summation over the cells.
This is the leading resonant nonlinear contribution.

Consequently, the quadratic term involving
$\mathsf T_{ijk}$ does not contribute to the projected equation,
whereas the fourth-order tensor
$\mathsf M_{ijk\ell}$
produces the leading nonlinear term
proportional to
$|A_c|^2A_c$.
Combining this contribution with the linear term gives
Eq.~\eqref{eq:Stuart_c}.

\clearpage
\onecolumn
\section{Expansion coefficients}\label{Ap:L_C}
We explicitly provide the expressions for $\lambda_E$ and $\chi_E$.
For concreteness, we take the edge wave vector along the
\(x\)-direction,
\(\bm k_E=(\pi/L)\bm e_x\).
The \(y\)-directed case follows analogously.
The Cartesian components of the generalized Hessian introduced in
Eq.~\eqref{D_{ij}} are defined as
\begin{align}
D_{ia,jb}(\bm{k})
:=
\sum_{\bm{\Delta}_{\bm{\mu\nu}}}
H_{ia,jb}^{\bm 0,-\bm{\Delta}_{\bm{\mu\nu}}}
e^{i\bm{k}\cdot\bm{L}_{\bm{\Delta}_{\bm{\mu\nu}}}}.
\end{align}
The eigenvalue at the edge wave vector $\bm{k}_E$ can be written as the Rayleigh quotient
\begin{align}
    \lambda_E &= \sum_{ i ,j } \sum_{a,b} D_{i a,jb}(\bm{k}_E)\hat{\varepsilon}_{i,a}^*(\bm{k}_E)\hat{\varepsilon}_{j,b}(\bm{k}_E),\notag\\
    & = \sum_{ i ,j } \sum_{\bm{\Delta}_{\bm{\mu\nu}}}  \sum_{a,b}H_{ia,jb}^{\bm 0,-\bm{\Delta}_{\bm{\mu\nu}}}e^{i\bm{k}_E\cdot\bm{L}_{\bm{\Delta}_{\bm{\mu\nu}}}} \hat{\varepsilon}_{i,a}^*(\bm{k}_E)\hat{\varepsilon}_{j,b}(\bm{k}_E).
\end{align}
At the edge wave vector, the eigenmode can be chosen to be real.
Thus, we omit the complex conjugate in the following.
Substituting the explicit pairwise expression for
$D_{i a,jb}(\bm{k}_E)$, we obtain
\begin{align}
   \lambda_E
      & = 
    \sum_{ i ,j }\sum_{\bm{\Delta}_{\bm{\mu\nu}}}\sum_{\langle\langle\bm{\zeta,\xi}\rangle\rangle}\sum_{\langle k,\ell\rangle}  \sum_{a,b}\bigg(\frac{d^2 U(r_{k\ell}^{\bm{\zeta}\bm{\xi}})}{d r_{k\ell}^{\bm{\zeta}\bm{\xi}\,2}}\delta_{k\ell,\bm{\zeta\xi}}^{i,\bm{0}} n_{k\ell,a}^{\bm{\zeta}\bm{\xi}}\delta_{k\ell,\bm{\zeta\xi}}^{j,-\bm{\Delta}_{\bm{\mu\nu}}} n_{k\ell,b}^{\bm{\zeta}\bm{\xi}} +   \frac{d U(r_{k\ell}^{\bm{\zeta}\bm{\xi}})}{d r_{k\ell}^{\bm{\zeta}\bm{\xi}}}\delta_{k\ell,\bm{\zeta\xi}}^{i,\bm{0}} \delta_{k\ell,\bm{\zeta\xi}}^{j,-\bm{\Delta}_{\bm{\mu\nu}}}  \frac{1}{r_{k\ell}^{\bm{\zeta}\bm{\xi}}}(\delta_{ab}-n_{k\ell,a}^{\bm{\zeta}\bm{\xi}}n_{k\ell,b}^{\bm{\zeta}\bm{\xi}}) \bigg)e^{i\bm{k}_E\cdot\bm{L}_{\bm{\Delta}_{\bm{\mu\nu}}}} \hat{\varepsilon}_{i,a}(\bm{k}_E)\hat{\varepsilon}_{j,b}(\bm{k}_E).
    \end{align}
We now rewrite this expression as a sum over contact pairs. To this end, we first carry out the contractions over the particle indices $i,j$ and the cell-offset index $\bm\Delta_{\bm{\mu\nu}}$.
Using the definition of
$\delta_{k\ell,\bm{\zeta\xi}}^{i,\bm0}$, we have
\begin{align}
    \sum_{\bm{\Delta}_{\bm{\mu\nu}}}\delta_{k\ell,\bm{\zeta\xi}}^{i,\bm{0}} \delta_{k\ell,\bm{\zeta\xi}}^{j,-\bm{\Delta}_{\bm{\mu\nu}}} e^{i\bm{k}_E\cdot\bm{L}_{\bm{\Delta}_{\bm{\mu\nu}}}} &= 
    \sum_{\bm{\Delta}_{\bm{\mu\nu}}}(\delta_{ik}\delta_{\bm0\bm\zeta} - \delta_{i\ell}\delta_{\bm0\bm\xi})(\delta_{jk}\delta_{-\bm\Delta_{\bm{\mu\nu}},\bm\zeta} - \delta_{j\ell}\delta_{-\bm\Delta_{\bm{\mu\nu}},\bm\xi})e^{i\bm{k}_E\cdot\bm{L}_{\bm{\Delta}_{\bm{\mu\nu}}}}\\
    & = (\delta_{ik}\delta_{\bm0\bm\zeta} - \delta_{i\ell}\delta_{\bm0\bm\xi})
    (\delta_{jk}e^{i\bm{k}_E\cdot\bm{L}_{-\bm{\zeta}}} - \delta_{j\ell}e^{i\bm{k}_E\cdot\bm{L}_{-\bm{\xi}}}).
\end{align}
Here, $\bm L_{\bm\Delta_{\bm{\mu\nu}}}=L\bm\Delta_{\bm{\mu\nu}}$, with $\bm\Delta_{\bm{\mu\nu}}=m_x\bm e_x+m_y\bm e_y$ and $(m_x,m_y)\in\{0,\pm1\}$. Similarly, $\bm L_{-\bm\zeta}=L(-\zeta_x\bm e_x-\zeta_y\bm e_y)=-\bm R^{\bm\zeta}$. The cell indices $\bm\zeta$ and $\bm\xi$ denote the cells containing particles $k$ and $\ell$, respectively.
Therefore, after performing the contraction over $\bm\Delta_{\bm{\mu\nu}}$, the Rayleigh quotient becomes
\begin{align} 
\lambda_E 
& = 
    \sum_{ i ,j }\sum_{\langle\langle\bm{\zeta,\xi}\rangle\rangle}\sum_{\langle k,\ell\rangle} \sum_{a,b} \bigg(\frac{d^2 U(r_{k\ell}^{\bm{\zeta}\bm{\xi}})}{d r_{k\ell}^{\bm{\zeta}\bm{\xi}\,2}}n_{k\ell,a}^{\bm{\zeta}\bm{\xi}}n_{k\ell,b}^{\bm{\zeta}\bm{\xi}} +   \frac{d U(r_{k\ell}^{\bm{\zeta}\bm{\xi}})}{d r_{k\ell}^{\bm{\zeta}\bm{\xi}}}  \frac{1}{r_{k\ell}^{\bm{\zeta}\bm{\xi}}}(\delta_{ab}-n_{k\ell,a}^{\bm{\zeta}\bm{\xi}}n_{k\ell,b}^{\bm{\zeta}\bm{\xi}}) \bigg)\hat{\varepsilon}_{i,a}(\bm{k}_E)\hat{\varepsilon}_{j,b}(\bm{k}_E)(\delta_{ik}\delta_{\bm0\bm\zeta} - \delta_{i\ell}\delta_{\bm0\bm\xi})
    (\delta_{jk}e^{-i\bm{k}_E\cdot\bm{R}^{\bm\zeta}} - \delta_{j\ell}e^{-i\bm{k}_E\cdot\bm{R}^{\bm\xi}})
    \notag
\\
& = \sum_{\langle\langle\bm{\zeta,\xi}\rangle\rangle}\sum_{\langle k,\ell\rangle} \bigg( \biggl(\frac{d^2 U(r_{k\ell}^{\bm{\zeta}\bm{\xi}})}{d r_{k\ell}^{\bm{\zeta}\bm{\xi}\,2}} -   \frac{d U(r_{k\ell}^{\bm{\zeta}\bm{\xi}})}{d r_{k\ell}^{\bm{\zeta}\bm{\xi}}}  \frac{1}{r_{k\ell}^{\bm{\zeta}\bm{\xi}}}  \biggr)\biggl(
(\bm{n}_{k\ell}^{\bm{\zeta}\bm{\xi}}\cdot \bm{\hat{\varepsilon}}_{k})(\bm{n}_{k\ell}^{\bm{\zeta}\bm{\xi}}\cdot \bm{\hat{\varepsilon}}_{k})\delta_{\bm0 \bm\zeta}e^{-i\bm{k}_E\cdot\bm{R}^{\bm\zeta}}+ (\bm{n}_{k\ell}^{\bm{\zeta}\bm{\xi}}\cdot \bm{\hat{\varepsilon}}_{\ell})(\bm{n}_{k\ell}^{\bm{\zeta}\bm{\xi}}\cdot \bm{\hat{\varepsilon}}_{\ell})
\delta_{\bm0 \bm\xi}e^{-i\bm{k}_E\cdot\bm{R}^{\bm\xi}}
\notag\\ & \quad 
- (\bm{n}_{k\ell}^{\bm{\zeta}\bm{\xi}}\cdot \bm{\hat{\varepsilon}}_{k})(\bm{n}_{k\ell}^{\bm{\zeta}\bm{\xi}}\cdot \bm{\hat{\varepsilon}}_{\ell})
\delta_{\bm0 \bm\zeta}e^{-i\bm{k}_E\cdot\bm{R}^{\bm\xi}}
- (\bm{n}_{k\ell}^{\bm{\zeta}\bm{\xi}}\cdot \bm{\hat{\varepsilon}}_{\ell})(\bm{n}_{k\ell}^{\bm{\zeta}\bm{\xi}}\cdot \bm{\hat{\varepsilon}}_{k})
\delta_{\bm0 \bm\xi}e^{-i\bm{k}_E\cdot\bm{R}^{\bm\zeta}}
\biggr)
\notag\\ & \quad 
+   \frac{d U(r_{k\ell}^{\bm{\zeta}\bm{\xi}})}{d r_{k\ell}^{\bm{\zeta}\bm{\xi}}}  \frac{1}{r_{k\ell}^{\bm{\zeta}\bm{\xi}}} 
\biggl((\bm{\hat{\varepsilon}}_{k}\cdot \bm{\hat{\varepsilon}}_{k})\delta_{\bm0 \bm\zeta}e^{-i\bm{k}_E\cdot\bm{R}^{\bm\zeta}} 
+  (\bm{\hat{\varepsilon}}_{\ell}\cdot \bm{\hat{\varepsilon}}_{\ell}) \delta_{\bm0 \bm\xi}e^{-i\bm{k}_E\cdot\bm{R}^{\bm\xi}}
    -  (\bm{\hat{\varepsilon}}_{k}\cdot \bm{\hat{\varepsilon}}_{\ell})\delta_{\bm0 \bm\zeta}e^{-i\bm{k}_E\cdot\bm{R}^{\bm\xi}}
    -  (\bm{\hat{\varepsilon}}_{\ell}\cdot \bm{\hat{\varepsilon}}_{k})\delta_{\bm0 \bm\xi}e^{-i\bm{k}_E\cdot\bm{R}^{\bm\zeta}}\biggr) \bigg).
\end{align}
We next perform the contraction over the cell indices. In doing so, we choose the representative particle pair with $k<\ell$ in order to avoid double counting of the particle labels.
For an arbitrary function $f$, the Kronecker deltas selecting the reference cell give
$\sum_{\langle \langle \bm{\zeta}, \bm{\xi} \rangle \rangle}f(r_{k\ell}^{\bm{\zeta}\bm{\xi}})\delta_{\bm 0 \bm \zeta}e^{-i\bm{k}_E\cdot\bm{R}^{\bm\zeta}} = \sum_{\bm{\xi} }f(r_{k\ell}^{\bm{0}\bm{\xi}})\text{exp}[-i \bm k_EL(0\bm e_x+ 0 \bm e_y)] = \sum_{\bm{\xi} }f(r_{k\ell}^{\bm{0}\bm{\xi}})$,
$\sum_{\langle \langle \bm{\zeta}, \bm{\xi} \rangle \rangle}f(r_{k\ell}^{\bm{\zeta}\bm{\xi}})\delta_{\bm 0 \bm \zeta}e^{-i\bm{k}_E\cdot\bm{R}^{\bm\xi}} = \sum_{\bm{\xi} }f(r_{k\ell}^{\bm{0}\bm{\xi}})\text{exp}[-i \bm k_EL(\xi_x\bm e_x+ \xi_y\bm e_y)] $.
Using these identities, we obtain
\begin{align} 
\lambda_E 
& =
\sum_{\langle k,\ell \rangle;\,k<\ell}
\biggl(
 \sum_{\bm\xi}\bigg(\frac{d^2 U(r_{k\ell}^{\bm{0}\bm{\xi}})}{d r_{k\ell}^{\bm{0}\bm{\xi}\,2}} -   \frac{d U(r_{k\ell}^{\bm{0}\bm{\xi}})}{d r_{k\ell}^{\bm{0}\bm{\xi}}}  \frac{1}{r_{k\ell}^{\bm{0}\bm{\xi}}}  \biggr)
(\bm{n}_{k\ell}^{\bm{0}\bm{\xi}}\cdot \bm{\hat{\varepsilon}}_{k})(\bm{n}_{k\ell}^{\bm{0}\bm{\xi}}\cdot \bm{\hat{\varepsilon}}_{k})+ 
\sum_{\bm\zeta}\bigg(\frac{d^2 U(r_{k\ell}^{\bm{\zeta}\bm{0}})}{d r_{k\ell}^{\bm{\zeta}\bm{0}\,2}} -   \frac{d U(r_{k\ell}^{\bm{\zeta}\bm{0}})}{d r_{k\ell}^{\bm{\zeta}\bm{0}}}  \frac{1}{r_{k\ell}^{\bm{\zeta}\bm{0}}}  \biggr)(\bm{n}_{k\ell}^{\bm{\zeta}\bm{0}}\cdot \bm{\hat{\varepsilon}}_{\ell})(\bm{n}_{k\ell}^{\bm{\zeta}\bm{0}}\cdot \bm{\hat{\varepsilon}}_{\ell})
\notag\\ & \quad 
- 
\sum_{\bm\xi}\bigg(\frac{d^2 U(r_{k\ell}^{\bm{0}\bm{\xi}})}{d r_{k\ell}^{\bm{0}\bm{\xi}\,2}} -   \frac{d U(r_{k\ell}^{\bm{0}\bm{\xi}})}{d r_{k\ell}^{\bm{0}\bm{\xi}}}  \frac{1}{r_{k\ell}^{\bm{0}\bm{\xi}}}  \biggr)(\bm{n}_{k\ell}^{\bm{0}\bm{\xi}}\cdot \bm{\hat{\varepsilon}}_{k})(\bm{n}_{k\ell}^{\bm{0}\bm{\xi}}\cdot \bm{\hat{\varepsilon}}_{\ell})
e^{-i\bm{k}_E\cdot\bm{R}^{\bm\xi}}
- \sum_{\bm\zeta}\bigg(\frac{d^2 U(r_{k\ell}^{\bm{\zeta}\bm{0}})}{d r_{k\ell}^{\bm{\zeta}\bm{0}\,2}} -   \frac{d U(r_{k\ell}^{\bm{\zeta}\bm{0}})}{d r_{k\ell}^{\bm{\zeta}\bm{0}}}  \frac{1}{r_{k\ell}^{\bm{\zeta}\bm{0}}}  \biggr)(\bm{n}_{k\ell}^{\bm{\zeta}\bm{0}}\cdot \bm{\hat{\varepsilon}}_{\ell})(\bm{n}_{k\ell}^{\bm{\zeta}\bm{0}}\cdot \bm{\hat{\varepsilon}}_{k})
e^{-i\bm{k}_E\cdot\bm{R}^{\bm\zeta}}
\notag  \\ & \quad + 
\sum_{\bm\xi}\frac{d U(r_{k\ell}^{\bm{0}\bm{\xi}})}{d r_{k\ell}^{\bm{0}\bm{\xi}}}  \frac{1}{r_{k\ell}^{\bm{0}\bm{\xi}}} (\bm{\hat{\varepsilon}}_{k}\cdot \bm{\hat{\varepsilon}}_{k})
+ \sum_{\bm\zeta} \frac{d U(r_{k\ell}^{\bm{\zeta}\bm{0}})}{d r_{k\ell}^{\bm{\zeta}\bm{0}}}  \frac{1}{r_{k\ell}^{\bm{\zeta}\bm{0}}} (\bm{\hat{\varepsilon}}_{\ell}\cdot \bm{\hat{\varepsilon}}_{\ell}) 
    -  \sum_{\bm\xi}\frac{d U(r_{k\ell}^{\bm{0}\bm{\xi}})}{d r_{k\ell}^{\bm{0}\bm{\xi}}}  \frac{1}{r_{k\ell}^{\bm{0}\bm{\xi}}}  (\bm{\hat{\varepsilon}}_{k}\cdot \bm{\hat{\varepsilon}}_{\ell})e^{-i\bm{k}_E\cdot\bm{R}^{\bm\xi}}
    -  \sum_{\bm\zeta} \frac{d U(r_{k\ell}^{\bm{\zeta}\bm{0}})}{d r_{k\ell}^{\bm{\zeta}\bm{0}}}  \frac{1}{r_{k\ell}^{\bm{\zeta}\bm{0}}} (\bm{\hat{\varepsilon}}_{\ell}\cdot \bm{\hat{\varepsilon}}_{k})e^{-i\bm{k}_E\cdot\bm{R}^{\bm\zeta}}\biggr) .
\end{align}

Furthermore, by shifting the cell indices, we use the identities $\sum_{\bm \zeta}f(r_{k\ell}^{\bm\zeta\bm0}) = 
\sum_{\bm \zeta}f(r_{k\ell}^{\bm0,-\bm\zeta}) = \sum_{\bm \xi}f(r_{k\ell}^{\bm0,\bm\xi})$, 
and $\sum_{\bm \zeta}f(r_{k\ell}^{\bm\zeta\bm0})e^{-i\bm k_E \cdot \bm R^\zeta} =\sum_{\bm \zeta}f(r_{k\ell}^{\bm0,-\bm\zeta})e^{-i\bm k_E \cdot \bm R^\zeta} = \sum_{\bm \xi}f(r_{k\ell}^{\bm0\bm\xi})e^{i\bm k_E \cdot \bm R^\xi}$. 
Then the expression can be written as
\begin{align} 
\lambda_E 
& =
\sum_{\langle k,\ell \rangle;\,k<\ell}
\sum_{\bm\xi}\biggl(
\biggl(
\bigg(\frac{d^2 U(r_{k\ell}^{\bm{0}\bm{\xi}})}{d r_{k\ell}^{\bm{0}\bm{\xi}\,2}} -   \frac{d U(r_{k\ell}^{\bm{0}\bm{\xi}})}{d r_{k\ell}^{\bm{0}\bm{\xi}}}  \frac{1}{r_{k\ell}^{\bm{0}\bm{\xi}}}  \biggr)
(\bm{n}_{k\ell}^{\bm{0}\bm{\xi}}\cdot \bm{\hat{\varepsilon}}_{k})(\bm{n}_{k\ell}^{\bm{0}\bm{\xi}}\cdot \bm{\hat{\varepsilon}}_{k})
+ 
\bigg(\frac{d^2 U(r_{k\ell}^{\bm{0}\bm{\xi}})}{d r_{k\ell}^{\bm{0}\bm{\xi}\,2}} -   \frac{d U(r_{k\ell}^{\bm{0}\bm{\xi}})}{d r_{k\ell}^{\bm{0}\bm{\xi}}}  \frac{1}{r_{k\ell}^{\bm{0}\bm{\xi}}}  \biggr)(\bm{n}_{k\ell}^{\bm{0}\bm{\xi}}\cdot \bm{\hat{\varepsilon}}_{\ell})(\bm{n}_{k\ell}^{\bm{0}\bm{\xi}}\cdot \bm{\hat{\varepsilon}}_{\ell})
\notag\\ & \quad 
- 
\bigg(\frac{d^2 U(r_{k\ell}^{\bm{0}\bm{\xi}})}{d r_{k\ell}^{\bm{0}\bm{\xi}\,2}} -   \frac{d U(r_{k\ell}^{\bm{0}\bm{\xi}})}{d r_{k\ell}^{\bm{0}\bm{\xi}}}  \frac{1}{r_{k\ell}^{\bm{0}\bm{\xi}}}  \biggr)(\bm{n}_{k\ell}^{\bm{0}\bm{\xi}}\cdot \bm{\hat{\varepsilon}}_{k})(\bm{n}_{k\ell}^{\bm{0}\bm{\xi}}\cdot \bm{\hat{\varepsilon}}_{\ell})
e^{-i\bm{k}_E\cdot\bm{R}^{\bm\xi}}
- \bigg(\frac{d^2 U(r_{k\ell}^{\bm{0}\bm{\xi}})}{d r_{k\ell}^{\bm{0}\bm{\xi}\,2}} -   \frac{d U(r_{k\ell}^{\bm{0}\bm{\xi}})}{d r_{k\ell}^{\bm{0}\bm{\xi}}}  \frac{1}{r_{k\ell}^{\bm{0}\bm{\xi}}}  \biggr)(\bm{n}_{k\ell}^{\bm{0}\bm{\xi}}\cdot \bm{\hat{\varepsilon}}_{\ell})(\bm{n}_{k\ell}^{\bm{0}\bm{\xi}}\cdot \bm{\hat{\varepsilon}}_{k})
e^{i\bm{k}_E\cdot\bm{R}^{\bm\xi}}
\notag \\ & \quad 
+ \frac{d U(r_{k\ell}^{\bm{0}\bm{\xi}})}{d r_{k\ell}^{\bm{0}\bm{\xi}}}  \frac{1}{r_{k\ell}^{\bm{0}\bm{\xi}}} (\bm{\hat{\varepsilon}}_{k}\cdot \bm{\hat{\varepsilon}}_{k})
+  \frac{d U(r_{k\ell}^{\bm{0}\bm{\xi}})}{d r_{k\ell}^{\bm{0},\bm{\xi}}}  \frac{1}{r_{k\ell}^{\bm{0},\bm{\xi}}} (\bm{\hat{\varepsilon}}_{\ell}\cdot \bm{\hat{\varepsilon}}_{\ell}) 
    -  \frac{d U(r_{k\ell}^{\bm{0}\bm{\xi}})}{d r_{k\ell}^{\bm{0}\bm{\xi}}}  \frac{1}{r_{k\ell}^{\bm{0}\bm{\xi}}}  (\bm{\hat{\varepsilon}}_{k}\cdot \bm{\hat{\varepsilon}}_{\ell})e^{-i\bm{k}_E\cdot\bm{R}^{\bm\xi}}
    -   \frac{d U(r_{k\ell}^{\bm{0}\bm{\xi}})}{d r_{k\ell}^{\bm{0},\bm{\xi}}}  \frac{1}{r_{k\ell}^{\bm{0},\bm{\xi}}} (\bm{\hat{\varepsilon}}_{\ell}\cdot \bm{\hat{\varepsilon}}_{k})e^{i\bm{k}_E\cdot\bm{R}^{\bm\xi}}\biggr) .
\end{align}

We next introduce a compact notation for the contact sum.
For a fixed representative particle pair satisfying $k<\ell$, the
relative cell index $\bm{\xi}$ specifies a contact between particle
$k$ in the reference cell and particle $\ell$ in the cell $\bm{\xi}$,
whenever such a contact exists.
The combined sum over all such contacts is denoted by
$
\sum_{\langle k,\ell\rangle_c}
:=
\sum_{\langle k,\ell\rangle;\,k<\ell}
\sum_{\bm{\xi}} .
$
We also define the cell displacement associated with each contact by
$
\bm L_{k\ell}:=-\bm R^{\bm\xi}.
$
For notational simplicity, we suppress the explicit cell indices and
introduce the abbreviations
$
r_{k\ell}:=r_{k\ell}^{\bm0\bm\xi}$, $
\bm n_{k\ell}:=\bm n_{k\ell}^{\bm0\bm\xi}$, $
U'_{k\ell}
:=
dU(r_{k\ell}^{\bm0\bm\xi})/
dr_{k\ell}^{\bm0\bm\xi} $,  and $
U''_{k\ell}
:=
d^2U(r_{k\ell}^{\bm0\bm\xi})/
d(r_{k\ell}^{\bm0\bm\xi})^2.
$
For the harmonic repulsive potential used in this study,
$
U''_{k\ell}=\kappa,
$
for every contacting pair.
With these definitions, the Rayleigh quotient can be written as
    \begin{align}
 \lambda_E
    & = 
    \sum_{\langle k, \ell \rangle_c }
    \Bigl(\Big(U_{k\ell}^{\prime \prime  } -\frac{U_{k\ell}^{\prime}}{r_{k\ell}}\Big)\bigl( (\bm{n}_{k\ell}\cdot \bm{\hat{\varepsilon}}_{k})(\bm{n}_{k\ell}\cdot \bm{\hat{\varepsilon}}_{k})+ (\bm{n}_{k\ell}\cdot \bm{\hat{\varepsilon}}_{\ell})(\bm{n}_{k\ell}\cdot \bm{\hat{\varepsilon}}_{\ell}) 
    - (\bm{n}_{k\ell}\cdot \bm{\hat{\varepsilon}}_{k})(\bm{n}_{k\ell}\cdot \bm{\hat{\varepsilon}}_{\ell})e^{i\bm{k}_E\cdot \bm{L}_{k\ell}}- (\bm{n}_{k\ell}\cdot \bm{\hat{\varepsilon}}_{\ell})(\bm{n}_{k\ell}\cdot \bm{\hat{\varepsilon}}_{k})e^{-i\bm{k}_E\cdot \bm{L}_{k\ell}})\bigr) \notag \\ 
    & \quad + \frac{U^{\prime}_{k\ell}}{r_{k\ell}} \big( (\bm{\hat{\varepsilon}}_{k}\cdot \bm{\hat{\varepsilon}}_{k}) +  (\bm{\hat{\varepsilon}}_{\ell}\cdot \bm{\hat{\varepsilon}}_{\ell}) 
    -  (\bm{\hat{\varepsilon}}_{k}\cdot \bm{\hat{\varepsilon}}_{\ell})e^{i\bm{k}_E\cdot \bm{L}_{k\ell}}- (\bm{\hat{\varepsilon}}_{\ell}\cdot \bm{\hat{\varepsilon}}_{k})e^{-i\bm{k}_E\cdot \bm{L}_{k\ell}}\big) \Bigr) .
\end{align}
For the edge wave vector, it is useful to introduce $\Delta_{k\ell,x} := L_{kl,x} /L =   - \xi_x$.
Since $\bm k_E=(\pi/L)\bm e_x$, we have
$e^{i\bm k_E\cdot\bm L_{k\ell}}
=
(-1)^{\Delta_{k\ell,x}}$.
Therefore,
 \begin{align}
 \lambda_E
    & = 
    \sum_{\langle k, \ell \rangle_c }
    \Bigl(\Big(U_{k\ell}^{\prime \prime  } -\frac{U_{k\ell}^{\prime}}{r_{k\ell}}\Big)\bigl( (\bm{n}_{k\ell}\cdot \bm{\hat{\varepsilon}}_{k})(\bm{n}_{k\ell}\cdot \bm{\hat{\varepsilon}}_{k})+ (\bm{n}_{k\ell}\cdot \bm{\hat{\varepsilon}}_{\ell})(\bm{n}_{k\ell}\cdot \bm{\hat{\varepsilon}}_{\ell}) 
    - 2(\bm{n}_{k\ell}\cdot \bm{\hat{\varepsilon}}_{k})(\bm{n}_{k\ell}\cdot \bm{\hat{\varepsilon}}_{\ell})(-1)^{\Delta_{k\ell,x}}\bigr) \notag \\ 
    & \quad + \frac{U^{\prime}_{k\ell}}{r_{k\ell}} \big( (\bm{\hat{\varepsilon}}_{k}\cdot \bm{\hat{\varepsilon}}_{k}) +  (\bm{\hat{\varepsilon}}_{\ell}\cdot \bm{\hat{\varepsilon}}_{\ell}) 
    -  2(\bm{\hat{\varepsilon}}_{k}\cdot \bm{\hat{\varepsilon}}_{\ell})(-1)^{\Delta_{k\ell,x}}\big) \Bigr) .
\end{align}

The nonlinear coefficient $\chi_E$ is obtained by projecting the fourth-order derivative tensor of the potential energy onto the eigenmode at $\bm{k}_E$.
Before carrying out the contractions, Eqs.~\eqref{eq:chiE_split}, \eqref{eq:chiE_in}, and \eqref{eq:chiE_inter} can be rewritten as
\begin{align}
    \chi_E & = 
    \sum_{ i ,j ,p,q}\sum_{\bm{\Delta}_{\bm{\mu\nu}},\bm{\Delta}_{\bm{\mu\zeta}},\bm{\Delta}_{\bm{\mu\xi}}}\sum_{\langle\langle\bm{\zeta,\xi}\rangle\rangle}\sum_{\langle k,\ell\rangle} \sum_{a,b,c,d}\biggl( \Bigl(  15 \frac{d^2 U(r_{k\ell}^{\bm{\zeta}\bm{\xi}})}{d r_{k\ell}^{\bm{\zeta}\bm{\xi}\,2}}\frac{1}{r_{k\ell}^{\bm{\zeta}\bm{\xi}\,2}} -15 \frac{d U(r_{k\ell}^{\bm{\zeta}\bm{\xi}})}{d r_{k\ell}^{\bm{\zeta}\bm{\xi}}} \frac{1}{r_{k\ell}^{\bm{\zeta}\bm{\xi}\,3}} \Bigr)n_{k\ell,a}^{\bm{\zeta}\bm{\xi}}n_{k\ell,b}^{\bm{\zeta}\bm{\xi}}n_{k\ell,c}^{\bm{\zeta}\bm{\xi}}n_{k\ell,d}^{\bm{\zeta}\bm{\xi}} \notag
      \\&
      + \Bigl( - 3\frac{d^2 U(r_{k\ell}^{\bm{\zeta}\bm{\xi}})}{d r_{k\ell}^{\bm{\zeta}\bm{\xi}\,2}}\frac{1}{r_{k\ell}^{\bm{\zeta}\bm{\xi}\,2}} + 3\frac{d U(r_{k\ell}^{\bm{\zeta}\bm{\xi}})}{d r_{k\ell}^{\bm{\zeta}\bm{\xi}}} \frac{1}{r_{k\ell}^{\bm{\zeta}\bm{\xi}\,3}} \Bigr)(\delta_{ab}n_{k\ell,c}^{\bm{\zeta}\bm{\xi}}n_{k\ell,d}^{\bm{\zeta}\bm{\xi}} +\delta_{bc}n_{k\ell,a}^{\bm{\zeta}\bm{\xi}}n_{k\ell,d}^{\bm{\zeta}\bm{\xi}} + \delta_{ca}n_{k\ell,b}^{\bm{\zeta}\bm{\xi}}n_{k\ell,d}^{\bm{\zeta}\bm{\xi}}
    + \delta_{da}n_{k\ell,b}^{\bm{\zeta}\bm{\xi}} n_{k\ell,c}^{\bm{\zeta}\bm{\xi}} + \delta_{db}n_{k\ell,c}^{\bm{\zeta}\bm{\xi}} n_{k\ell,a}^{\bm{\zeta}\bm{\xi}} + \delta_{dc}n_{k\ell,a} ^{\bm{\zeta}\bm{\xi}}n_{k\ell,b}^{\bm{\zeta}\bm{\xi}})\notag
    \\& + \Bigl(\frac{d^2 U(r_{k\ell}^{\bm{\zeta}\bm{\xi}})}{d r_{k\ell}^{\bm{\zeta}\bm{\xi}\,2}}\frac{1}{r_{k\ell}^{\bm{\zeta}\bm{\xi}\,2}}  -\frac{d U(r_{k\ell}^{\bm{\zeta}\bm{\xi}})}{d r_{k\ell}^{\bm{\zeta}\bm{\xi}}} \frac{1}{r_{k\ell}^{\bm{\zeta}\bm{\xi}\,3}} \Bigr)
    (\delta_{ab}\delta_{dc} + \delta_{ac}\delta_{db} + \delta_{ad}\delta_{bc})\biggr)
      \delta_{k\ell,\bm{\zeta}\bm{\xi}}^{i,\bm{0}}\delta_{k\ell,\bm{\zeta}\bm{\xi}}^{j,-\bm{\Delta}_{\bm{\mu\nu}}} \delta_{k\ell,\bm{\zeta}\bm{\xi}}^{p,-\bm{\Delta}_{\bm{\mu\zeta}}} \delta_{k\ell,\bm{\zeta}\bm{\xi}}^{q,-\bm{\Delta}_{\bm{\mu\xi}}}
      e^{i\bm{k}_E\cdot\bm{L}_{\bm{\Delta}_{\bm{\mu\nu}}}} e^{i\bm{k}_E\cdot\bm{L}_{\bm{\Delta}_{\bm{\mu\zeta}}}} e^{i\bm{k}_E\cdot\bm{L}_{\bm{\Delta}_{\bm{\mu\xi}}}} \hat{\varepsilon}_{i,a} \hat{\varepsilon}_{j,b} \hat{\varepsilon}_{p,c} \hat{\varepsilon}_{q,d} .
      \end{align}
      Using the same contact sum notation as for $\lambda_E$, we carry out the contractions and obtain
      \begin{align}
     \chi_E
      & = 
    \sum_{\langle k, \ell \rangle_c }
    \biggl(\Bigl(  15 \frac{U^{\prime \prime }_{k\ell} }{r^2_{k\ell}} -15 \frac{U^{\prime }_{k\ell} }{r^3_{k\ell}}  \Bigr)
      \bigl(
      (\bm{n}_{k\ell}\cdot\hat{\bm{\varepsilon}}_{k})^4 + 6(\bm{n}_{k\ell}\cdot\hat{\bm{\varepsilon}}_{\ell})^2(\bm{n}_{k\ell}\cdot\hat{\bm{\varepsilon}}_{k})^2 \notag
      \\
      & \quad 
      -4  (\bm{n}_{k\ell}\cdot\hat{\bm{\varepsilon}}_{\ell})(\bm{n}_{k\ell}\cdot\hat{\bm{\varepsilon}}_{k})^3 (-1)^{\Delta_{k\ell,x}}
      -4 (\bm{n}_{k\ell}\cdot\hat{\bm{\varepsilon}}_{\ell})^3 (\bm{n}_{k\ell}\cdot\hat{\bm{\varepsilon}}_{k})(-1)^{\Delta_{k\ell,x}}
      + (\bm{n}_{k\ell}\cdot\hat{\bm{\varepsilon}}_{\ell})^4 
      \bigr) \notag
      \\
      &  + \Bigl( - 3\frac{U^{\prime \prime }_{k\ell} }{r^2_{k\ell}} + 3\frac{U^{ \prime }_{k\ell} }{r^3_{k\ell}}\Bigr)6\Bigl(
      (\hat{\bm{\varepsilon}}_k \cdot \hat{\bm{\varepsilon}}_k)(
      \bm{n}_{k\ell}\cdot\hat{\bm{\varepsilon}}_k)^2+(\hat{\bm{\varepsilon}}_\ell \cdot \hat{\bm{\varepsilon}}_\ell)(
      \bm{n}_{k\ell}\cdot\hat{\bm{\varepsilon}}_k)^2
      -2(\hat{\bm{\varepsilon}}_k \cdot \hat{\bm{\varepsilon}}_\ell)(
    \bm{n}_{k\ell}\cdot\hat{\bm{\varepsilon}}_k)^2(-1)^{\Delta_{k\ell,x}}\notag
      \\ & \quad
      -2(\hat{\bm{\varepsilon}}_k \cdot \hat{\bm{\varepsilon}}_k)(
      \bm{n}_{k\ell}\cdot\hat{\bm{\varepsilon}}_k)(
      \bm{n}_{k\ell}\cdot\hat{\bm{\varepsilon}}_\ell)(-1)^{\Delta_{k\ell,x}}
      -2(\hat{\bm{\varepsilon}}_\ell \cdot \hat{\bm{\varepsilon}}_\ell)(
      \bm{n}_{k\ell}\cdot\hat{\bm{\varepsilon}}_k)(
      \bm{n}_{k\ell}\cdot\hat{\bm{\varepsilon}}_\ell)(-1)^{\Delta_{k\ell,x}}
      +4(\hat{\bm{\varepsilon}}_k \cdot \hat{\bm{\varepsilon}}_\ell)(
      \bm{n}_{k\ell}\cdot\hat{\bm{\varepsilon}}_k)(
      \bm{n}_{k\ell}\cdot\hat{\bm{\varepsilon}}_\ell) \notag
      \\ & \quad
      +(\hat{\bm{\varepsilon}}_k\cdot \hat{\bm{\varepsilon}}_k)(
      \bm{n}_{k\ell}\cdot\hat{\bm{\varepsilon}}_\ell)^2+(\hat{\bm{\varepsilon}}_\ell \cdot \hat{\bm{\varepsilon}}_\ell)(
      \bm{n}_{k\ell}\cdot\hat{\bm{\varepsilon}}_\ell)^2
      -2(\hat{\bm{\varepsilon}}_k \cdot \hat{\bm{\varepsilon}}_\ell)(
      \bm{n}_{k\ell}\cdot\hat{\bm{\varepsilon}}_\ell)^2(-1)^{\Delta_{k\ell,x}}
      \Bigr) \notag\\
    & + \Bigl(\frac{U^{\prime \prime }_{k\ell} }{r^2_{k\ell}}  -\frac{U^{ \prime }_{k\ell} }{r^3_{k\ell}}\Bigr)3
    \bigl(
    (\hat{\bm{\varepsilon}}_k \cdot \hat{\bm{\varepsilon}}_k)^2
      +2(\hat{\bm{\varepsilon}}_\ell\cdot\hat{\bm{\varepsilon}}_\ell)(\hat{\bm{\varepsilon}}_k \cdot \hat{\bm{\varepsilon}}_k)
      -4(\hat{\bm{\varepsilon}}_k\cdot\hat{\bm{\varepsilon}}_\ell)(\hat{\bm{\varepsilon}}_k \cdot \hat{\bm{\varepsilon}}_k)(-1)^{\Delta_{k\ell,x}}
      +4(\hat{\bm{\varepsilon}}_k \cdot \hat{\bm{\varepsilon}}_\ell)^2
      +(\hat{\bm{\varepsilon}}_\ell \cdot \hat{\bm{\varepsilon}}_\ell)^2
      -4(\hat{\bm{\varepsilon}}_k\cdot\hat{\bm{\varepsilon}}_\ell)(\hat{\bm{\varepsilon}}_\ell \cdot \hat{\bm{\varepsilon}}_\ell)(-1)^{\Delta_{k\ell,x}}
    \bigr) \biggr) .
\end{align}

\twocolumn

\section{Multiple Unstable Eigenmodes}\label{Ap:multi}
\begin{figure}[h]
    \centering
        \centering
        \includegraphics[width=0.3\textwidth,trim=1 29 355 1, clip]{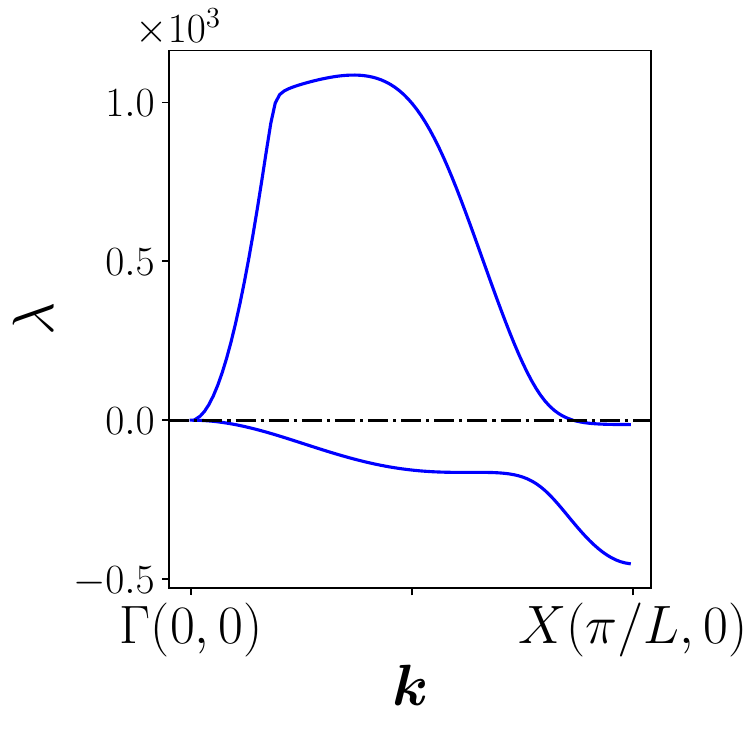}    
    \caption{
    An example of the dispersion relation with two negative eigenvalues for a system with particle number $N = 128$ and pressure $P = 5 \times 10^{-4} \kappa$.
    }
    \label{fig:dis_Ap}
\end{figure}
In this section, we consider cases where multiple branches with negative eigenvalues appear at a certain wavenumber in the dispersion relation (Fig.~\ref{fig:dis_Ap}). In such cases, mixing of multiple eigenmodes occurs.
In this case, eqn.~\eqref{eq:Fourier_Inv_transf_Lat} can be rewritten as 
\begin{equation}
    \delta \bm{r}^{\bm \mu} \approx A_{1,E}(t)\hat{\bm{\varepsilon}}_1(\bm{k}_E)(-1)^{\mu_\parallel} +  A_{2,E}(t)\hat{\bm{\varepsilon}}_2(\bm{k}_E)(-1)^{\mu_\parallel} .
\end{equation}
Here, $A_{1,E}$ and $A_{2,E}$ denote the amplitudes associated with $\hat{\bm{\varepsilon}}_1(\bm{k}_E) $ and $\hat{\bm{\varepsilon}}_2(\bm{k}_E)$, respectively.
Following the same procedure as in the single-mode case, we obtain
\begin{align}
    \eta \dot{A}_{1,E} & = \lambda_{1,E}^\prime A_{1,E}   -  \frac{1}{6 }(\chi_{1:111,E}A_{1,E}^3 
     +3\chi_{1:221,E}A_{1,E}A_{2,E}^2  \notag \\
    & \quad
    + 3\chi_{1:112,E}A_{2,E}A_{1,E}^2 + \chi_{1:222,E}A_{2,E}^3 ) \label{eq:keke_3}
    \\
    \eta \dot{A}_{2,E} & = \lambda_{2,E}^\prime A_{2,E}  -\frac{1}{6 }(\chi_{2:111,E}A_{1,E}^3 + 3\chi_{2:221,E}A_{1,E}A_{2,E}^2\notag  \\
    & \quad + 3\chi_{2:112,E}A_{2,E}A_{1,E}^2 + \chi_{2:222,E}A_{2,E}^3 ), \label{eq:keke_4}
\end{align}
where
$\lambda^\prime_{n,E} = -\lambda_n(\bm k_E)$, and
\begin{align}
\chi_{n:mpq,E}
&:=
\sum_{
\bm\Delta_{\bm\mu\bm\nu},\bm\Delta_{\bm\mu\bm\zeta},\bm\Delta_{\bm\mu\bm\xi}\in\mathcal N}
\sum_{i,j,k,\ell}
\hat{\bm\varepsilon}_{n,i}(\bm k_E)
\mathsf M_{ijk\ell}^{\bm0,-\bm\Delta_{\bm\mu\bm\nu},-\bm\Delta_{\bm\mu\bm\zeta},-\bm\Delta_{\bm\mu\bm\xi}}
\notag\\
&\quad\times
\hat{\bm\varepsilon}_{m,j}(\bm k_E)
\hat{\bm\varepsilon}_{p,k}(\bm k_E)
\hat{\bm\varepsilon}_{q,\ell}(\bm k_E)
(-1)^{\bm e_\parallel\cdot\bm\Delta_{\bm\mu\bm\nu}}
(-1)^{\bm e_\parallel\cdot\bm\Delta_{\bm\mu\bm\zeta}}
(-1)^{\bm e_\parallel\cdot\bm\Delta_{\bm\mu\bm\xi}}.
\label{eq:chi_nmpq}
\end{align}
Thus, owing to the nonlinear terms, even when a perturbation is initially
applied along only one of the unstable eigenmodes, the other unstable
mode can be induced, resulting in mode-mode coupling.

\begin{figure*}[h]
\begin{minipage}[t]{0.48\textwidth}
    \centering
        \centering
        \includegraphics[width=\linewidth]{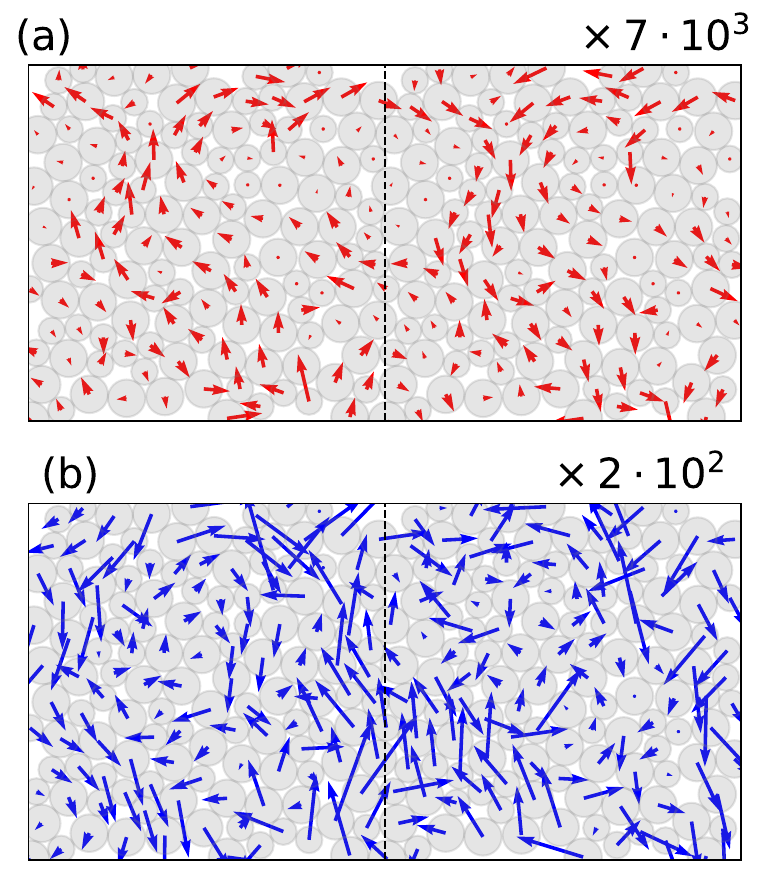}
    \caption{The system’s response to perturbations corresponding to the two negative eigenmodes at the X-point in Fig\ref{fig:dis_Ap}.
Panel (a) shows the deformation associated with the lowest eigenmode and its corresponding response in panel (b). The initial amplitude is \(A_{E,0}=10^{-4}\sqrt{N}\).
}
\label{fig:vec_Neg2_1}
\end{minipage}
%
\hfill
%
\begin{minipage}[t]{0.48\textwidth}
    \centering
        \centering
        \includegraphics[width=\linewidth]{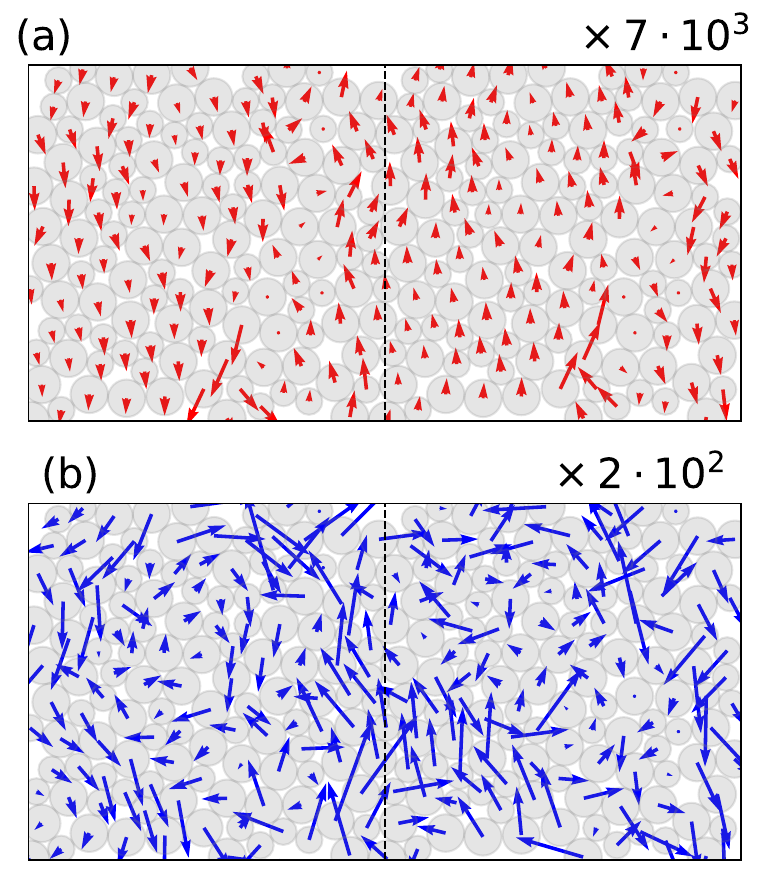}
    \caption{The system’s response to perturbations corresponding to the two negative eigenmodes at the X-point in Fig\ref{fig:dis_Ap}.
Panel (a) shows the deformation associated with the second-lowest eigenmode and its corresponding response in panel (b).
A mixed mode similar to that shown in Fig.~\ref{fig:vec_Neg2_1} also emerges in this response. The initial amplitude is \(A_{E,0}=10^{-4}\sqrt{N}\).
}
\label{fig:vec_Neg2_2}
\end{minipage}
\end{figure*}

To examine this mode coupling numerically, we separately apply
perturbations along the two unstable eigenmodes with the same initial
amplitude \(A_{E,0}\). Specifically, we consider the two initial
conditions
$
(A_{1,E}(0),A_{2,E}(0))=(A_{E,0},0)
$ (see Fig. \ref{fig:vec_Neg2_1} (a)), and $
(A_{1,E}(0),A_{2,E}(0))=(0,A_{E,0})
$ (see Fig. \ref{fig:vec_Neg2_2} (a)).
Figures~\ref{fig:vec_Neg2_1} (b) and~\ref{fig:vec_Neg2_2} (b)
show the corresponding responses, where the resulting displacement fields are mixed states of both unstable eigenmodes.
To quantify this mode mixing, we define
\begin{equation}
    C_{ij}
    =
    \frac{
    \left|
    \hat{\bm{\varepsilon}}_i(\bm k_E)\cdot\bm{u}^{(j)}
    \right|
    }{
    \left|\bm{u}^{(j)}\right|
    },
    \qquad
    W_{ij}=C_{ij}^{2},
\end{equation}
where $\bm{u}^{(i)}$ denotes the response displacement obtained when a perturbation is applied along $\hat{\bm{\varepsilon}}_i(\bm k_E)$.
For the cases shown in Figs.~\ref{fig:vec_Neg2_1} and~\ref{fig:vec_Neg2_2}, we obtain $W_{11} = 0.41$,$ W_{21} = 0.39 $,$ W_{12} = 0.43 $, and $ W_{22} = 0.37$. Thus, approximately $80\%$ of the squared norm of each relaxed
displacement field is contained in its projection onto the subspace
spanned by the two unstable eigenmodes.

These results demonstrate that the single-mode Landau equation is
not, in general, sufficient when multiple negative eigenvalues coexist
at the same wave vector. In such cases, mode-mode coupling, as described
by Eqs.~\eqref{eq:keke_3} and~\eqref{eq:keke_4}, must be taken into
account in describing the nonlinear response.

\section{System-size dependence of the theoretical prediction for $|\delta V_\infty|$}
\label{Ap:dV_Ndepend}

\begin{figure}[h]
    \centering
    \begin{subfigure}[b]{0.44\textwidth}
        \centering
        \begin{overpic}[width=\textwidth]{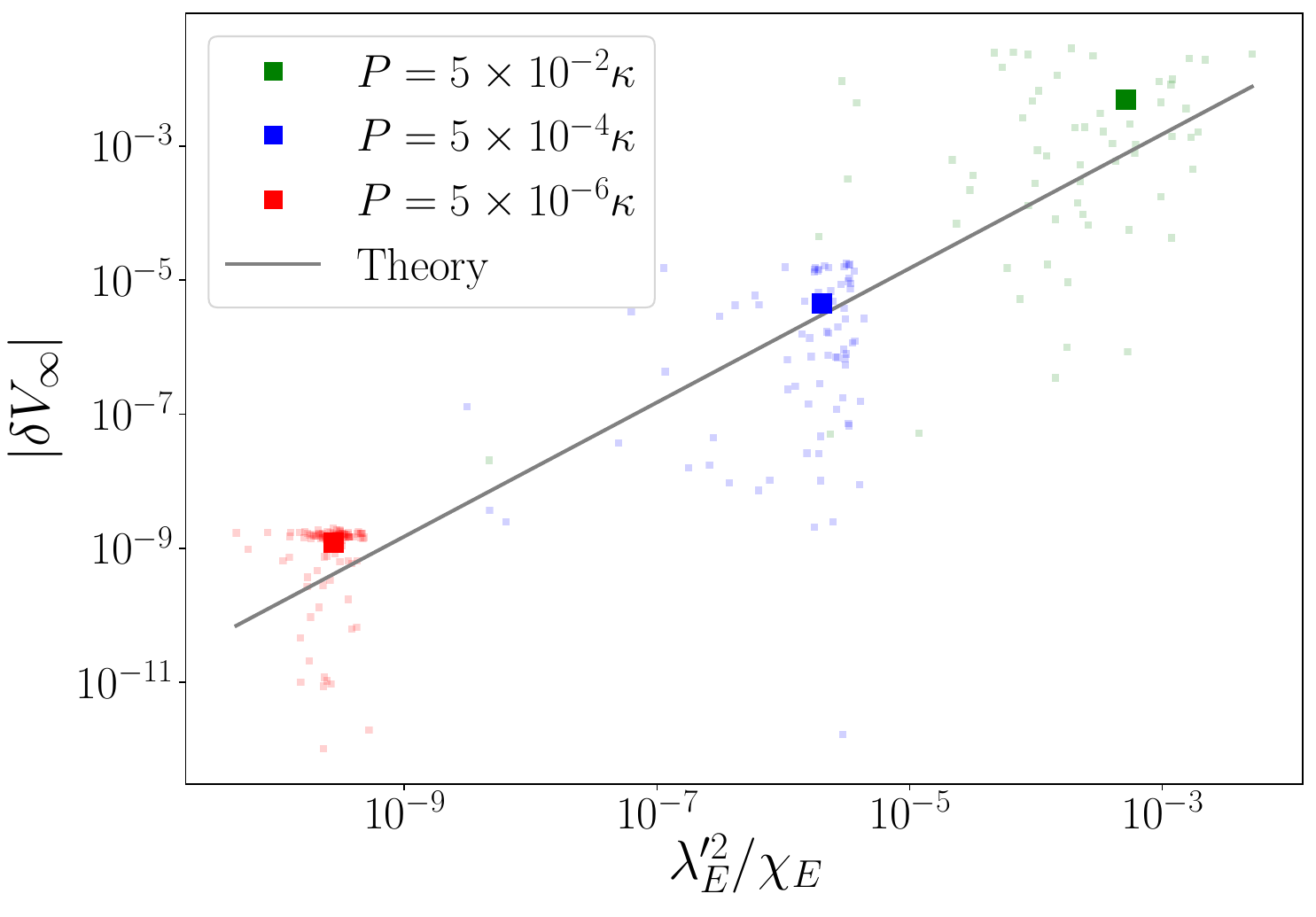}
            \put(10,70){\large\textbf{(a) $N=32$}}
        \end{overpic}       \label{fig:scatter_N_dependence_N32}
    \end{subfigure}
    \hfill
    \begin{subfigure}[b]{0.44\textwidth}
        \centering
        \begin{overpic}[width=\textwidth]{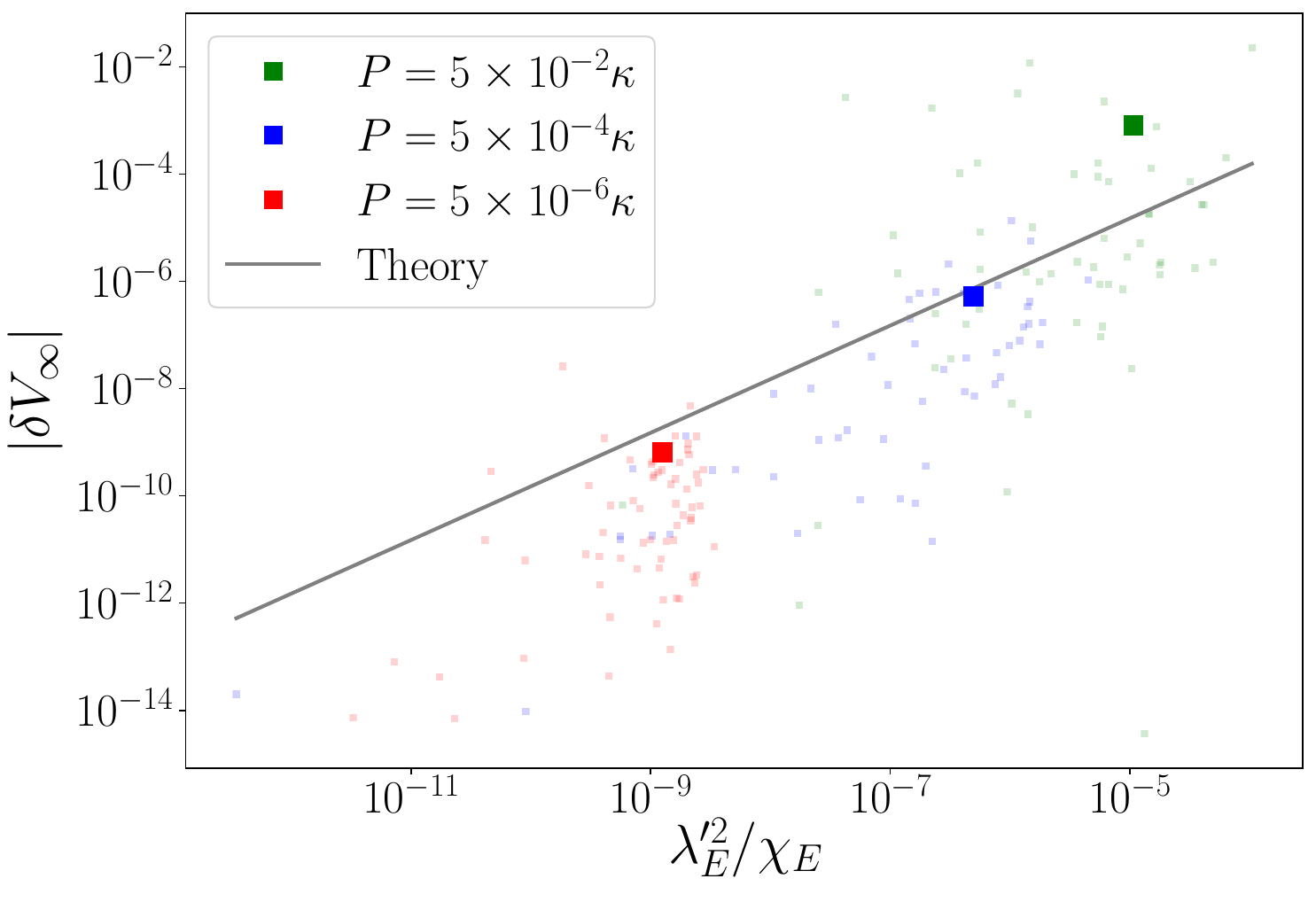}
            \put(10,70){\large\textbf{(b) $N=512$}}
        \end{overpic}     \label{fig:scatter_N_dependence_N512}
    \end{subfigure}
    \caption{
System-size dependence of the relation between $\lambda_E^{\prime 2}/\chi_E$ and $|\delta V_\infty|$.
Panels (a) and (b) show the results for $N=32$ and $N=512$, respectively.
The data are shown for pressures $P = 5\times 10^{-2}\kappa$ (green), $5\times 10^{-4}\kappa$ (blue), and $5\times 10^{-6}\kappa$ (red). 
Only samples with a negative edge-mode eigenvalue
\(\lambda_E<0\) are included. 
The small transparent points represent individual samples, while the large square markers represent arithmetic averages over these samples. 
The solid line, $y=3x/2$, represents the theoretical prediction given by eqn.~\eqref{eq:dU_LC}.
}
    \label{fig:scatter_N_dependence}
\end{figure}
In the main text, we examined the relation between 
$\lambda_E^{\prime 2}/\chi_E$ and the final potential-energy drop
$|\delta V_\infty|$ for systems with $N=128$ particles per cell. Figure~\ref{fig:scatter_N_dependence} presents the corresponding results
for $N=32$ and $N=512$.
For both system sizes, the data show a trend similar to that observed for
$N=128$ in the main text. 
Although broad sample-to-sample variations are present, the overall distribution is consistent with the statistical trend discussed in the main text.

\section{Notations used in this paper}\label{notations}

In this section, we summarize the notations used in this paper in Table~\ref{tab:notation}. 

\begin{table}[htb]
\centering
\caption{Notation and index conventions used throughout this paper.}
\label{tab:notation}
\begin{tabular}{ll}
\hline
Symbol / index & Meaning \\
\hline
$\bm{\mu,\nu,\zeta,\xi}$ & cell labels on $\mathbb{Z}^2$ \\
$i,j,k,\ell$ & particle labels within a cell \\
$a,b,c,d$ & Cartesian components ($x,y$) \\
$\bm{r}_i^{\bm\mu}$ & position of particle $i$ in cell $\mu$ \\
$\delta \bm{r}_i^{\bm \mu}$ & displacement from the reference configuration \\
$\bm{R}^{\bm\mu}$ & position of cell $\bm\mu$ \\
$\bm{k}$ & wave vector \\
\hline
\end{tabular}
\end{table}


\end{CJK} 
\end{document}